\documentclass{elsart}

\usepackage[square,comma]{natbib}
\usepackage{graphicx}
\usepackage{pxfonts}
\usepackage{lineno}

\usepackage{amssymb}

\journal{}

\begin{document}
\thispagestyle{empty}
\begin{Large}
\textbf{DEUTSCHES ELEKTRONEN-SYNCHROTRON}

\textbf{\large{Ein Forschungszentrum der
Helmholtz-Gemeinschaft}\\}
\end{Large}

DESY 10-033

March 2010

\begin{eqnarray}
\nonumber &&\cr \nonumber && \cr \nonumber &&\cr
\end{eqnarray}
\begin{eqnarray}
\nonumber
\end{eqnarray}
\begin{center}
\begin{Large}
\textbf{Scheme for generation of highly monochromatic X-rays from
a baseline XFEL  undulator}
\end{Large}
\begin{eqnarray}
\nonumber &&\cr \nonumber && \cr
\end{eqnarray}

\begin{large}
Gianluca Geloni,
\end{large}
\textsl{\\European XFEL GmbH, Hamburg}
\begin{large}

Vitali Kocharyan and Evgeni Saldin
\end{large}
\textsl{\\Deutsches Elektronen-Synchrotron DESY, Hamburg}
\begin{eqnarray}
\nonumber
\end{eqnarray}
\begin{eqnarray}
\nonumber
\end{eqnarray}
ISSN 0418-9833
\begin{eqnarray}
\nonumber
\end{eqnarray}
\begin{large}
\textbf{NOTKESTRASSE 85 - 22607 HAMBURG}
\end{large}
\end{center}
\clearpage
\newpage

\begin{frontmatter}



\title{Scheme for generation of highly monochromatic X-rays from a baseline
XFEL  undulator}


\author[XFEL]{Gianluca Geloni\thanksref{corr},}
\thanks[corr]{Corresponding Author. Tel: ++49 40 8998 5450. Fax: ++49 40 8998 1905. E-mail address: gianluca.geloni@xfel.eu}
\author[DESY]{Vitali Kocharyan,}
\author[DESY]{and Evgeni Saldin}

\address[XFEL]{European XFEL GmbH, Hamburg, Germany}
\address[DESY]{Deutsches Elektronen-Synchrotron (DESY), Hamburg,
Germany}

\begin{abstract}
One goal of XFEL facilities is the production of narrow bandwidth
X-ray radiation. The self-seeding scheme was proposed to obtain a
bandwidth narrower than that achievable with conventional X-ray
SASE FELs. A self-seeded FEL is composed of two undulators
separated by a monochromator and an electron beam bypass that must
compensate for the path delay of X-rays in the monochromator. This
leads to a long bypass, with a length in the order of $40-60$ m,
which requires modifications of the baseline undulator
configuration. As an attempt to get around this obstacle, together
with a study of the self-seeding scheme for the European XFEL,
here we propose a novel technique based on a pulse doubler
concept. Using a crystal monochromator installed within a short
magnetic chicane in the baseline undulator, it is possible to
decrease the bandwidth of the radiation well beyond the XFEL
design down to $10^{-5}$. The magnetic chicane can be installed
without any perturbation of the XFEL focusing structure, and does
not interfere with the baseline mode of operation. We present a
feasibility study and we make exemplifications with the parameters
of the SASE2 line of the European XFEL.
\end{abstract}

%
%

\end{frontmatter}



\section{\label{sec:intro} Introduction}

The quality of the output radiation of X-ray SASE FELs is far from
ideal because of the short longitudinal coherence length. This is
a consequence of the fact that the process of amplification in a
SASE FEL starts up from noise. The output radiation thus consists
of a number of independent wavepackets (or spikes). As a rule, the
length of each wavepacket is much shorter than the radiation pulse
length and there is no phase correlation between the wavepackets.
However, the improvement of the spectral brightness of an X-ray
SASE FEL is of great practical importance, and has always been one
of the goals of XFEL facilities. Due to the poor longitudinal
coherence of the output radiation of conventional XFELs
\cite{tdr-2006}-\cite{SPRIN}, an increase up to two order of
magnitude (i.e. up to full longitudinal coherence) is, in
principle, possible.


%

The self-seeding scheme \cite{SELF}, first proposed in the VUV and
soft X-ray region makes use of a grating monochromator, and allows
one to monochromatize radiation within an active
frequency-filtering process, where the bandwidth is narrowed
during, and not after the amplification process. The
implementation of such scheme for the Angstrom wavelength range
would take advantage of a crystal monochromator \cite{SXFE}. The
presence of the monochromator introduces a path delay with respect
to the straight path, which has to be compensated with the
introduction of a long electron beam bypass in the order of
$40-60$ m for the case of an XFEL.  As a result, this self-seeding
scheme is not compatible with the baseline XFEL design presented
in technical design reports \cite{tdr-2006,LCLS1}, and cannot be
implemented in the very initial stage of operation.

In this paper we propose a way to obtain a similar result without
interfering with the baseline mode of operation of the XFEL. We
will show how, based on a pulse doubler concept, one can use a
crystal monochromator installed within a short magnetic chicane in
the baseline undulator to decrease the bandwidth of radiation well
beyond the XFEL design, down to $10^{-5}$. The scheme can also
work in combination with a fresh bunch technique, both for short
($6$ fs) and long ($60$ fs) pulse mode of operation. In the
following section we will revisit the self-seeding scheme proposed
in \cite{SXFE} (which will be named here "single bunch
self-seeding scheme") for the case of beam parameters
experimentally demonstrated \cite{LCLS2,DING} at LCLS, and we will
discuss our novel method (which will be named here "double bunch
self-seeding scheme"). Further on, we will discuss a feasibility
study for our new double bunch self-seeding technique, which may
be used to enable the single bunch self-seeding technique as well.
We present studies both for short and long pulse mode of
operation.

%

\section{\label{sec:meth} Methods for controlling the X-ray pulse linewidth}

\subsection{Single bunch self-seeding scheme}

\begin{figure}[tb]
\includegraphics[width=1.0\textwidth]{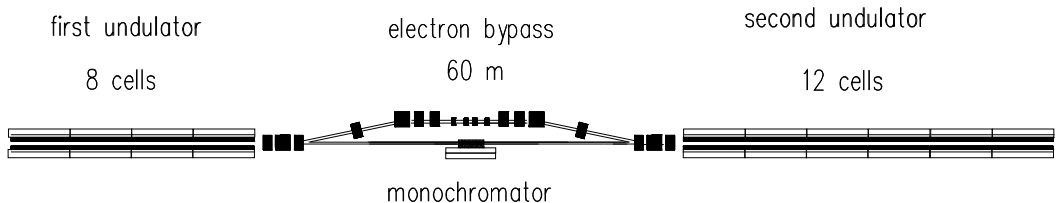}
\caption{Design of the undulator system for  narrow bandwidth mode
operation. The scheme is based on the use of single bunch self
seeded technique} \label{m99}
\end{figure}

A self-seeded FEL \cite{SELF}-\cite{SCOM} is composed of two
undulators separated by a monochromator and an electron beam
bypass, which washes out the microbunching and compensates for the
path delay of X-rays in the monochromator, without significantly
increasing the overall length of the bunch. Radiation in the
linear regime is filtered through the monochromator, as the
electron beam passes through the bypass. Further on, the
monochromatized radiation seeds the washed out electron beam in
the second undulator, and narrowband radiation, beyond the XFEL
design, is produced. A schematic of the setup for the European
XFEL is shown in Fig. \ref{m99}. In the original VUV-soft X-ray
case, a grating monochromator was proposed \cite{SELF}. In the
hard X-ray case a crystal monochromator should be used instead
\cite{SXFE}.

In the wavelength range around $0.1$ nm, radiation undergoes an
extra-path length in the order of at least a centimeter, due to
the minimal transverse displacement between crystals in the
monochromator. Such extra-path length is in the same order of that
foreseen for the VUV self-seeding option at FLASH \cite{STTF}. As
mentioned above, the bypass both compensates for this path delay
and washes out the electron beam microbunching.

\begin{figure}[tb]
\includegraphics[width=1.0\textwidth]{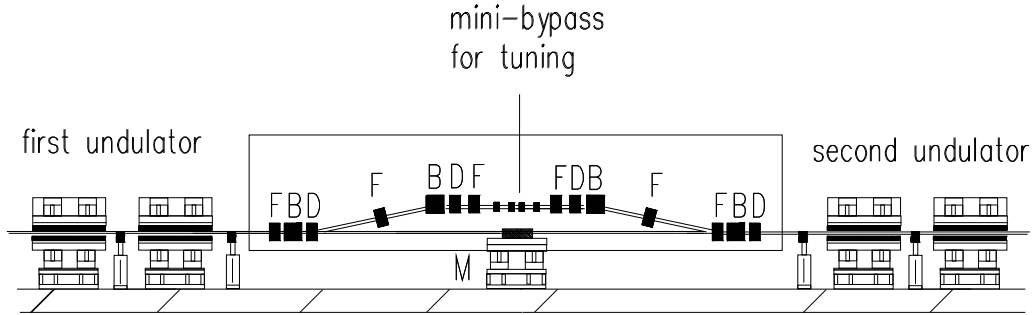}
\caption{Layout of the electron bypass. F - focusing quadrupole, D
- defocusing quadrupole, B - bypass bending magnets, M - crystal
monochromator.} \label{m100}
\end{figure}

For an energy spread of order $\Delta \gamma/\gamma \sim 0.01 \%$,
and wavelengths in the order of $0.1$ nm, the electron beam
microbunching is washed out already for an $R_{56}$ in the order
of a few microns. In the case of a four-magnet chicane with extra
path length $\delta L$ the dispersion is simply given by $R_{56} =
2 \delta L$. This result does not depend on the distance between
second and third magnet, which, due to technical reasons, can be
comparable with the distance between first and second magnet. In
our case, this distance is about $20$ m. Without a special
focusing system, the bypass should have a very large $R_{56}$ in
the order of a few centimeters. The bypass system should be nearly
dispersion free, with matched beta functions at the entrance and
at the exit, and should include the possibility of fine tuning the
extra path length. Such setup, adapted from \cite{STTF} for the
XFEL case, is schematically presented in Fig. \ref{m100}. Given
the bending angle of the bypass magnets of about $2$ degrees, the
total length of the bypass is about $60$ m.

\begin{figure}[tb]
\includegraphics[width=1.0\textwidth]{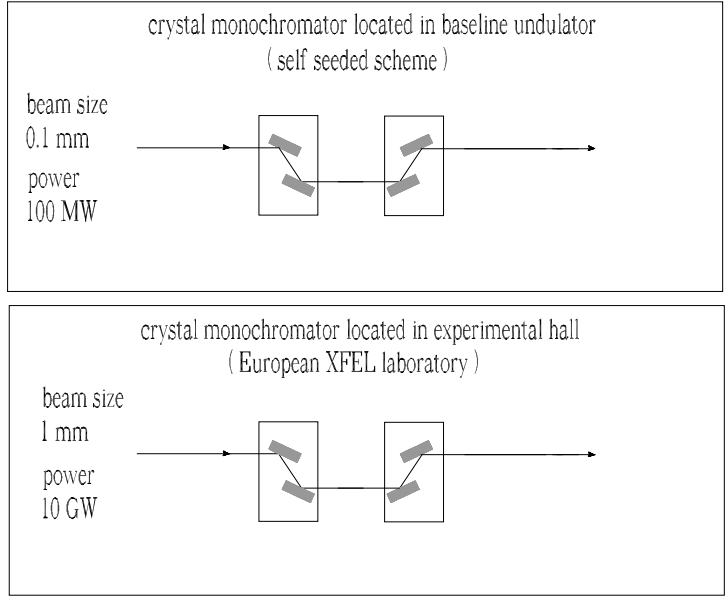}
\caption{An X-ray crystal monochromator will be required to reduce
the bandwidth of the FEL for various applications, in particular
for spectroscopy. A monochromator can be located in the baseline
undulator (self-seeding scheme, upper picture) or in the
experimental hall (bottom picture), with the self-seeding
monochromatization improving the spectral brightness of two orders
of magnitude. Comparing  both geometries for the European XFEL
case in relation with heat-loading problem, one finds that the
power density in the experimental hall is practically the same as
the power density in the self-seeding scheme. } \label{m65}
\end{figure}
\begin{figure}[tb]
\includegraphics[width=1.0\textwidth]{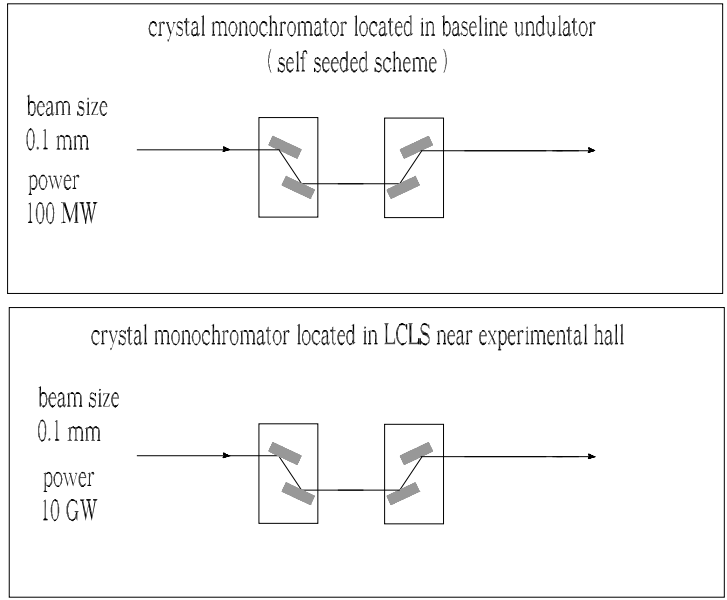}
\caption{Comparing  the same geometries in Fig. \ref{m65} for the
LCLS case in relation with heat-loading problems. The
monochromatization in the self-seeding scheme (upper picture) is
performed at much lower power density level with respect to
monochromatization in the near experimental hall (lower picture).}
\label{m75}
\end{figure}
\begin{figure}[tb]
\includegraphics[width=1.0\textwidth]{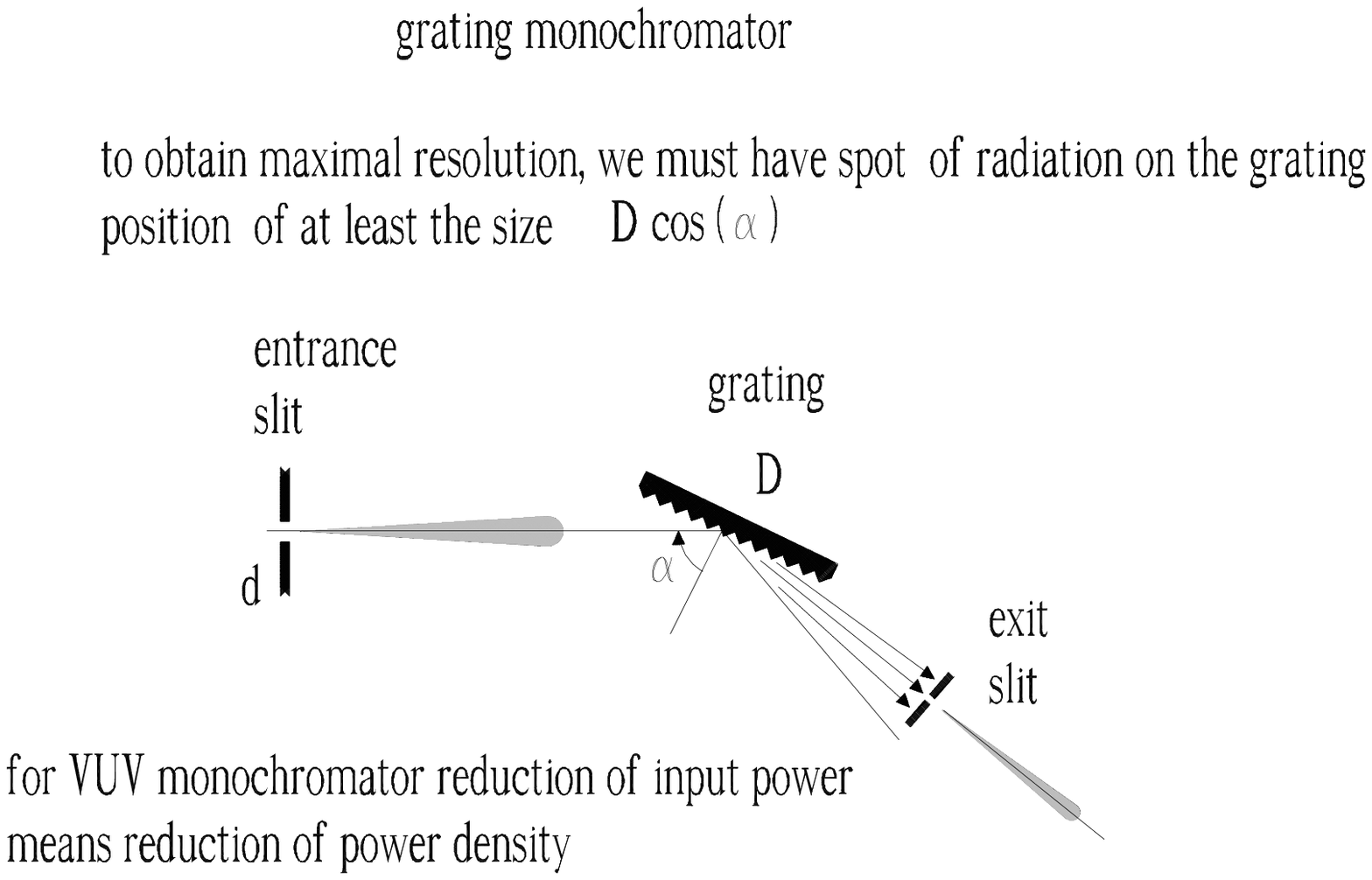}
\caption{Illustration of a grating monochromator principle. In the
VUV range, the heat-load (both power and power density) on the
grating monochromator is two or three orders of magnitude less
than on a monochromator installed in the experimental hall,
independently of the distance between undulator and experimental
hall.} \label{m55}
\end{figure}
An important point to be considered when studying the
implementation of the self-seeding scheme is that heat-loading
problems are not automatically avoided when a crystal
monochromator is used at the European XFEL. In fact, while the
power incident on the crystal is reduced of two orders of
magnitude, the beam area inside the undulator is smaller of two
orders of magnitude, as is illustrated in Fig. \ref{m65}, leading
to the same power density on the crystal. The situation changes
when one considers e.g. the case of the LCLS near experimental
hall, Fig. \ref{m75}. First, in general, at LCLS one deals with a
much lower repetition rate. Moreover, in the near hall the
radiation spot size is still comparable with the spot size inside
the undulator. Therefore, the power density on the crystal is
reduced of two orders of magnitude if the monochromator is placed
inside the undulator. Thus, heat-loading problems are completely
avoided using the self-seeding scheme. From this viewpoint, it
should also be noted how the situation differs in the VUV case,
where a grating monochromator is used as explained in \cite{SELF}.
In this case the resolution depends on the spot-size on the
grating, Fig. \ref{m55}. Once the resolution is fixed, the
spot-size on the grating is fixed too, independently of the
distance between undulator and experimental hall. As a result, in
the self-seeding scheme for the VUV range, heat loading issues are
relaxed of two or three orders of magnitude. Therefore, installing
a crystal monochromator in the undulator always leads to an
improvement in terms of spectral brightness, but advantages
concerning heat loading problems depend on the situation. There
are always advantages when considering the VUV case, but not
always when dealing with XFELs.

%



\subsection{Double bunch self-seeding scheme}

\begin{figure}[tb]
\includegraphics[width=1.0\textwidth]{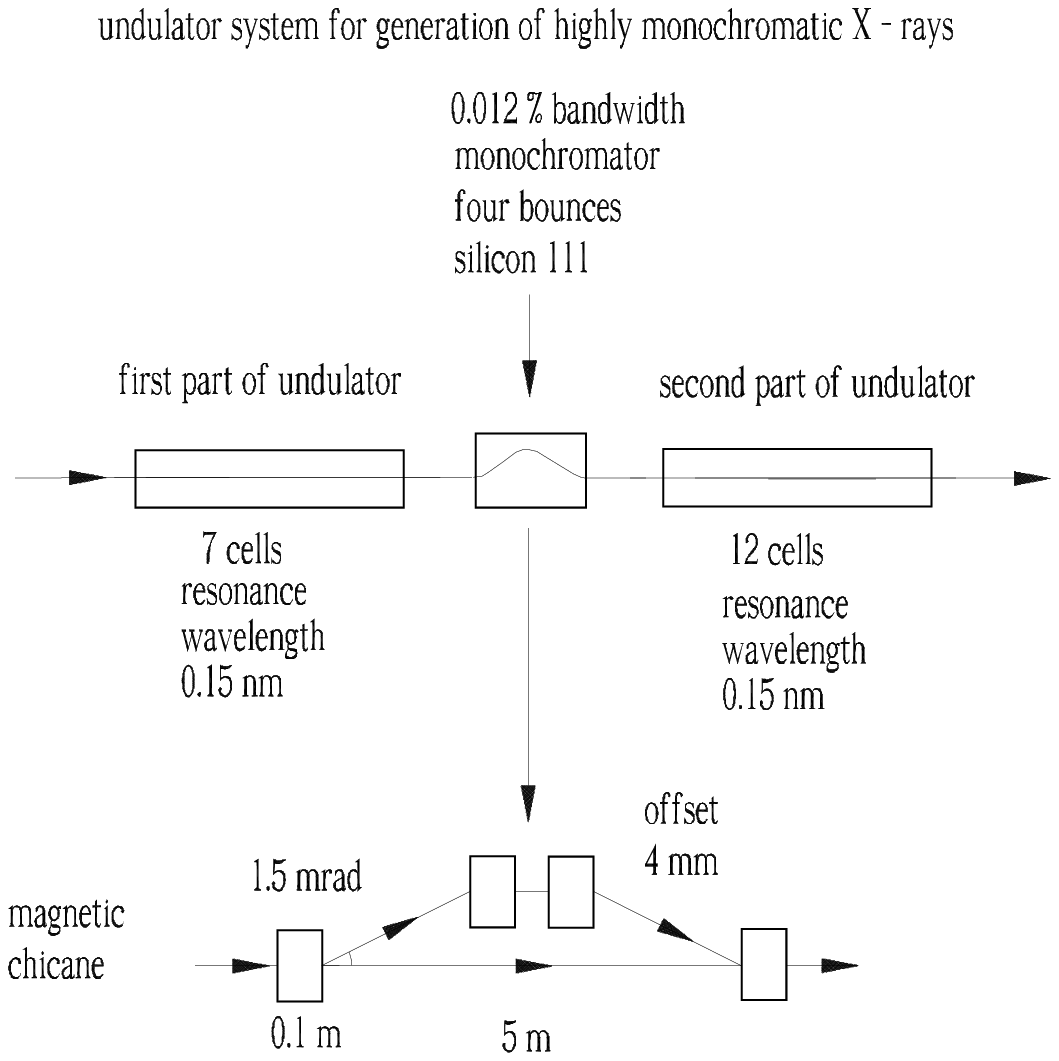}
\caption{Design of an undulator system for highly monochromatic
X-ray source} \label{m2}
\end{figure}
\begin{figure}[tb]
\includegraphics[width=1.0\textwidth]{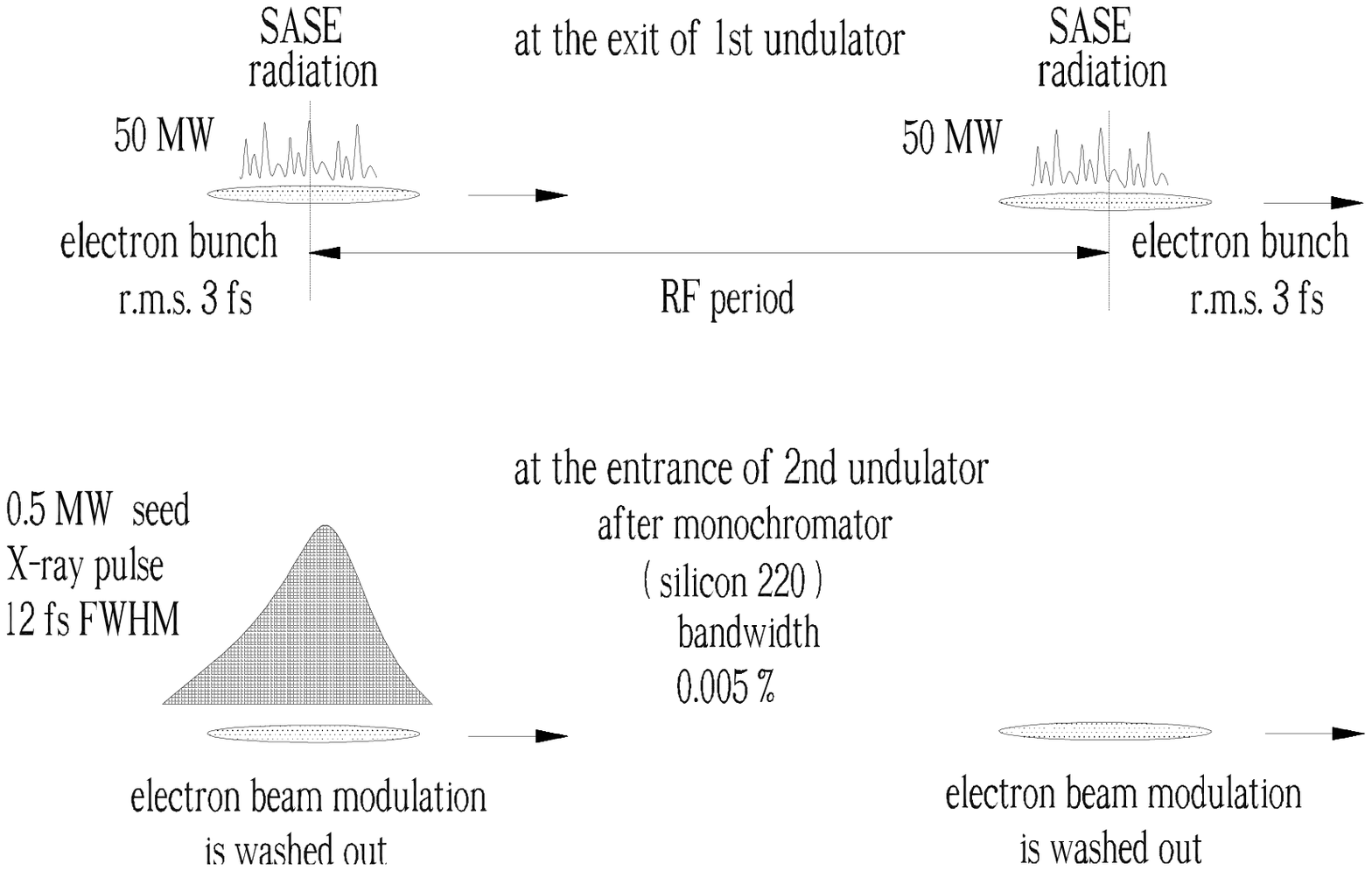}
\caption{Sketch of highly monochromatic X-ray pulse synthesization
through double electron bunch generation and spectral filtering.}
\label{m19}
\end{figure}
\begin{figure}[tb]
\includegraphics[width=1.0\textwidth]{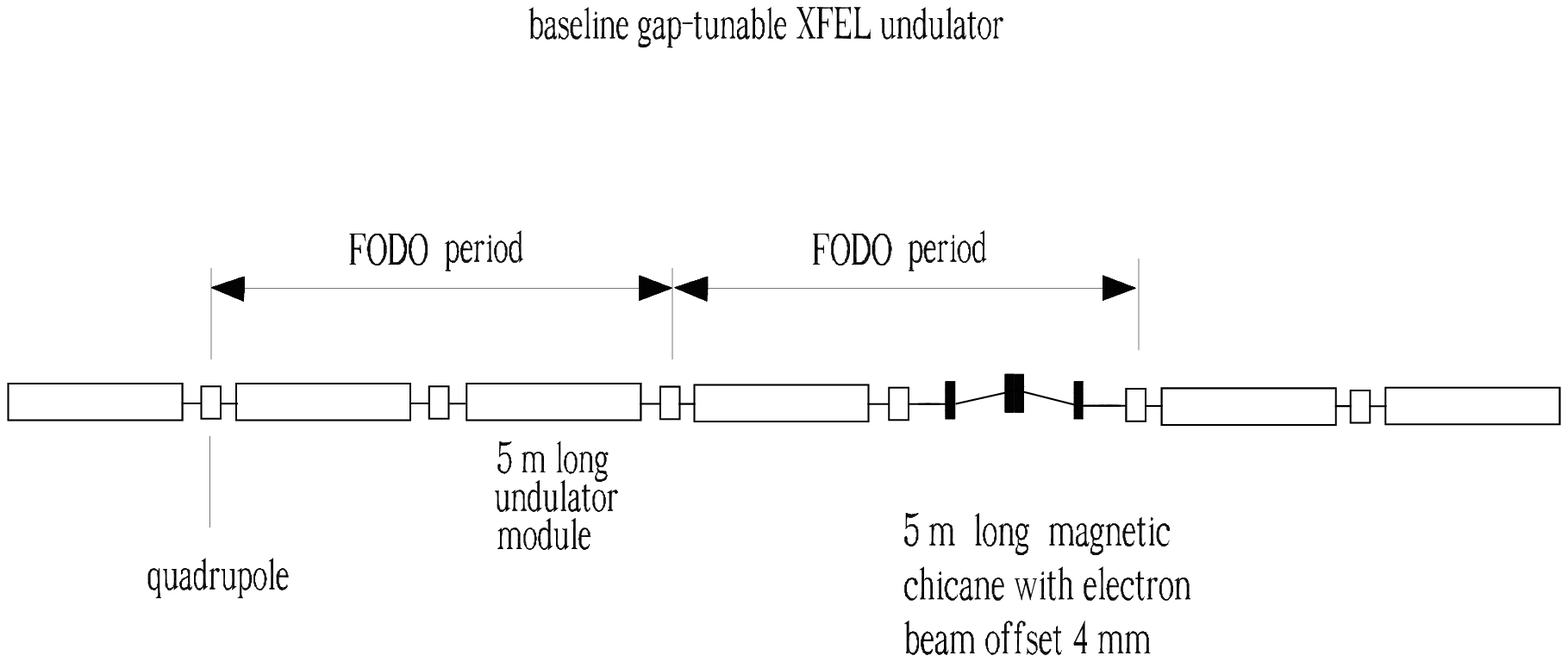}
\caption{Installation of the magnetic chicane in the baseline XFEL
undulator. The magnetic chicane absolves two tasks. First, it
allows for the installation of the monochromator. Second, the
strength of the magnetic chicane as a dispersive section is
sufficient for suppression of the beam modulation.} \label{m21}
\end{figure}

As an alternative to the single-bunch self-seeding scheme we
suggest a novel technique based on the creation of two identical
electron bunches separated by the RF period $T_0$ of the
accelerating system used.
The scheme requires limited hardware, which can be installed in
place of a single $5$ m-long undulator module. Therefore, it does
not perturb the baseline mode of operation of the facility. In the
case of short bunches\footnote{The method also works for long
electron bunches as discussed later on.}, the overall idea is
sketched in Fig. \ref{m2} and Fig. \ref{m19}, while the
installation position of the magnetic chicane in the baseline
undulator of an XFEL is shown in Fig. \ref{m21}. Two short ($3$ fs
rms) bunches, separated one from another of a temporal interval
$T_0$, travel through the first part of the undulator ($7$-cells
long at a resonance wavelength of $0.15$ nm in the example in the
figure), and start lasing from shot noise in the linear regime,
producing pulses in the $50$ MW level. Subsequently, they pass
through a short magnetic chicane installed in place of a single
module. The chicane has a deflection angle of $1.5$ mrad and a
length of $5$m and has two purposes\footnote{A third effect is to
delay the electron bunches relative to the radiation pulse of a
short time (in the $10$ fs range) which may be useful later on to
finely tune the system by effect.}. First, it washes out the beam
microbunching ($R_{56} \simeq 11 \mu$m is more than enough to this
purpose). Second  it creates an offset of $4$ mm, where we plan to
install a four-bounce crystal monochromator. The monochromator
filters the radiation from the first electron bunch, and it
introduces a long path difference corresponding to the time $T_0$.
Thus, at the entrance of the second undulator, the second electron
bunch is seeded with an almost longitudinally-coherent seed.

\begin{figure}[tb]
\includegraphics[width=1.0\textwidth]{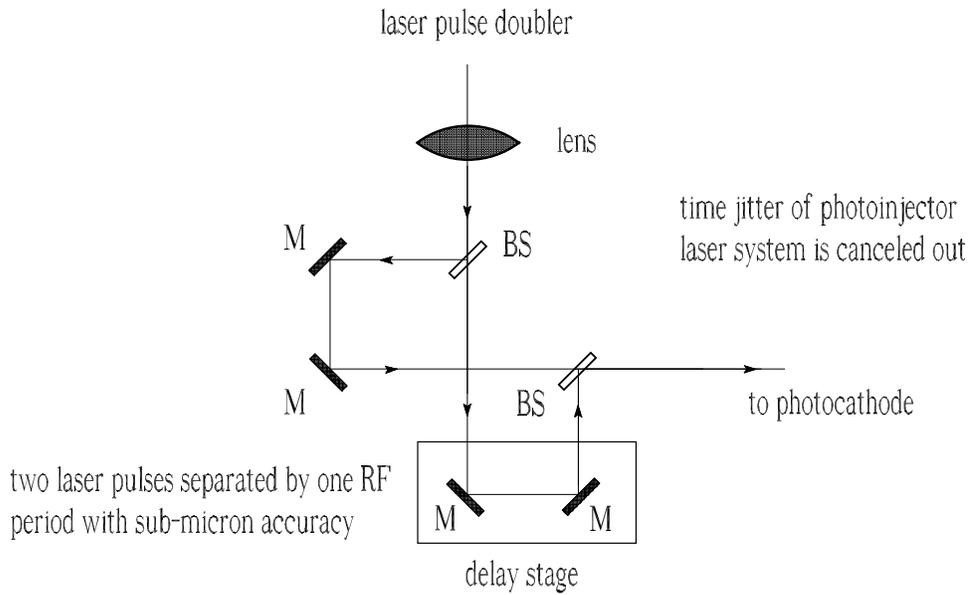}
\caption{Concept of laser pulse doubler.  A Michelson
interferometer provides control of the delay between two copies of
the input laser pulse with sub-micron accuracy. Pulse separation
is the RF period $T_0$.} \label{m18}
\end{figure}
\begin{figure}[tb]
\includegraphics[width=1.0\textwidth]{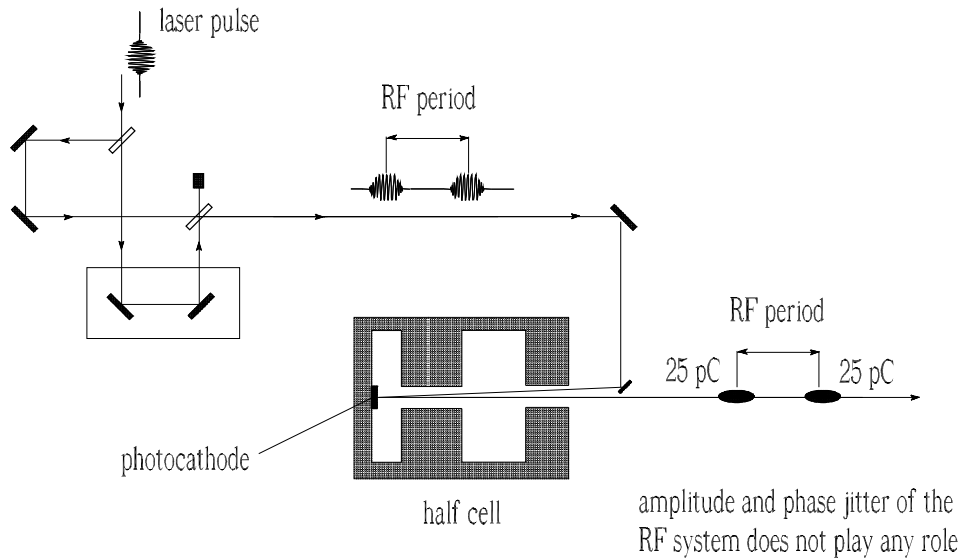}
\caption{Illustrative view of the proposed photoinjector setup
using a laser pulse doubler.} \label{m102}
\end{figure}
The method requires the development of a photoinjector setup
making use of a laser-pulse doubler concept like the one sketched
in Fig. \ref{m18}. The laser pulse to be used is split and
delayed. A Michelson interferometer can be used to provide
sub-micron accuracy control of the delay between the two split
pulses, and enables a separation delay $T_0$. The two pulses are
then sent to the photoinjector setup, as in Fig. \ref{m102}, and
two identical electron bunches, delayed of $T_0$, are created. A
pioneering experiment for integrating the laser pulse doubler with
an XFEL photoinjector was performed at FLASH. Results, which
should be considered as first steps into a novel direction of XFEL
technology, are reported in \cite{DOUB}.

Usually, two main sources of time jitter between bunches are
present: first, phase and amplitude jitter of the RF system.
Second, time jitter of the photo injector laser system. Since
bunches are only one RF period apart, the jitter of the RF system
does not play any role. Also, in our scheme the jitter of the
photo injector laser system is cancelled out because we produce
two laser pulses by splitting a single input laser pulse. As a
result, the distance between the two laser pulses can be fixed
with sub-micron optical accuracy. Moreover, due to the
single-pulse splitting, we automatically have identical
longitudinal pulse shapes, and the identity of the energy and of
the transverse shape can be controlled with optical accuracy too,
at least up to $0.1\%$.  Finally, the low charge mode of operation
works with charges ($25$ pC) which are $40$ times smaller compared
with the design value ($1$ nC). This allows one to neglect wakes
of the first bunch on the second. However, we will demonstrate
that our scheme enables a relatively large jitter acceptance up to
tens rms of the electron bunch length.

The low charge mode of operation includes an additional technical
advantage, because the photoinjector laser system can work, in
this case, with laser pulses yielding an energy an order of
magnitude smaller than those needed for the baseline mode of
operation. therefore, no upgrade of the photoinjector laser system
is needed, meaning that we propose a cost effective solution of
the line-width control problem.

\begin{figure}[tb]
\includegraphics[width=1.0\textwidth]{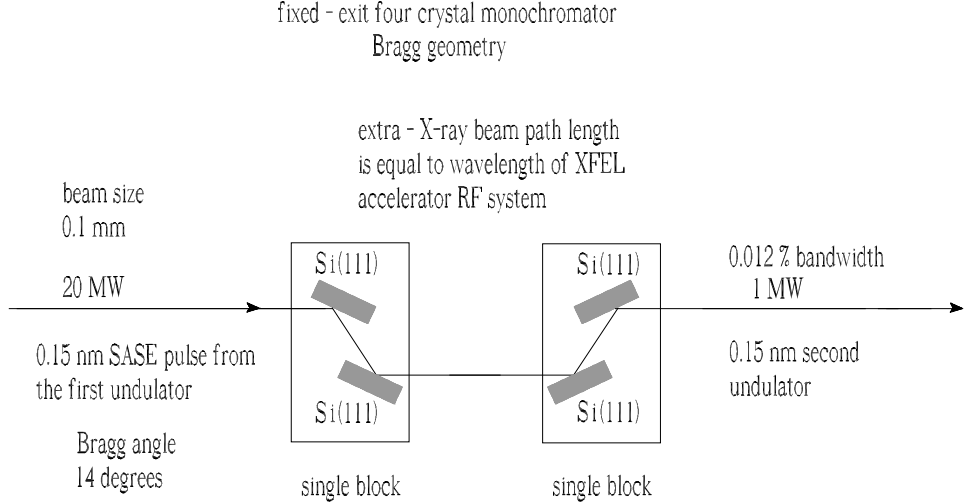}
\caption{Fixed-exit four-crystal monochromator: Silicon 111
reflection. The additional path length acquired by the X-rays in
the monochromator is equal to wavelength of XFEL accelerator RF
system.} \label{m1}
\end{figure}

\begin{figure}[tb]
\includegraphics[width=1.0\textwidth]{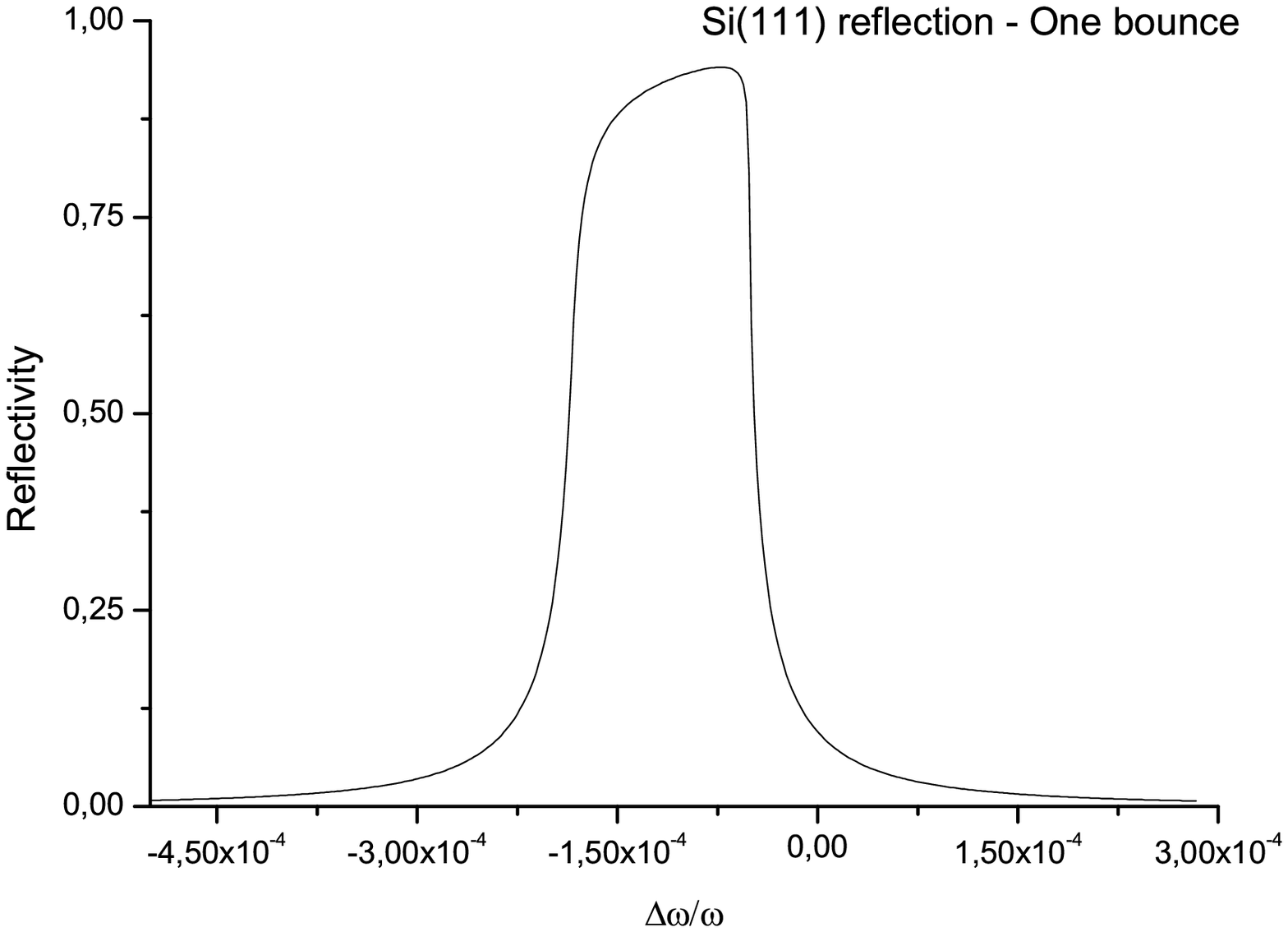}
\caption{Reflectivity curve for a thick absorbing crystal in Bragg
geometry. Si(111) reflection of 0.15 nm X-rays. One bounce.}
\label{m003}
\end{figure}
\begin{figure}[tb]
\includegraphics[width=1.0\textwidth]{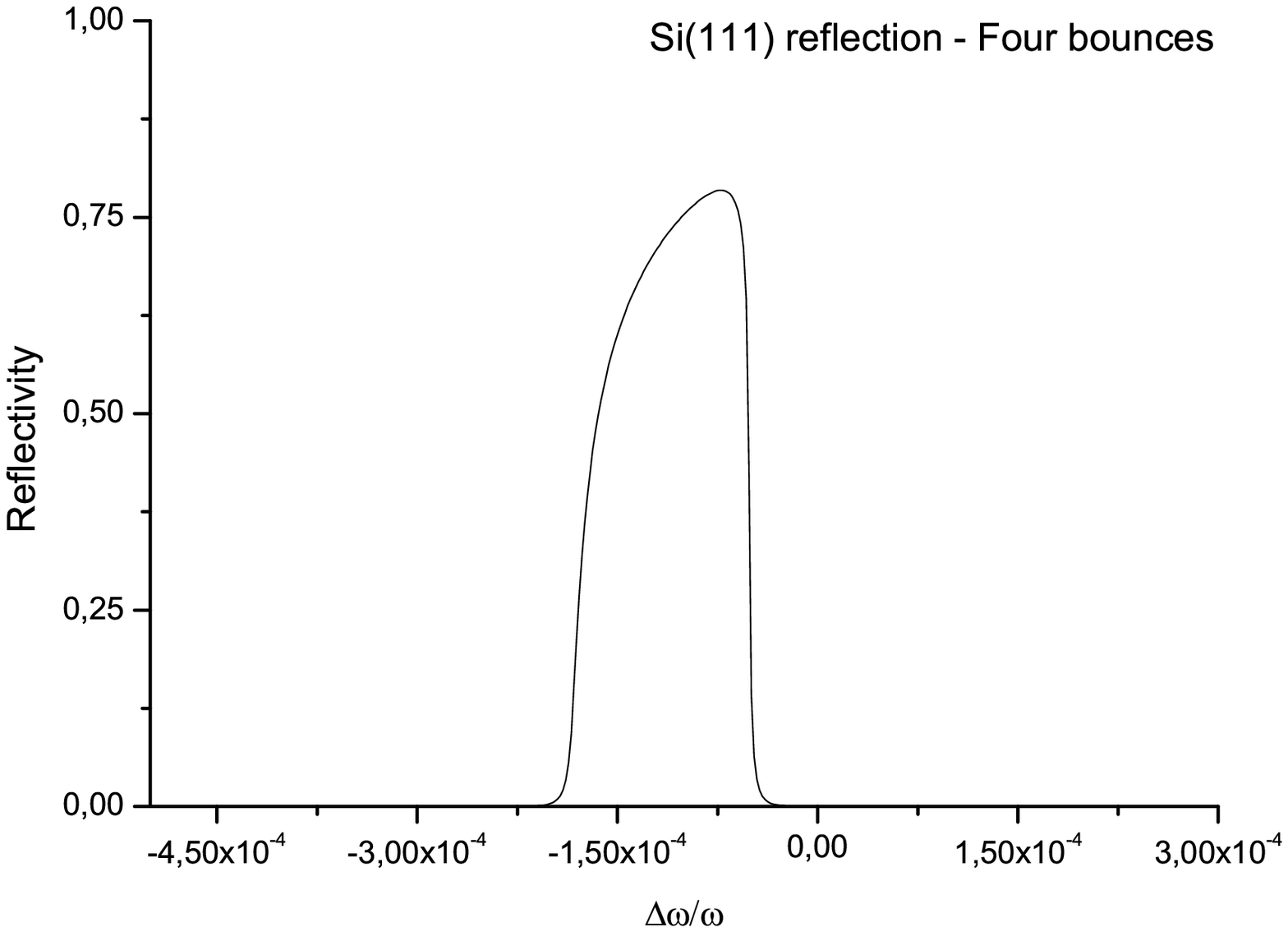}
\caption{Reflectivity curve for a thick absorbing crystal in Bragg
geometry. Si(111) reflection of 0.15 nm X-rays. Four bounces.}
\label{m004}
\end{figure}
A key component of the proposed setup is the X-ray monochromator.
We have chosen a four-bounce scheme as shown in Fig. \ref{m1}.
This solution is advantageous because it allows one to keep the
exit direction and position of the X-ray beam equal to the
entrance direction and position. In the monochromator, the X-ray
pulse acquires a path delay given by $\Delta L = 2 H
\tan(\theta_{nkl})$, where H is the beam shift, and $\theta_{nkl}$
is the Bragg angle of the reflections in the monochromator. Note
that this relation is valid also in the case of back-reflections,
i.e. for $2 \theta_{nkl} > \pi/2$. By varying H one can keep the
delay $\Delta L$ constant in the whole tunability range of the
monochromator. $\Delta L$ should be chosen in such way that the
total time delay of the radiation with respect to the electron
beam, inclusive of corrections due to the presence of the magnetic
chicane, equals $T_0$.

Multiple reflections result in a reduction of the tails of the
reflectivity curve. Namely, if $I(\omega)$ is the reflectivity
curve for the single crystal, the reflectivity after $N$
reflections is given by $I^{N}(\omega)$. Fig. \ref{m003} and Fig.
\ref{m004} show the reflectivity curves for Silicon 111,
respectively after one and four reflections, at $\lambda = 0.15$
nm. The tails of the reflectivity curve decrease very rapidly with
the number of reflections, due to the fact that the tails behave
asymptotically as $(\Delta \lambda/\lambda)^{-2}$, after one
reflection and as $(\Delta \lambda/\lambda)^{-8}$ after four.
However, the width of the curve remains constant. It should be
noticed that here we discuss about silicon crystals only. There
are two reasons for this. First, the main advantage of silicon is
the availability of almost perfect synthetic monocrystals with
high-reflectivity, thanks to semiconductor-industry demands.
Second, silicon has the advantage of a high heat-conductivity and
a high damage threshold, and can thus achieve good performance at
high power load. As is well known, if the crystal plate is cooled,
the thermal deformation results in a slope error composed of two
components: a bending and a bump. The thermal deformation of the
crystal induced by heat load depends on the ratio $\alpha/k$,
$\alpha$ being the thermal expansion coefficient and $k$ the
thermal conductivity of the crystal, and for silicon it strongly
depends on temperature, being zero at $125$ K. This is the best
high-power working point for the crystal. Lowering the temperature
of silicon from room temperature to liquid-nitrogen temperatures
improves the ratio $k/\alpha$ by a factor $50$.

The advantages of our scheme with respect to the single-bunch
self-seeding method are obvious. The hardware needed here can be
installed without perturbation of the baseline parameters, and
switching to the normal SASE mode of operation is achieved by
switching off the magnetic chicane and retracting the
monochromator. Also, although it requires the development of a
pulse doubler, it avoids complications and costs associated with
the development of a dispersion-free $60$-m long bypass.
Wakefields problems, due to the influence of the passage of the
first bunch on the second should be negligible for a low-charge
mode of operation, while they deserve better attention in the
long-pulse mode. Note that the time jitter acceptance for the
double bunch scheme increases when the bandwidth of the
monochromator decreases. Simultaneously, the seeding power
decreases too, but it should be much larger than the shot noise
power. Since the effective shot noise power is about $3$ kW and
the seeding power should be larger than $100$ kW,  for the short
pulse mode of operation we are limited to Si(111) and Si(220)
reflections. The situation changes combining the double bunch
scheme with other recently introduced techniques, as discussed
below.


\subsection{Combination of self-seeding scheme and fresh bunch technique}

\begin{figure}[tb]
\includegraphics[width=1.0\textwidth]{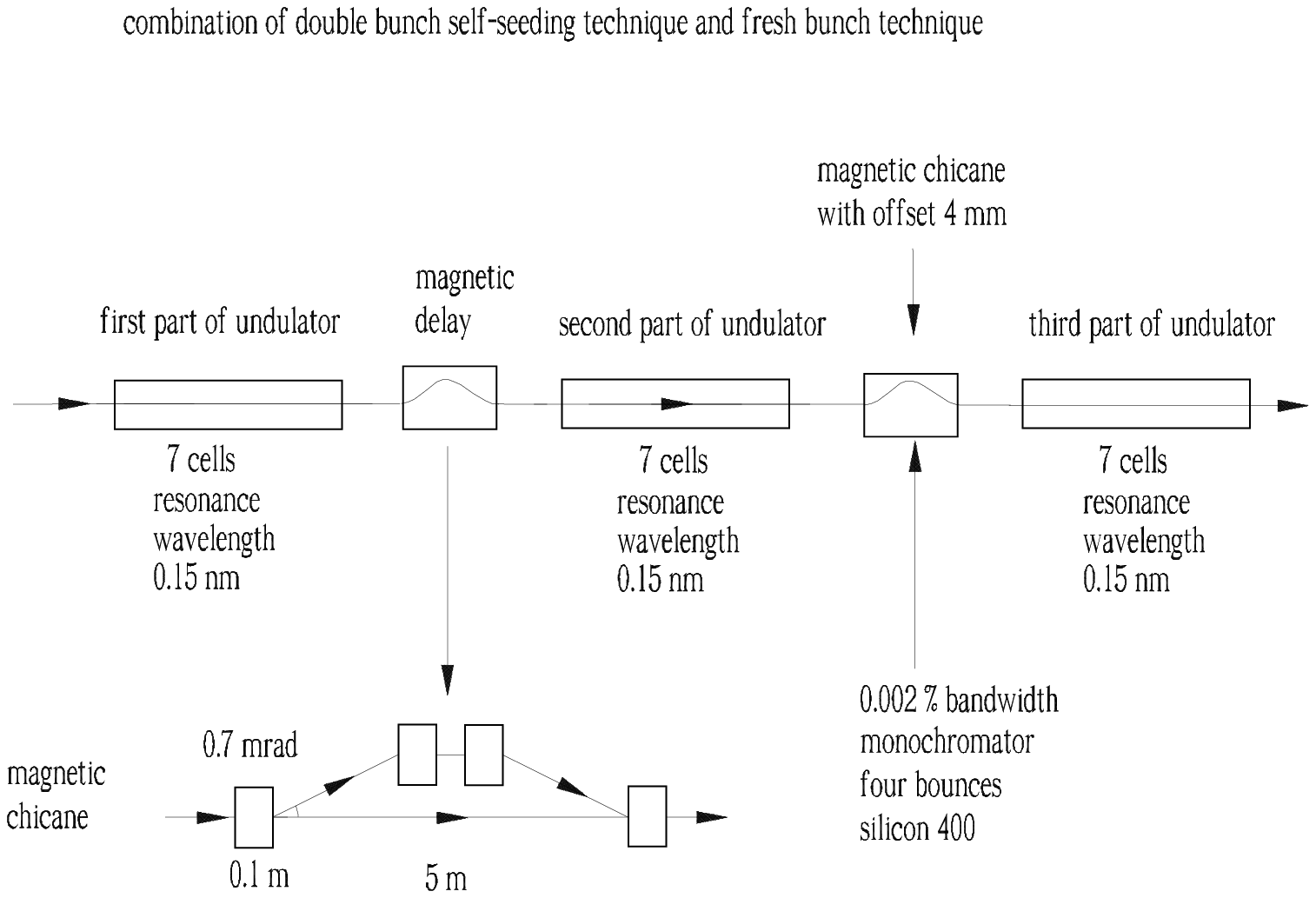}
\caption{Design of an undulator system for generation of highly
monochromatic X-ray pulses. The method is based on a combination
of the double-bunch self-seeding scheme and of the fresh-bunch
technique.} \label{m15}
\end{figure}
\begin{figure}[tb]
\includegraphics[width=1.0\textwidth]{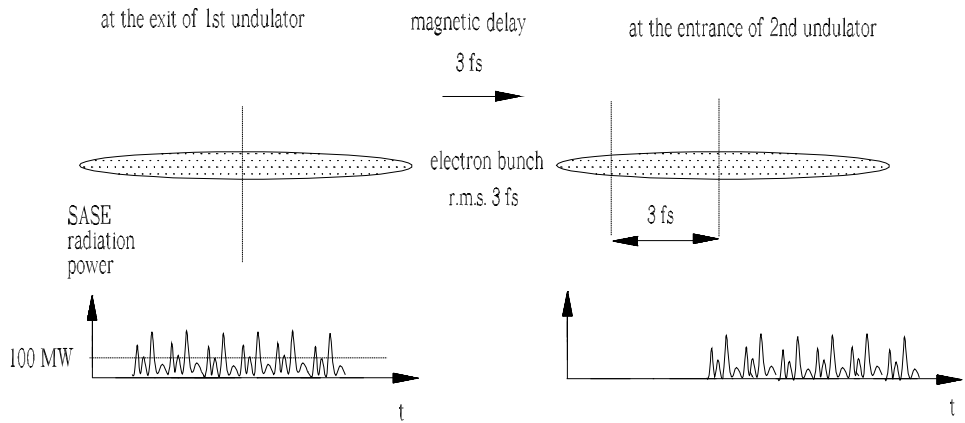}
\caption{Sketch of principle of fresh bunch technique for short
($6$ fs) pulse mode operation.} \label{m16}
\end{figure}
\begin{figure}[tb]
\includegraphics[width=1.0\textwidth]{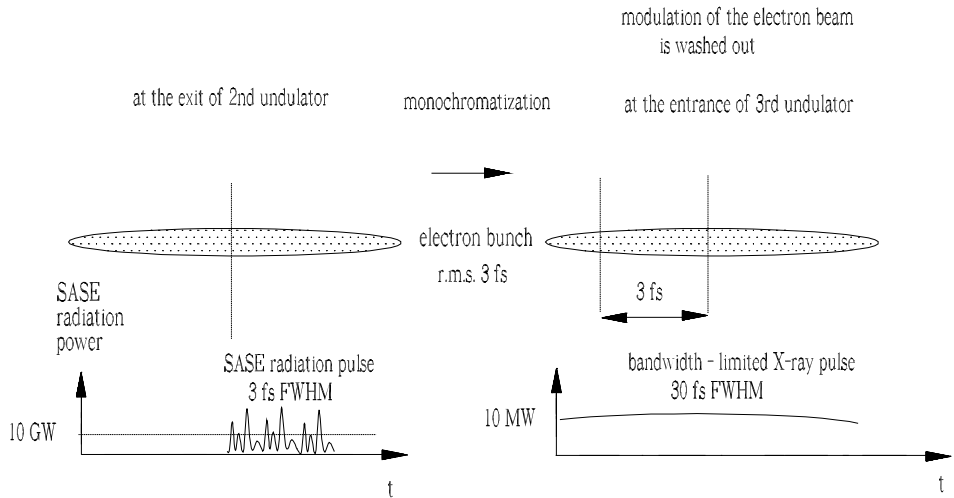}
\caption{Sketch of the X-ray monochromatization principle from
second to third undulator. By using a combination of self-seeding
and fresh bunch techniques, a $30$ fs, monochromatic radiation can
be produced behind the monochromator.} \label{m17}
\end{figure}
The method described above may be combined with a fresh bunch
technique \cite{HUAYU}-\cite{OUR05}. The idea is sketched in Fig.
\ref{m15}, Fig. \ref{m16} and Fig. \ref{m17}. Compared to the
previous scheme, this method makes use of the space occupied by
two undulator segments, Fig. \ref{m15}. After the first undulator
part, where they undergo the SASE process in the linear regime,
the two short ($6$ fs) electron bunches created with the pulse
doubler described above are sent to a weak chicane, which washes
out the microbunching and introduces a relative delay of the
electron bunch compared to the radiation, Fig. \ref{m16}. Then,
half of the delayed electron bunches are seeded by their own
radiation, and go up to saturation in the second undulator part.
Finally, the strong radiation pulse at the end of the second
undulator part is stretched by the crystal monochromator, see Fig.
\ref{m17} and seeds the remaining, fresh part of the second
electron bunch.

\begin{figure}[tb]
\includegraphics[width=1.0\textwidth]{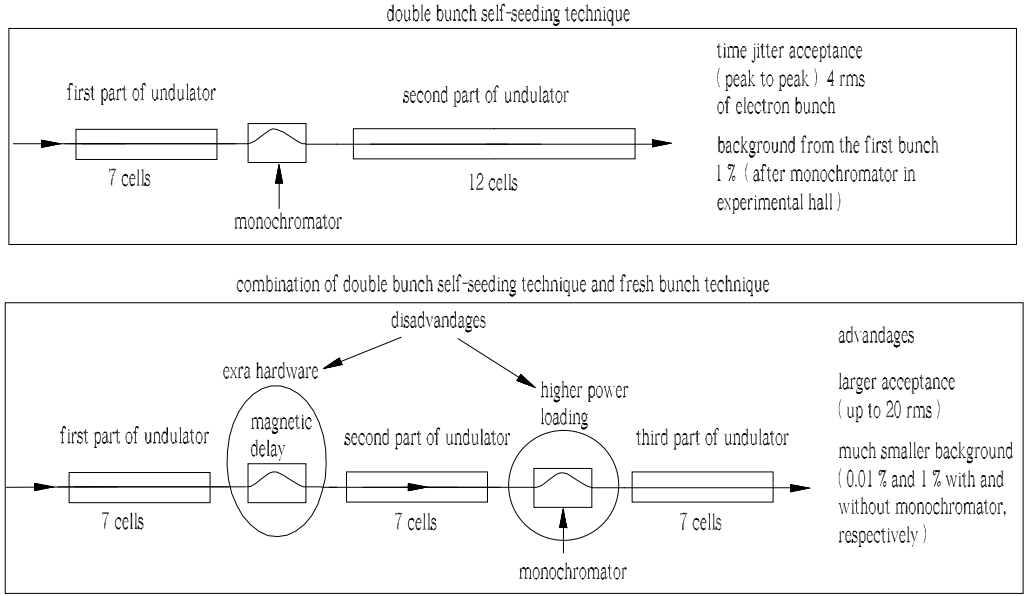}
\caption{Comparison of the two self-seeding schemes considered in
this paper. The scheme based on the additional use of the
fresh-bunch technique has two significant advantages. It is
insensitive to non-ideal effects like time-jitter between two
electron bunches, and presents much smaller background from the
first (non monochromatic) X-ray pulse. However, in this scheme the
first crystal in the monochromator operates at much more severe
heat-loading conditions.} \label{m20}
\end{figure}
The combination of self-seeding and fresh-bunch technique brings
mainly two advantages when dealing with the short pulse mode of
operation. These are summarized in Fig. \ref{m20}, which compares
the two self-seeding schemes considered in this paper. The first
advantage is constituted by the fact that the seeding power is
increased due to the use of the fresh-bunch technique. Therefore,
one can operate with monochromators with the smallest bandwidths
(Silicon 400 and Silicon 440). This brings the advantage of
increasing the jitter acceptance up to $20$ rms lengths of the
short bunch.

The second advantage is constituted by the fact that the
background SASE radiation from the first bunch now carries much
smaller power with respect to the monochromatic pulse, at variance
with what happens when no fresh bunch technique is used. Note that
users may also want to use an additional monochromator in the
experimental hall (for example using Silicon 111 or Silicon 220).
In this case, the second pulse (already monochromatized) would
simply pass through without any change, and monochromatization
would affect the first pulse only. In the case of a short pulse,
the third undulator part only consists of $7$ cells. The
contribution of the first pulse will be a few per cent of the
monochromatized pulse even without filtering, and would reduce to
about $0.01 \%$ after filtering. Note that head-loading problems
will be severe at the European XFEL (but not at LCLS) in the case
of a long-pulse mode of operation. Yet, this scheme allows, in
principle, for bandwidth limited pulse mode operation for $50$ fs
pulses too. Such mode of operation ultimately leads to an increase
in peak brilliance of two orders of magnitude.

It should also be remarked, that the possibility of combining
self-seeding scheme and fresh-bunch technique would be of great
importance during the commissioning stage of our scheme. The
inclusion of an extra chicane is not expensive and may find many
other applications \cite{OUR01}-\cite{OUR05}. Moreover, it adds on
to the jitter budget, and to the accuracy of the extra-path length
in the monochromator and in the pulse doubler, as one has a few
orders of magnitude more input power than strictly needed. During
commissioning, all non ideal effects can be reduced to a minimum
and, finally, the fresh bunch scheme may be reduced to the
previously discussed scheme.

\begin{figure}[tb]
\includegraphics[width=1.0\textwidth]{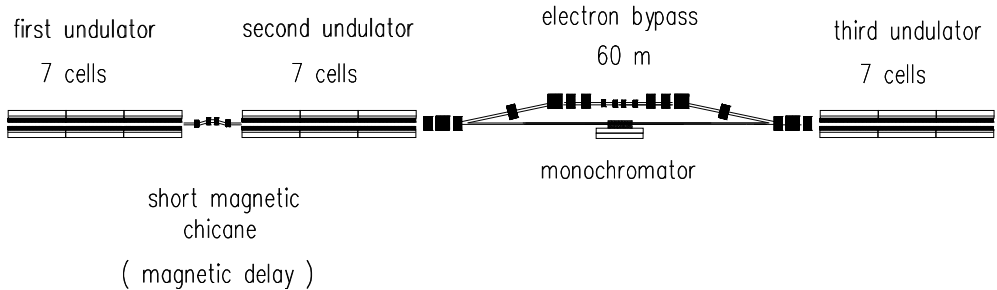}
\caption{Design of an undulator system for narrow bandwidth mode
of operation. The method is based on a combination of single bunch
self-seeding scheme and fresh bunch technique.} \label{m170}
\end{figure}
It should be mentioned that the fresh bunch technique may also be
combined with the single-bunch self-seeding option, as described
in Fig. \ref{m170}. This will allow one to work with Si(400) and
Si(444) reflections. The extra-available seeding power allows for
the production, at the exit of the third undulator, of bandwidth
limited X-ray pulses with small level of noise contribution. This
can be important for further manipulation of electron bunch and
X-ray pulse, for example attempting to increase the efficiency of
an XFEL by tapering.

Finally note that, when considering single-bunch self-seeding
schemes, one should account for diffraction of the radiation along
the $60$ m free-flight between first and second undulator.
Simulations show that this leads to a degradation of the effective
seeding power of about a factor ten. In this case one should
straightforwardly upgrade the original setup to include two more
cells (from $7$ to $9$) in the first undulator, which we did not
consider in this paper. However, increase of the total energy from
the first stage of a factor ten also implies an increase of a
factor ten in heat-load of the monochromator. It follows that heat
loading may become an issue for the implementation of some variant
(long pulse, single-bunch self-seeding in combination with fresh
bunch technique) of our technique at the European XFEL.


\section{Feasibility study for short pulse mode of operation}

Following the introduction of the proposed methods, in the present
and in the following sections we report on a feasibility study
performed with the help of the FEL simulation code GENESIS 1.3
\cite{GENE} running on a parallel machine.

Simulations were performed in the following way: first, we
calculated the three-dimensional field distribution at the exit of
the first undulator, and downloaded the field file. Subsequently,
we performed a Fourier transformation followed by filtering
through the monochromator, by using the reflectivity curve for the
specific four-bounces Bragg reflection. Finally, we performed an
inverse Fourier transformation, and used the field file as seed at
the entrance of the second undulator. The electron microbunching
is washed out by the presence of non-zero $R_{56}$, and for the
second undulator we used a beam file with energy and energy spread
introduced by the FEL amplification process in the first
undulator. The amplification process in the second undulator
starts from the seed-field file. Shot noise initial condition was
included. It is understood that these simulations also apply for
the single bunch self seeding scheme.

\subsection{Two-stage scheme}

\subsubsection{First stage}

\begin{figure}[tb]
\includegraphics[width=1.0\textwidth]{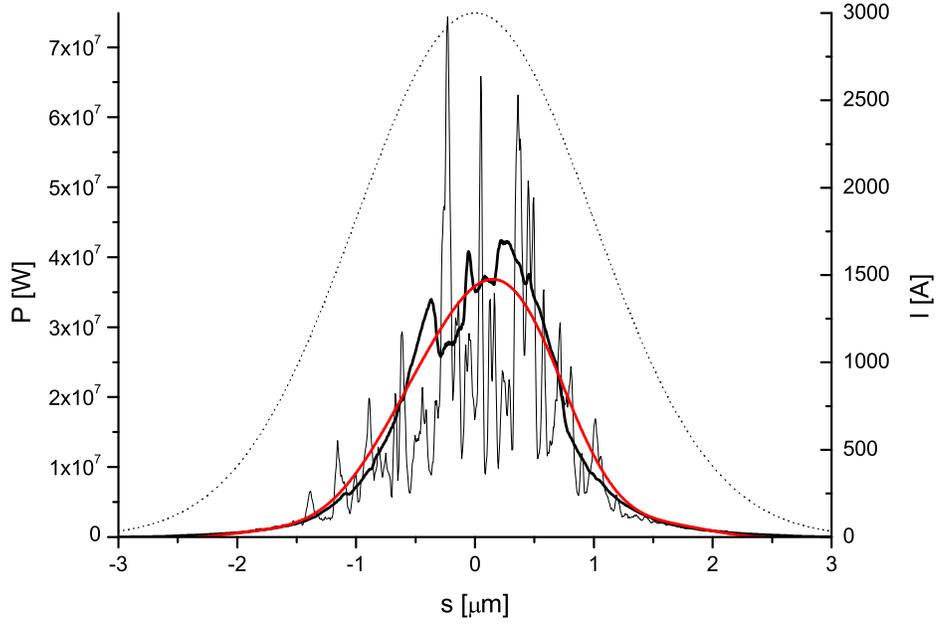}
\caption{Average (bold) and typical single-shot temporal structure
of the SASE radiation pulse at the exit of the first undulator
with a length of 42 m (7 cells). A smoothed average line is also
presented. Calculations have been performed over 100 shots. The
dashed line presents the corresponding distribution of the
electron beam current. } \label{m9}
\end{figure}
\begin{figure}[tb]
\includegraphics[width=1.0\textwidth]{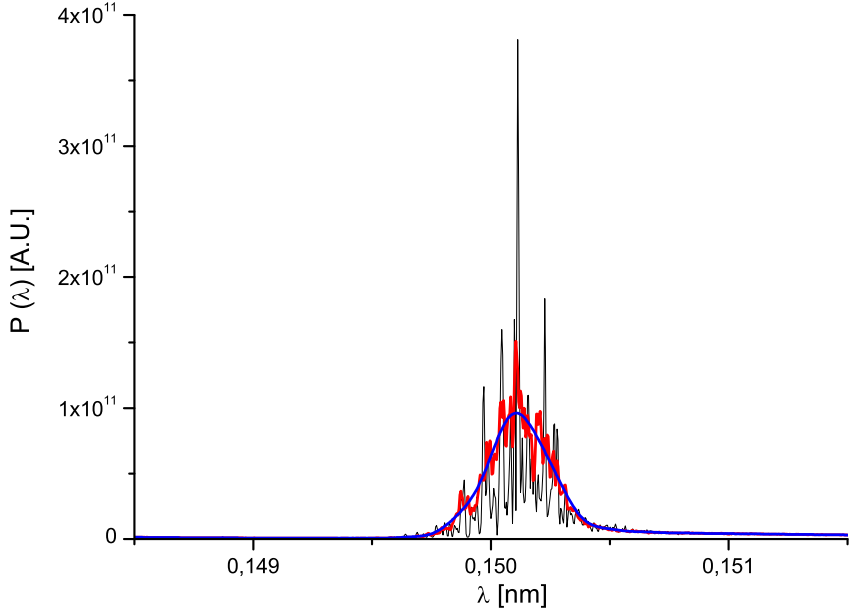}
\caption{Average (bold) and typical single-shot spectrum of SASE
radiation at the exit of  the first undulator with a length of
$42$ m ($7$ cells). A smoothed average line is also presented.
Calculations have been performed over 100 shots.} \label{m14}
\end{figure}
\begin{figure}[tb]
\includegraphics[width=1.0\textwidth]{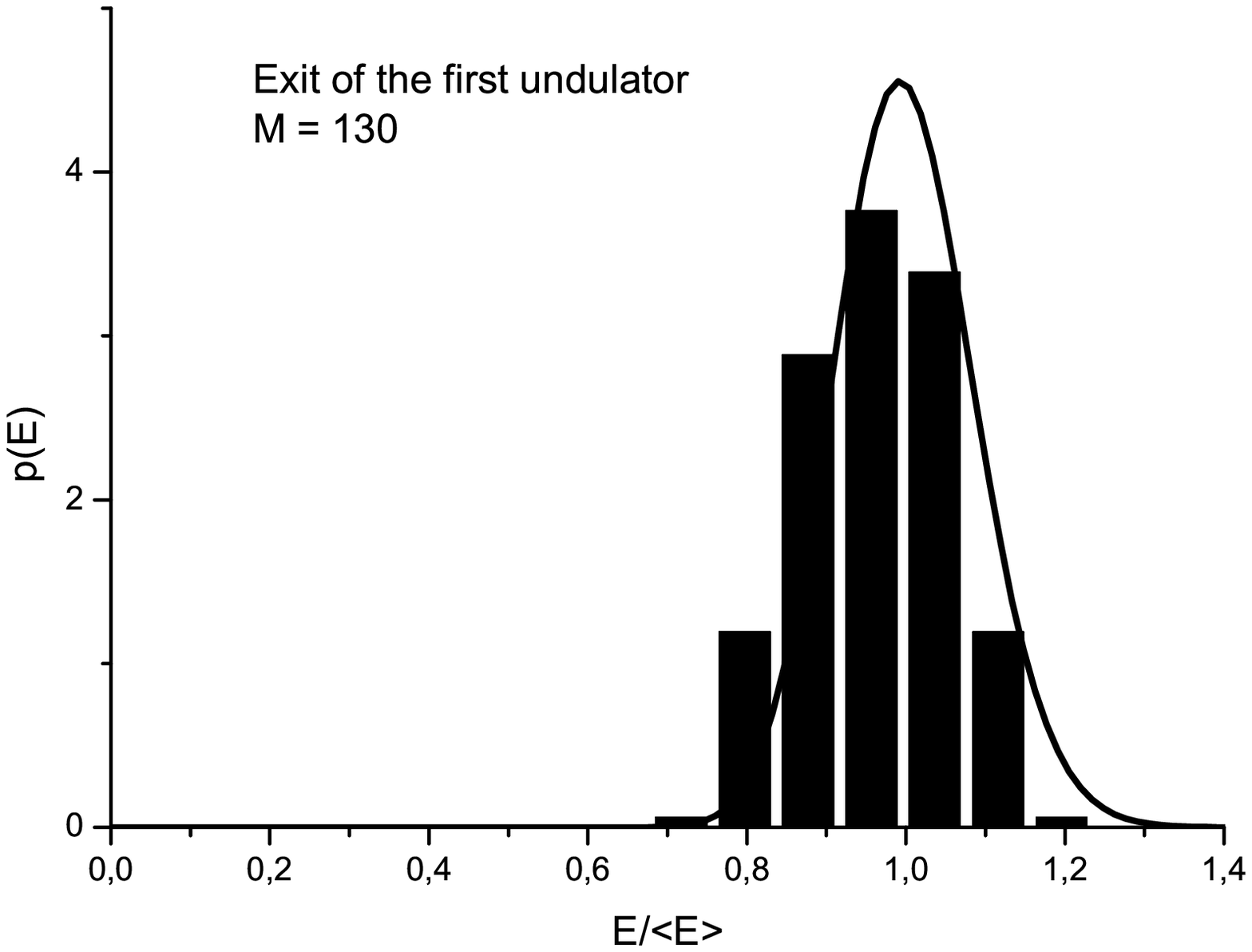}
\caption{Shot-to-shot fluctuations of the SASE radiation pulse at
the exit of  the first undulator with a length of $42$ m ($7$
cells). Calculations have been performed over 200 shots. Here
$<E>$ denotes the average energy. The solid curve represents the
gamma distribution, Eq. (\ref{aftmono}), where the value $M$ has
been calculated from the simulation data according to Eq.
(\ref{Mdef}).} \label{m10}
\end{figure}
In this section we present results of numerical studies of the
first stage of operation of the self-seeding setup. Parameters
chosen are reported in Table \ref{tt1}. Fig. \ref{m9} shows a
typical single-shot temporal profile of a SASE pulse, together
with the average temporal structure and the electron beam profile
just after the undulator. The corresponding spectra are shown in
Fig. \ref{m14}. Fig. \ref{m10} shows the shot-to-shot fluctuations
of the SASE pulse after the exit of the first undulator.  Since
the first stage is in the linear regime, the radiation process
obeys Guassian statistics.

\begin{table}
\caption{Parameters for the short pulse mode of operation used in
this paper.}

\begin{small}\begin{tabular}{ l c c}
\hline & ~ Units &  ~ \\ \hline
Undulator period      & mm                  & 48     \\
K parameter (rms)     & -                   & 2.516  \\
Wavelength            & nm                  & 0.15   \\
Energy                & GeV                 & 17.5   \\
Charge                & nC                  & 0.025 \\
Bunch length (rms)    & $\mu$m              & 1.0    \\
Normalized emittance  & mm~mrad             & 0.4    \\
Energy spread         & MeV                 & 1.5    \\
\hline
\end{tabular}\end{small}
\label{tt1}
\end{table}
%
Using well-known results obtained in the framework of statistical
optics, we can state that the distribution of the radiation energy
$E$ after the monochromator is described rather well by the gamma
probability density function $p(E)$:

\begin{eqnarray}
p(E) = \frac{M^M}{\Gamma(M)} \left(\frac{E}{\left\langle
E\right\rangle}\right)^{M-1} \frac{1}{\left\langle E\right\rangle}
\exp\left(-M\frac{E}{\left\langle E\right\rangle}\right),
\label{aftmono}
\end{eqnarray}
where $\Gamma(M)$ is gamma function with argument M, and

\begin{eqnarray}
M = \frac{1}{\sigma^2_E}~, \label{Mdef}
\end{eqnarray}
with $\sigma_E$ the rms energy fluctuation. This distribution
provides correct values for the mean $Z<E>$ and for the variance
$\sigma_E^2 = 1/M$,

\begin{eqnarray}
\int_0^\infty E ~p(E) dE  = <E>~,~~~~ \int_0^\infty
\frac{(E-<E>)^2}{<E>^2}p(E) dE = 1/M~. \label{pEdef}
\end{eqnarray}
The parameter M can be interpreted as an average number of degrees
of freedom (or modes) in the radiation pulse. We calculated $<E>$
and $\sigma_E^2$ from the simulation data for different
reflections which yields $M \simeq 130$. Then, $p(E)$ calculated
according to Eq. (\ref{aftmono}) could be plotted in Fig.
\ref{m11} together with the histogram of the probability density
distribution of radiation, also independently obtained from the
data. It is seen that it fits the gamma distribution well.

\subsubsection{Second stage}


\begin{figure}[tb]
\includegraphics[width=0.5\textwidth]{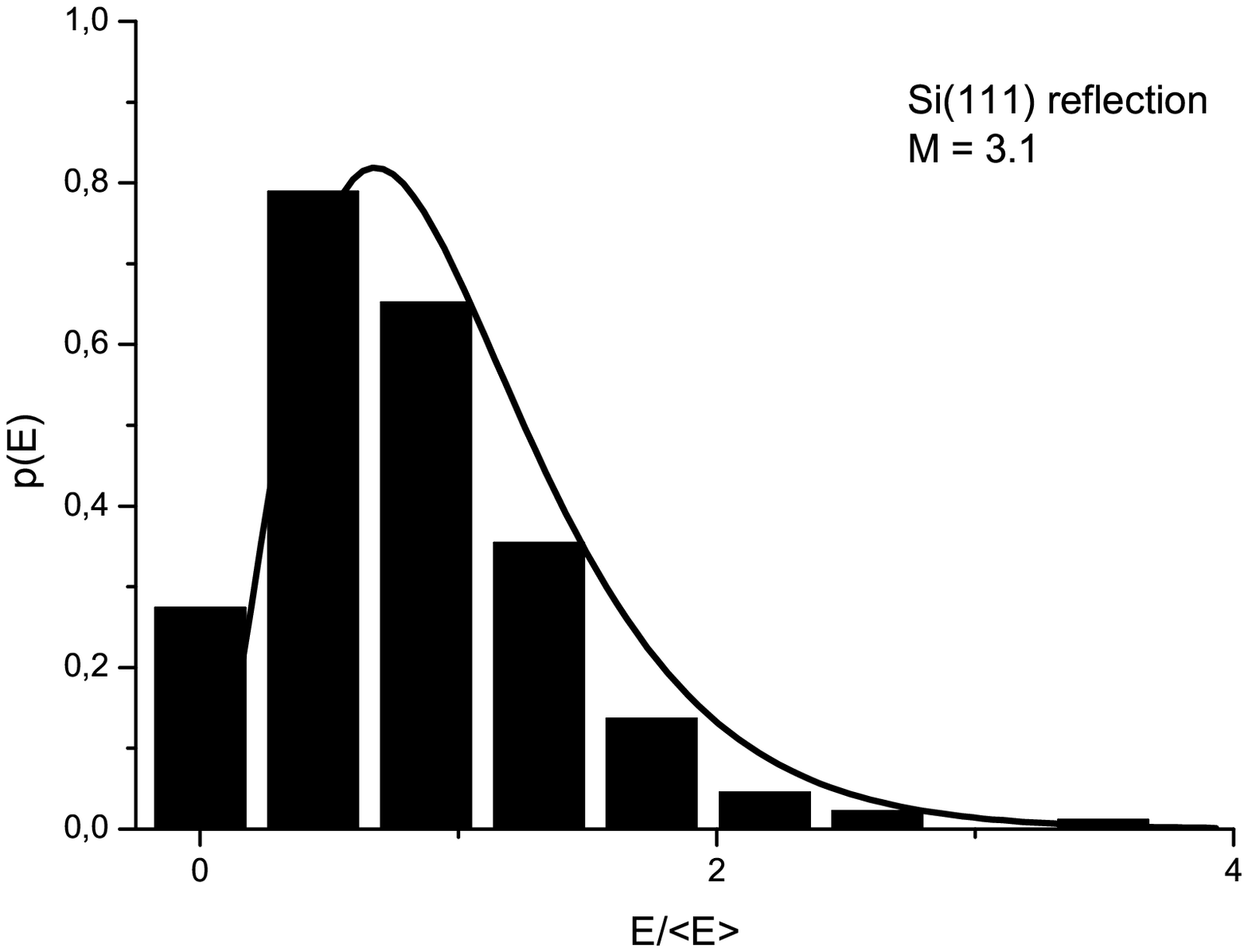}
\includegraphics[width=0.5\textwidth]{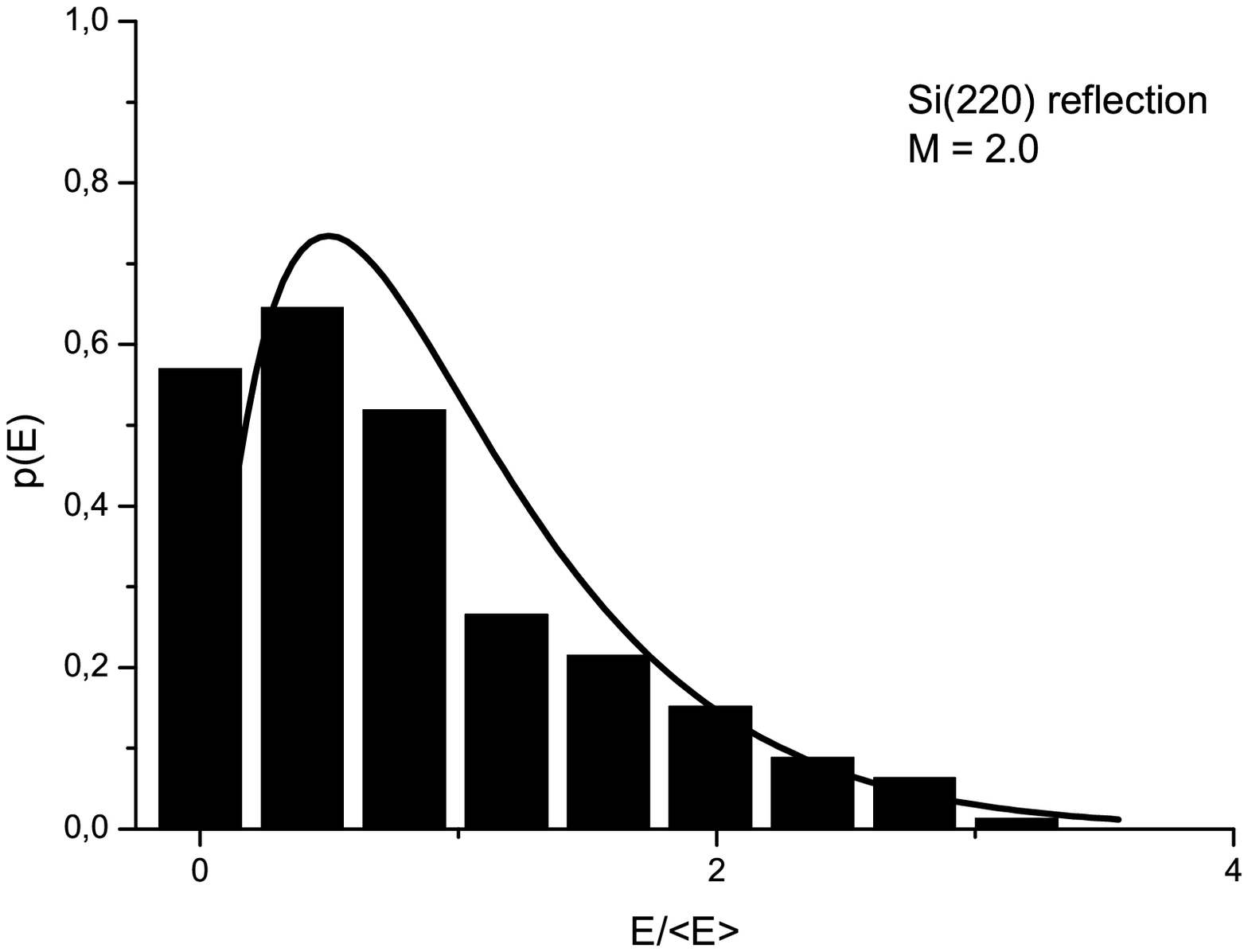}
\caption{Histogram of the probability density distribution $p(E)$
of radiation energy after the four-bounce monochromator. Si(111)
(left) and Si(220) (right) reflection cases. Calculations have
been performed over 200 shots. Here $<E>$ denotes the average
energy. The solid curve represents the gamma distribution, Eq.
(\ref{aftmono}), where the value $M$ has been calculated from the
simulation data according to Eq. (\ref{Mdef}).} \label{m11}
\end{figure}
\begin{figure}[tb]
\includegraphics[width=0.5\textwidth]{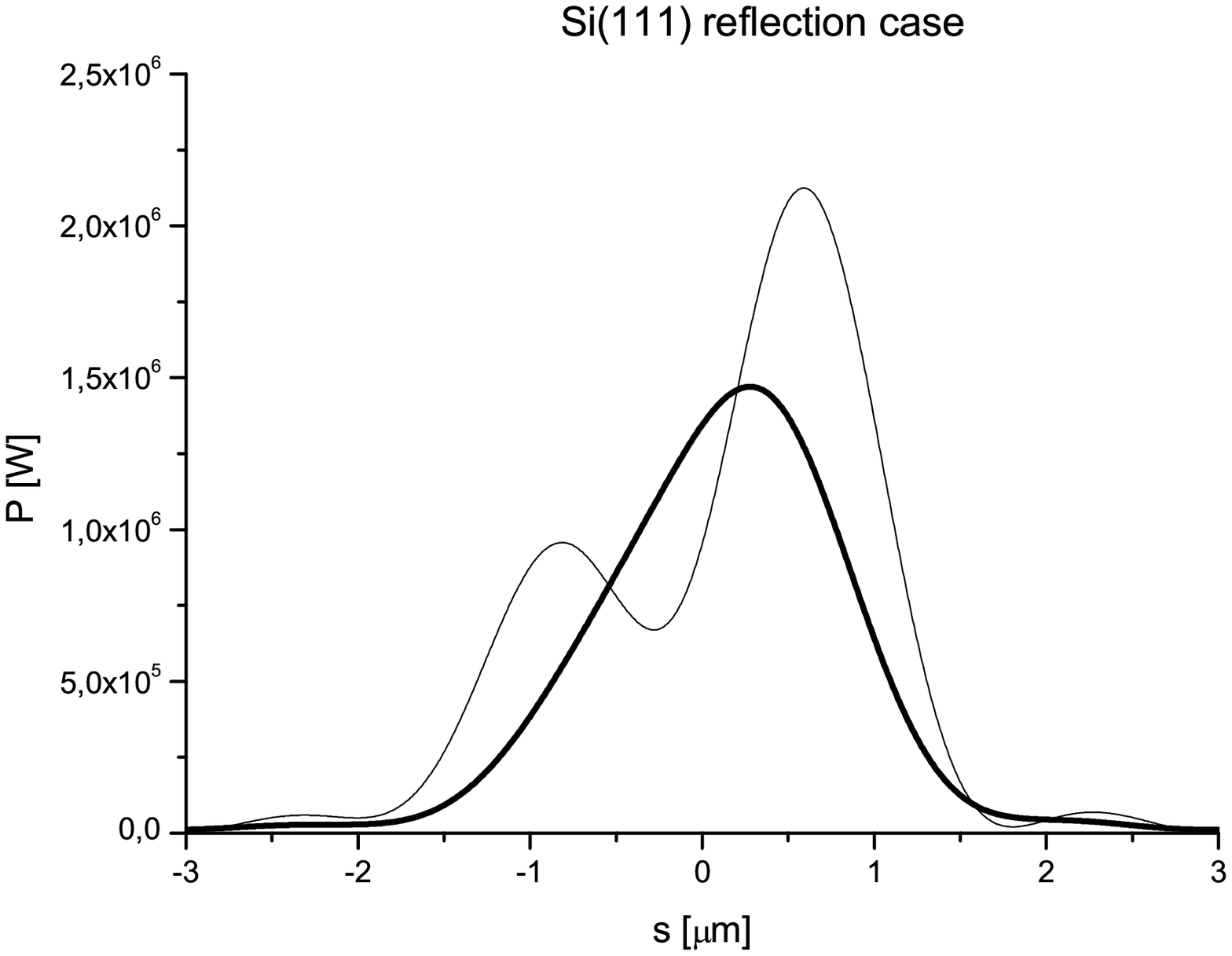}
\includegraphics[width=0.5\textwidth]{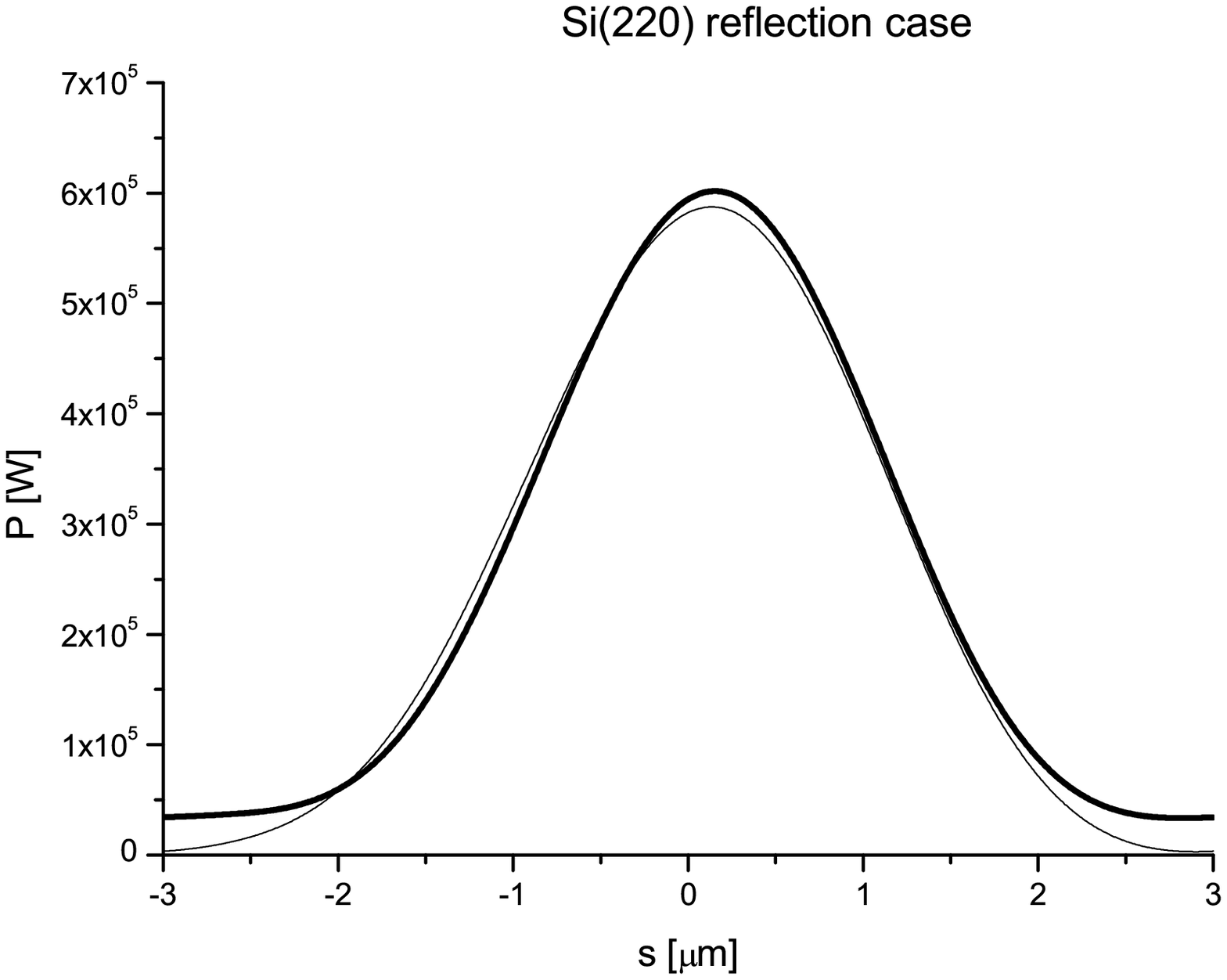}
\caption{Average (bold) and typical single-shot temporal structure
of the radiation pulse after the four-bounce monochromator.
Si(111) (left) and Si(220) (right) reflection cases.} \label{m12}
\end{figure}
\begin{figure}[tb]
\includegraphics[width=1.0\textwidth]{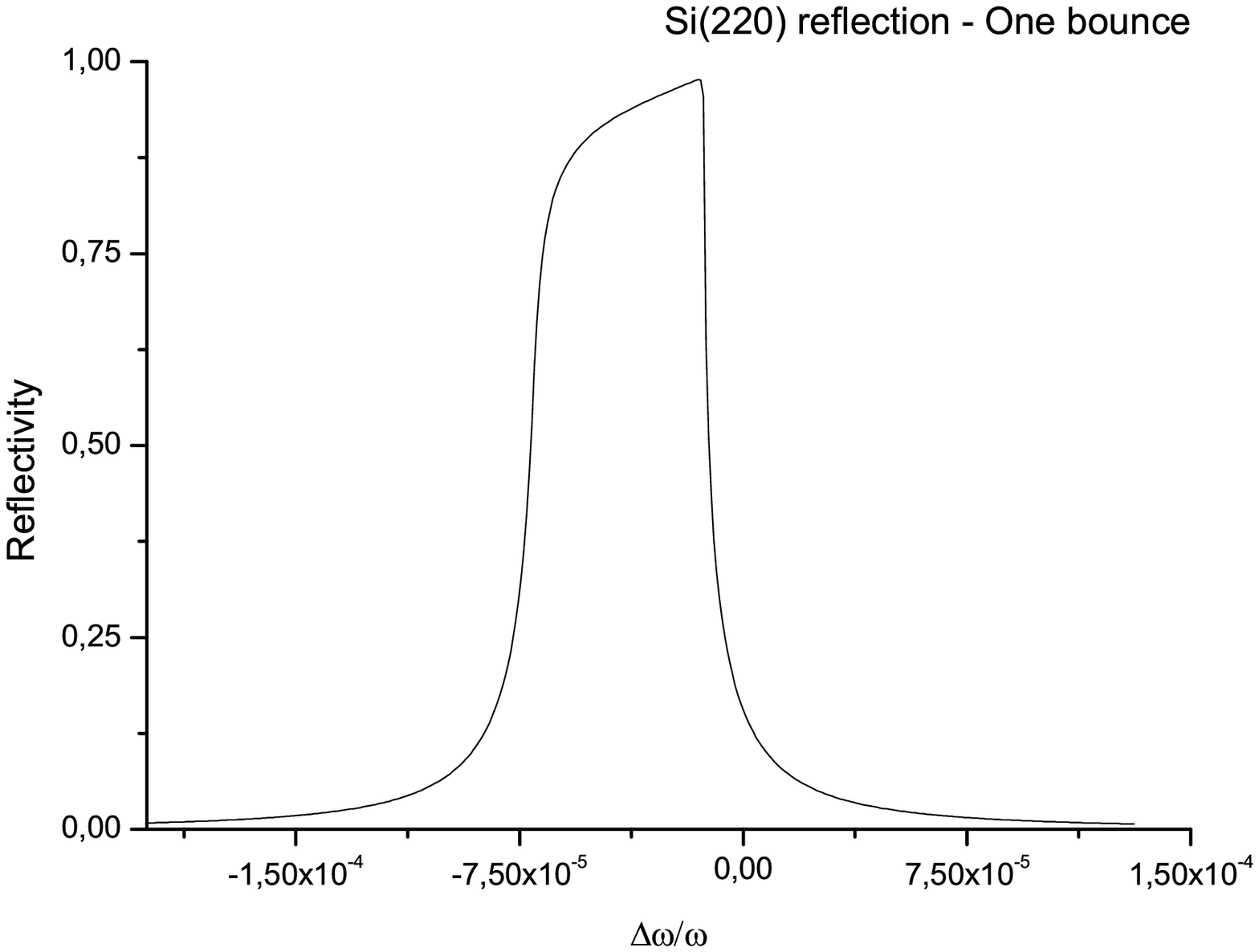}
\caption{Reflectivity curve for a thick absorbing crystal in Bragg
geometry. Si(220) reflection of 0.15 nm X-rays. One bounce.}
\label{m114}
\end{figure}
\begin{figure}[tb]
\includegraphics[width=1.0\textwidth]{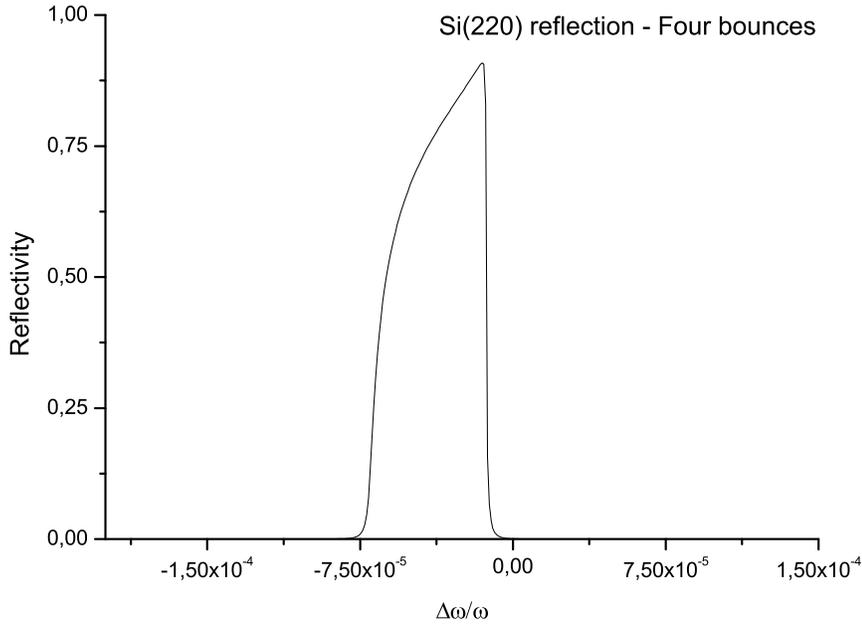}
\caption{Reflectivity curve for a thick absorbing crystal in Bragg
geometry. Si(220) reflection of 0.15 nm X-rays. Four bounces.}
\label{m115}
\end{figure}
\begin{figure}[tb]
\includegraphics[width=1.0\textwidth]{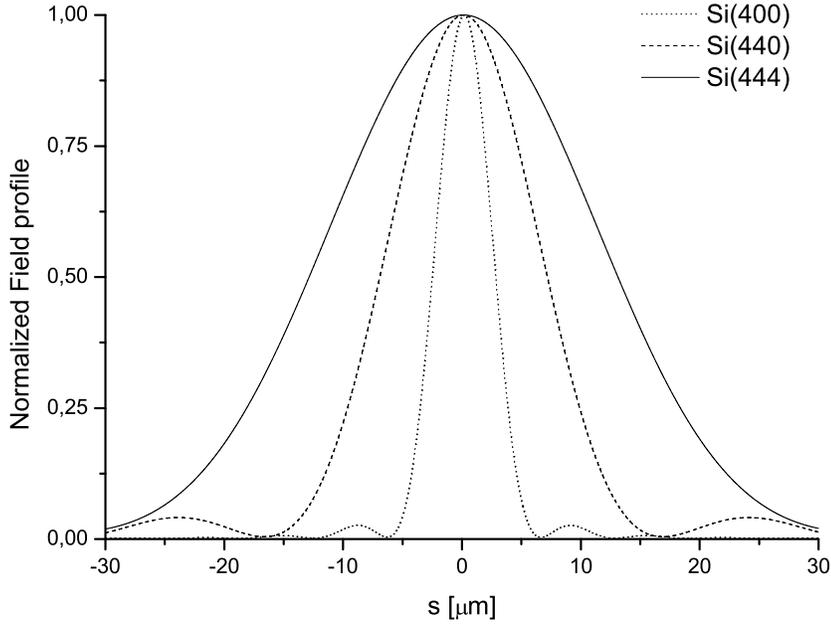}
\caption{Normalized field profile after the four-bounce
monochromator for the short bunch case, single shot.} \label{m34}
\end{figure}

After the first stage, radiation is monochromatized. Since the
monochromator acts as a linear filter, the radiation process still
obeys Guassian statistics. Fig. \ref{m11} shows the shot-to-shot
fluctuations of the SASE pulse energy. Fig. \ref{m12} shows a
typical single-shot temporal profile of a SASE pulse, together
with the average temporal structure and the electron beam profile
after a four-bounce monochromator. The reflectivity curves for the
Si(220) reflection are shown in Fig. \ref{m114} and Fig.
\ref{m115}, for one bounce and four bounces, respectively. The
reflectivity curves for the Si(111) reflection have already been
shown in Fig. \ref{m003} and Fig. \ref{m004}, also for one bounce
and four bounces. The field profiles after different
monochromators are shown in Fig. \ref{m34}. These plots illustrate
the properties of the seed signal for different monochromator
resolutions. If the monochromator resolution is good enough, the
radiation pulse is temporally stretched in such way that one only
sees the characteristics of the monochromator. This is the case
for Si(400), Si(440) and Si(444) reflections. The jitter
acceptance is obviously linked to the FWHM width of these plots,
i.e. to the temporal line width of the monochromator. Note that in
Fig. \ref{m34}, as everywhere in this paper, we treated the
monochromator as a real filter. The effects of non-zero phase is
discussed in Appendix.

\begin{figure}[tb]
\includegraphics[width=1.0\textwidth]{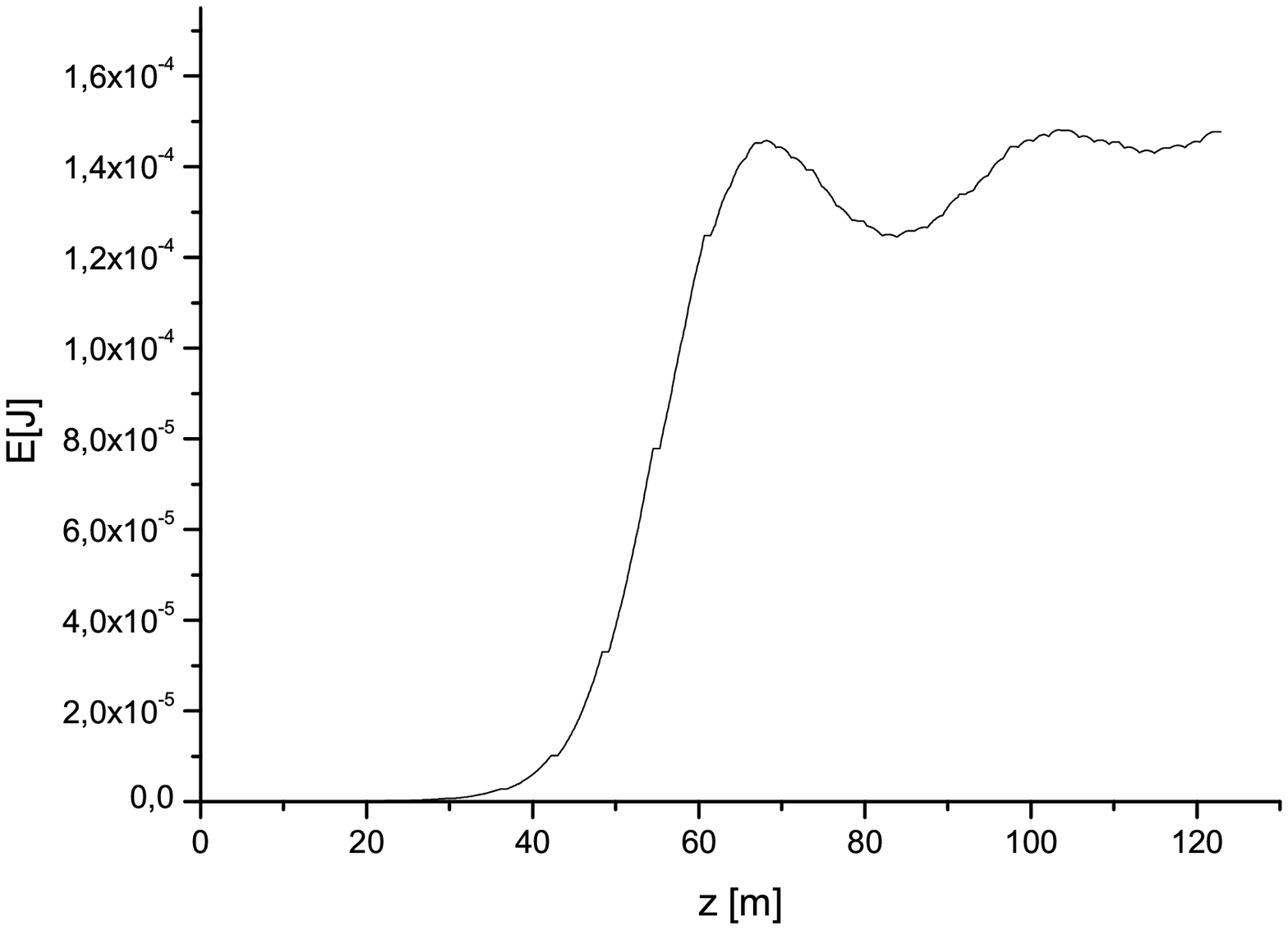}
\caption{Average pulse energy as a function of the position along
the second undulator for the Si(111) reflection case.} \label{m22}
\end{figure}
\begin{figure}[tb]
\includegraphics[width=1.0\textwidth]{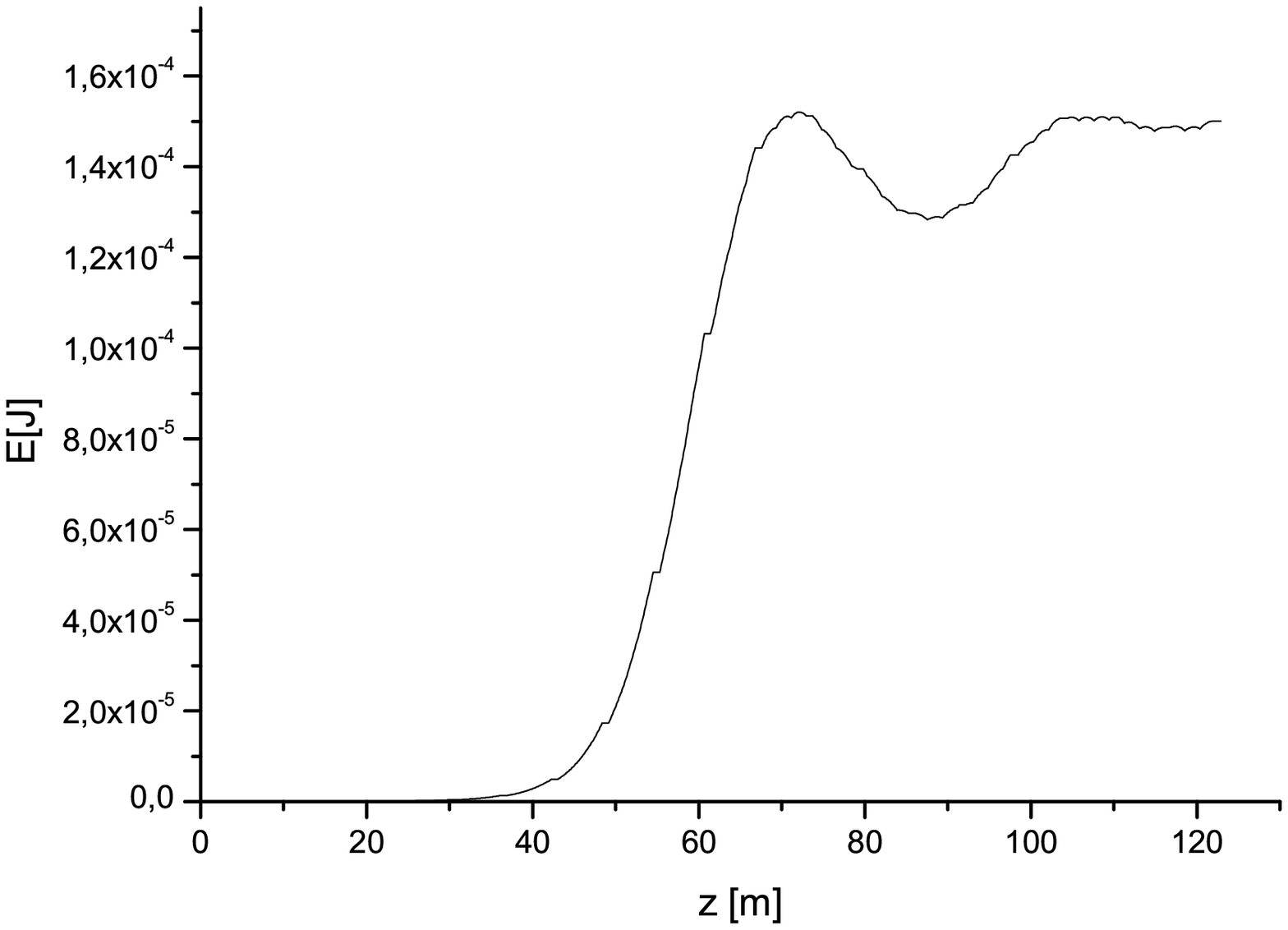}
\caption{Average pulse energy as a function of the position along
the second undulator for the Si(220) reflection case.} \label{m28}
\end{figure}
%

%

\begin{figure}[tb]
\includegraphics[width=1.0\textwidth]{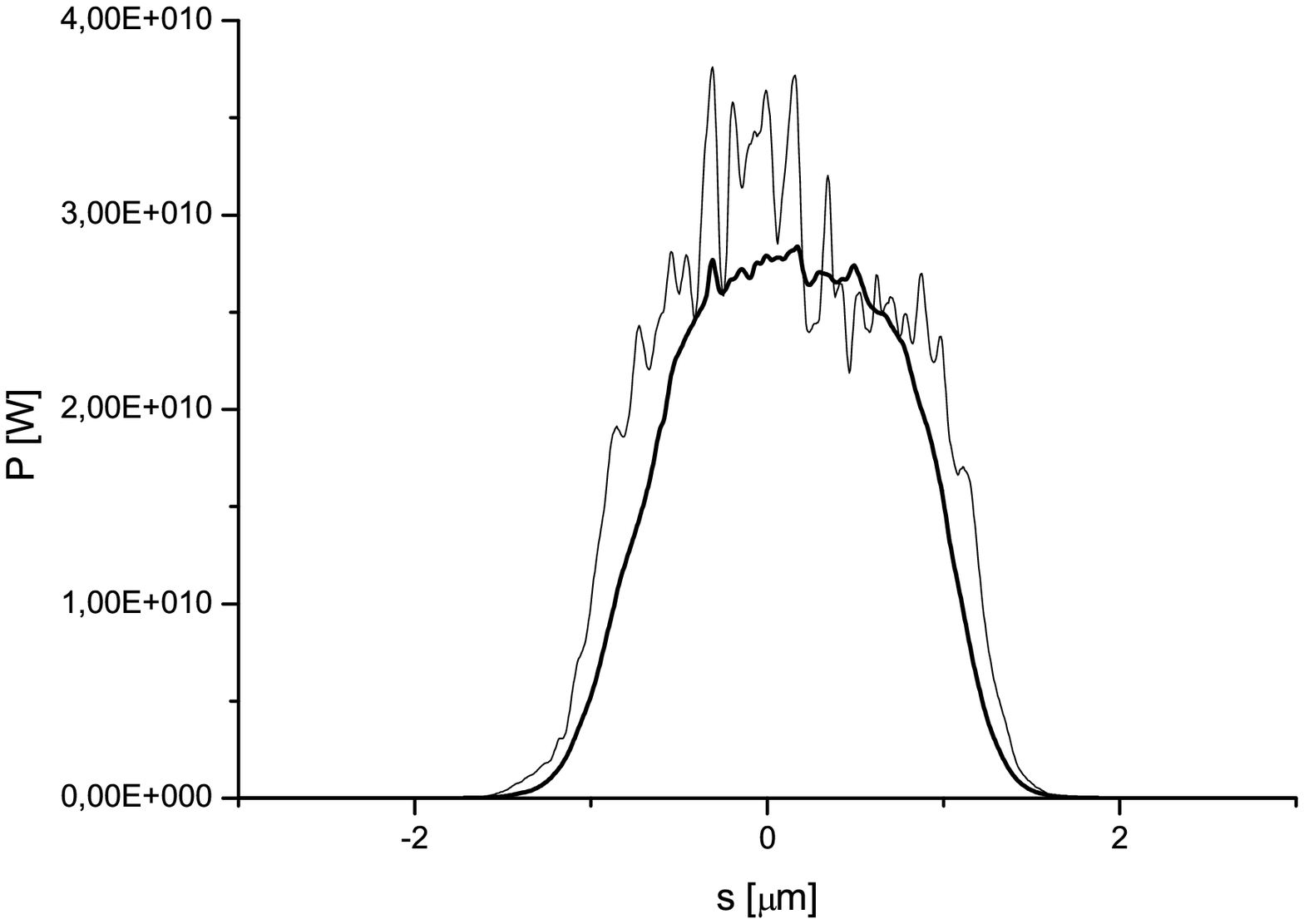}
\caption{Average (bold) and typical single-shot temporal structure
of the radiation pulse at the second undulator length of 72 m (12
cells) for the Si(111) reflection case.} \label{m24}
\end{figure}

\begin{figure}[tb]
\includegraphics[width=1.0\textwidth]{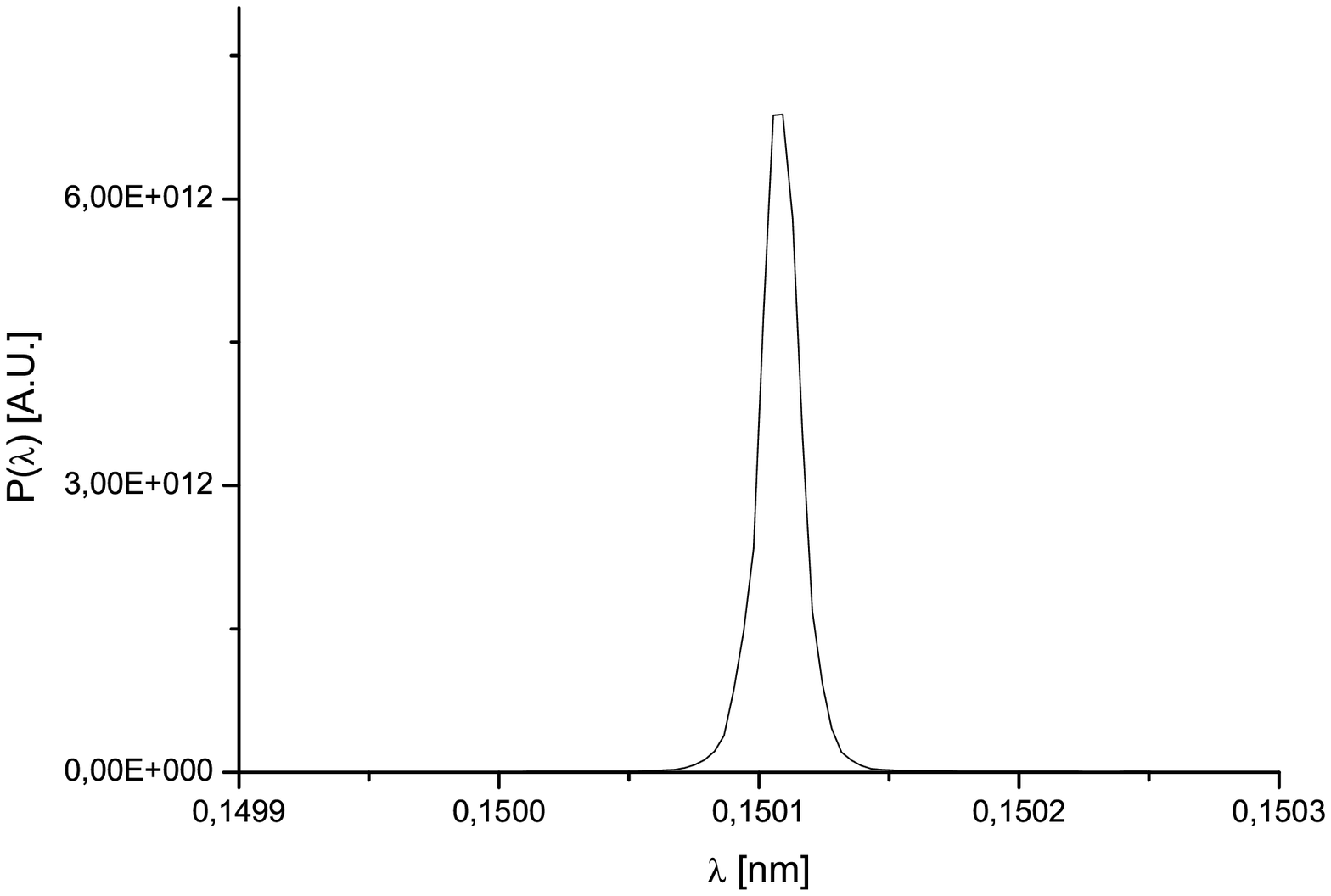}
\caption{Averaged spectrum of radiation at the second undulator
length of 72 m (12 cells) for the Si(111) reflection case.}
\label{m25}
\end{figure}

\begin{figure}[tb]
\includegraphics[width=1.0\textwidth]{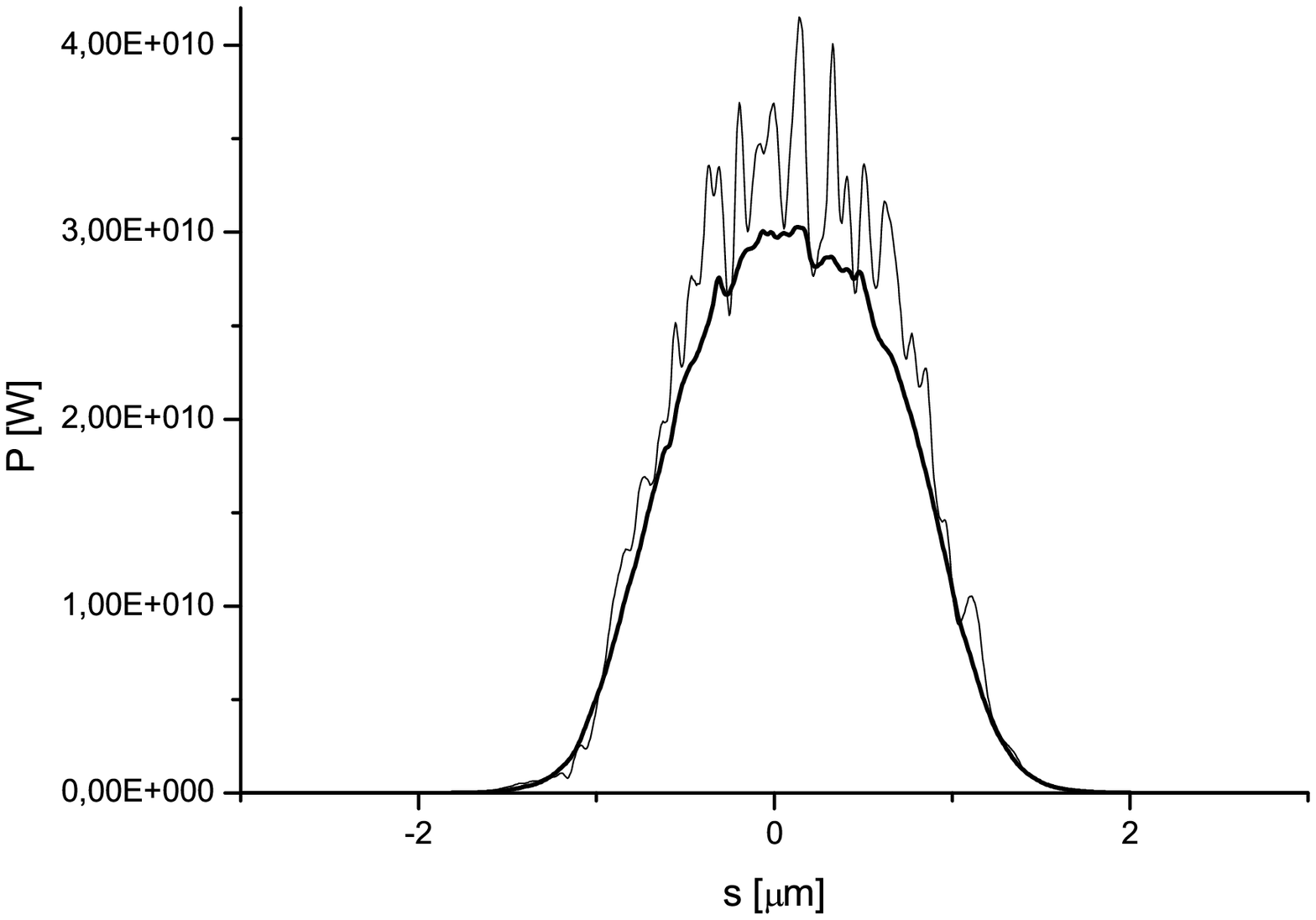}
\caption{Average (bold) and typical single-shot temporal structure
of the radiation pulse at the second undulator length of 72 m (12
cells) for the Si(220) reflection case.} \label{m30}
\end{figure}

\begin{figure}[tb]
\includegraphics[width=1.0\textwidth]{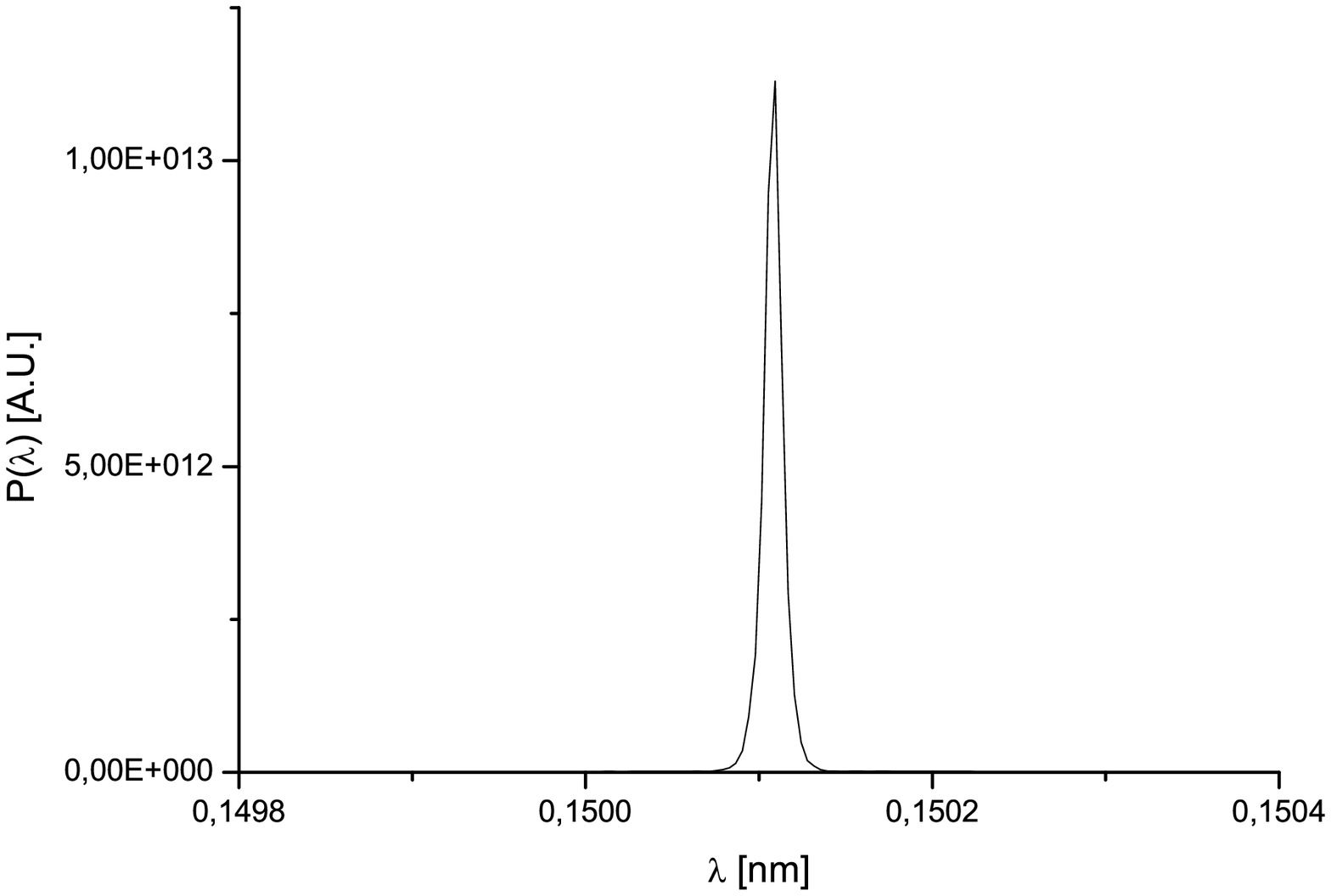}
\caption{Averaged spectrum of radiation at the second undulator
length of 72 m (12 cells) for the Si(220) reflection case.}
\label{m31}
\end{figure}
%


\begin{figure}[tb]
\includegraphics[width=1.0\textwidth]{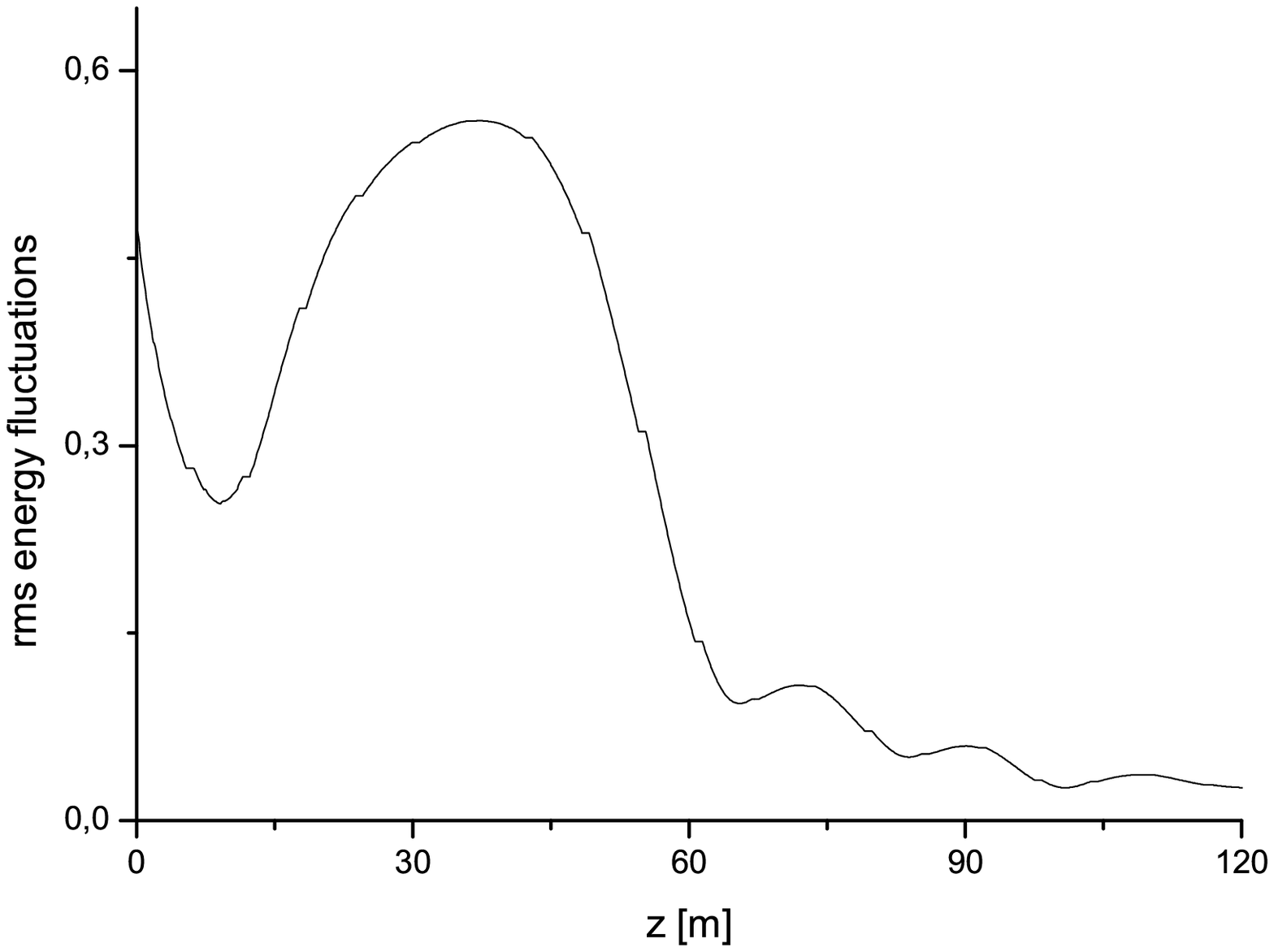}
\caption{RMS pulse energy fluctuations along the second undulator
for the Si(111) reflection case.} \label{m27}
\end{figure}

\begin{figure}[tb]
\includegraphics[width=1.0\textwidth]{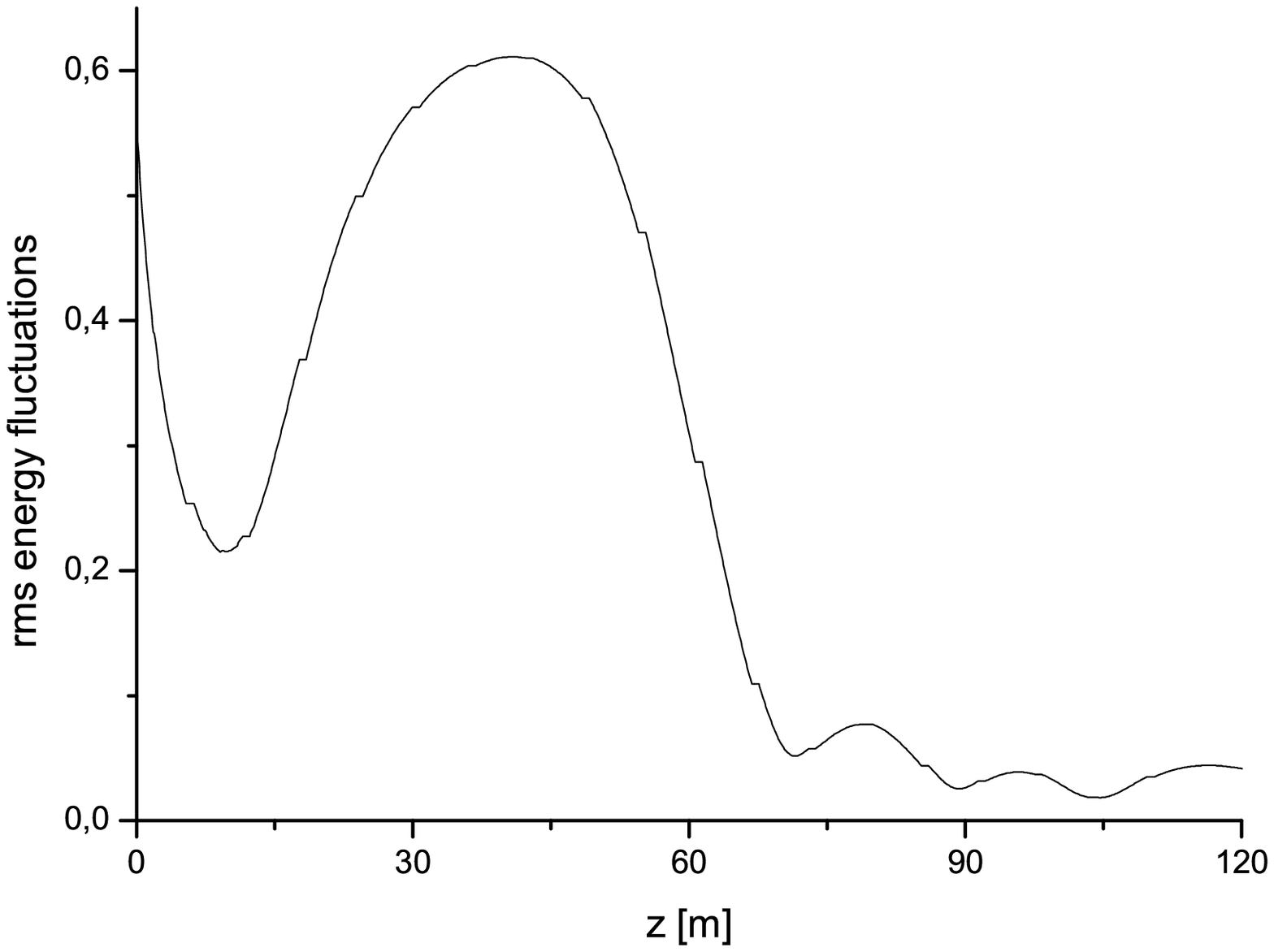}
\caption{RMS pulse energy fluctuations along the second undulator
for the Si(220) reflection case.} \label{m33}
\end{figure}

Following the monochromator, we consider the outcome from the
second undulator part, either using the Si(111) or the Si(220)
reflection. These results constitute the output of our scheme, and
demonstrate that we can obtain nearly Fourier-limited, ten-GW
power level pulses of SASE radiation in the Angstrom wavelength
range. The pulse energy, averaged over 100 shots, is presented as
a function of the position along the second undulator in Fig.
\ref{m22} for the case of the Si(111) reflection, and in Fig.
\ref{m28} for the case of the Si(220) reflection. One may see that
there is an optimal length of the second undulator. In both cases,
saturation is expected around $70$ m. We therefore consider the
output after $12$ cells. The shot-to-shot fluctuations of the
radiation pulse after $12$ cells was already studied in Fig.
\ref{m11}.  Typical single-shot and averaged (over 200 shots)
temporal structure and spectra of the radiation pulse at the
undulator length of $72$ m (corresponding to 12 cells) are
presented in Fig. \ref{m24} and Fig. \ref{m25} for the Si(111)
case, and in Fig. \ref{m30} and Fig. \ref{m31} for the Si(220)
case. It can be seen that the contribution of the shot noise to
the total power is small. An important characteristic of the
radiation is the width of the radiation spectrum. The FWHM of the
radiation spectra at the undulator length of $72$ m for the
Silicon 111 and the Silicon 220 are about $10^{-4}$ and $8\cdot
10^{-5}$ respectively, and are mainly defined by the finite pulse
duration.

It is interesting to discuss the origin of the asymmetric field
profile in Fig. \ref{m12} for the Si(111) reflection. Such
asymmetry cannot be explained in terms of phase of the
monochromator filter (which is discussed in Appendix), as the
filter is considered real, in our calculations. It cannot be
explained in terms of a spurious result due to finite statistics
either. Therefore, we ascribe this effect to a physical
phenomenon. The first stage operates in the deep linear regime.
Therefore, analytical methods can be applied to describe the
situation. In particular one can use the analytical solution for
the initial value problem in the high-gain limit and in the
frequency domain (the FEL Green's function). The growing root of
the eigenvalue equation can be expanded near exact resonance. As a
result, the imaginary part of the FEL Green's function includes a
term proportional to the square of the detuning parameter. The
asymmetry takes place due to this quadratic dependence of the
imaginary part of Green's function on the frequency. Effects
related to the FEL Green's function phase have been studied in
\cite{SCOM} for both HGHG and self-seeding cases. It should be
added that such effect is only important when the width of the
filter is comparable with the interval of spectral coherence, and
can be neglected when the spectral width of the filter is a few
times shorter than the spectral coherence interval.  Therefore,
this effect is only important for the Si(111) reflection and for
the case of a short pulse, while it is much weaker in all other
cases.

A typical figure of merit used to characterize the radiation from
third and fourth generation light sources is the brilliance. One
can estimate the foreseen brilliance for the cases discussed above
by inspecting Figs. \ref{m25}-\ref{m31} to obtain the FWHM
temporal duration and spectral width of the radiation pulses. Once
the pulse duration ($4$ fs FWHM) and spectral width ($0.008 \%$
FWHM) are calculated, knowing the number of photons per pulse
($1.1 \cdot 10^{11}$) and assuming perfect transverse coherence
one estimates a brilliance in the order of $5\cdot 10^{34}
\mathrm{ph/s/mm^2/mrad^2/0.1\% BW}$, which is about one order of
magnitude larger than the design parameter.

It is interesting to discuss the behavior of the rms pulse energy
fluctuations along the undulator, which is shown are Fig.
\ref{m27} and Fig. \ref{m33} for the Si(111) and for the Si(220)
case respectively. As one may see, at the beginning of the
undulator the fluctuations of energy per pulse apparently drop.
This can be explained considering the fact that the Genesis output
consists of the total power integrated over the full grid up to an
artificial boundary, i.e. there is no spectral selection.
Therefore, Fig. \ref{m27} includes a relatively large spontaneous
emission background, which has a much larger spectral width with
respect to the amplification bandwidth and which fluctuates
negligibly from shot to shot. Since there is a long lethargy of
the seeded radiation at the beginning of the FEL amplifier, one
observes an apparent decrease of fluctuations. Then, when lethargy
ends, the seed pulse gets amplified and fluctuations effectively
return to the same level as after monochromator\footnote{Actually
fluctuations return to a slightly higher level compared to that
after the monochromator. This can be explained considering the
fact that we deal with a parametric amplifier, where the
properties of the active medium, i.e. the electron beam, depend on
time. The duration of the electron bunch is shorter compare to the
duration of the seed pulse. Therefore, the amplified pulse has
effectively higher coherence compared to the full radiation pulse
after the monochromator, hence the higher fluctuation level.}. The
energy fluctuations are drastically reduced when the FEL amplifier
operates in the nonlinear regime. The minimal value of the
fluctuations\footnote{The oscillations of the energy fluctuations
after this point are connected with the growth of spikes in the
nonlinear medium, since we have taken into account the shot noise
in the electron beam. The spectrum gets broader due to such spikes
growth. The total power can even increase, while the maximum value
of the spectral density decreases.}, around $10 \%$, occurs at the
undulator length of $72$ m (12 cells).

Finally, let us briefly analyze the heat loading problem. For the
short pulse mode of operation and for the European XFEL case we
have an average power of 5 mW ($150$ nJ $\times 3000$ pulses/train
$\times 10$ trains/s). This corresponds to a normal incident power
density\footnote{We consider a transverse rms dimension of the
bunch of about $20 \mu$m. Assuming, with some approximation, that
radiation is distributed as the electron bunch, we obtain an area
of $2.4 \cdot 10^{-3} \mathrm{mm^2}$.}  of $2 \mathrm{W/mm^2}$ at
the position of the monochromator located in baseline undulator,
Fig. \ref{m2}. For the double bunch scheme, this number should be
increased of a factor two, up to $4 \mathrm{W/mm^2}$. This is more
than two orders of magnitude smaller compared with the power at
monochromators of third generation synchrotron sources ($200
\mathrm{W/mm^2}$).  The problem of heat loading obviously does not
exist at all for LCLS, due to the much lower repetition rate, and
all our results can be applied for LCLS without any limitations
concerning heat loading of the monochromator.

\clearpage

\subsection{Three-stage scheme: combination of self-seeding scheme and fresh bunch technique}

Combination of our self-seeding scheme with a fresh bunch
technique has been schematically illustrated in Figs.
\ref{m15}-\ref{m17}. The first undulator part is identical as
before. However, now, between first and second part of the
undulator a magnetic chicane is introduced, which washes out the
electron beam microbunching and provides a relative delay between
electron beam and radiation. Then, the head of the electron bunch
is seeded by the radiation from the first undulator part and
saturates faster in the second part of the undulator.

\begin{figure}[tb]
\includegraphics[width=1.0\textwidth]{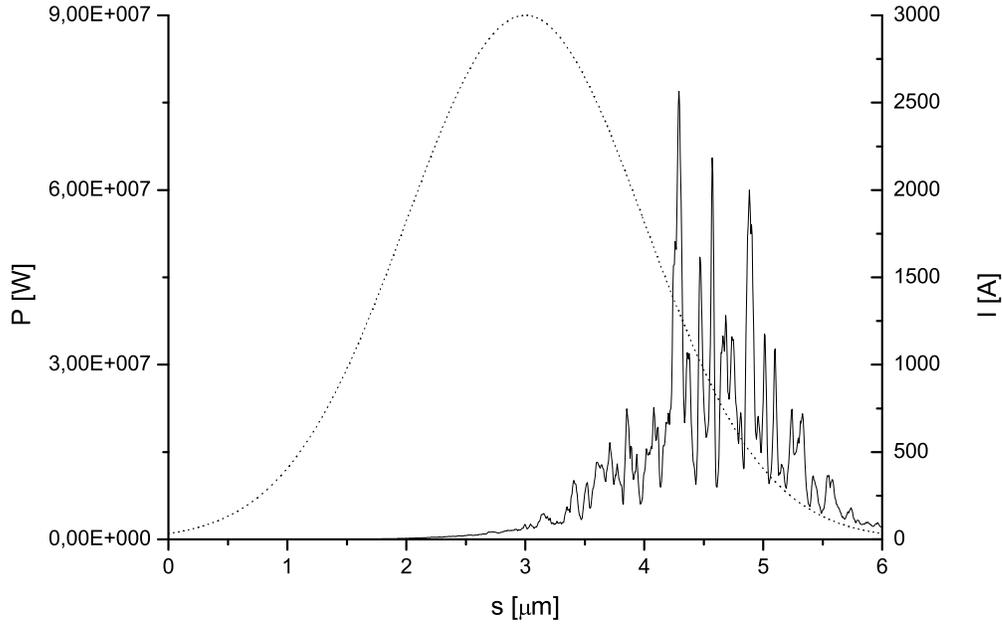}
\caption{Short pulse mode operation, combination of self-seeding
and fresh bunch techniques. Power distribution after the first
undulator part (7 cells) and the magnetic delay. Radiation is used
as input for the second undulator part (7 cells).} \label{mf1}
\end{figure}
\begin{figure}[tb]
\includegraphics[width=1.0\textwidth]{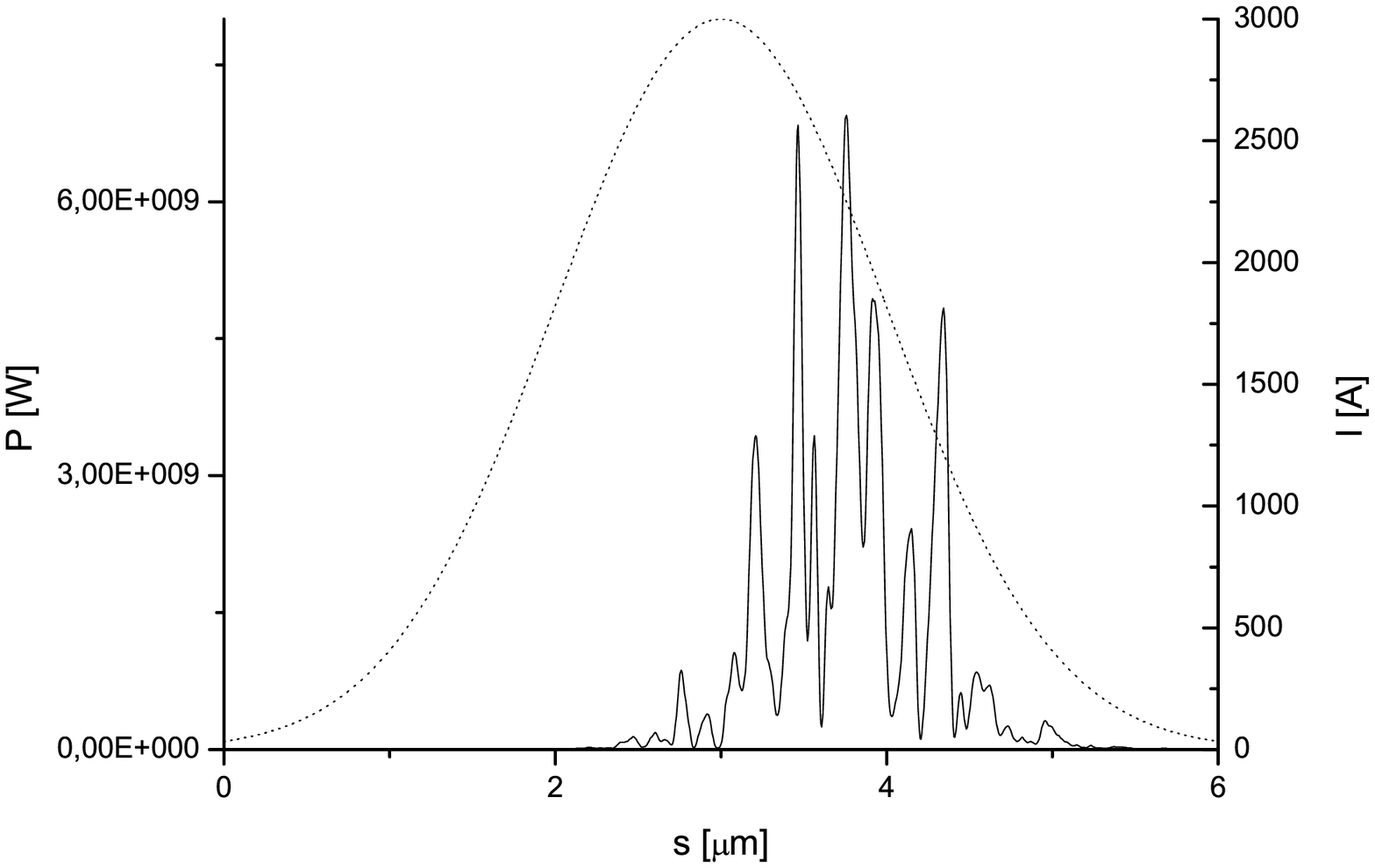}
\caption{Short pulse mode operation, combination of self-seeding
and fresh bunch techniques. Output power at the end of the second
stage, 7 cells long (42 m).} \label{mf2}
\end{figure}
\begin{figure}[tb]
\includegraphics[width=1.0\textwidth]{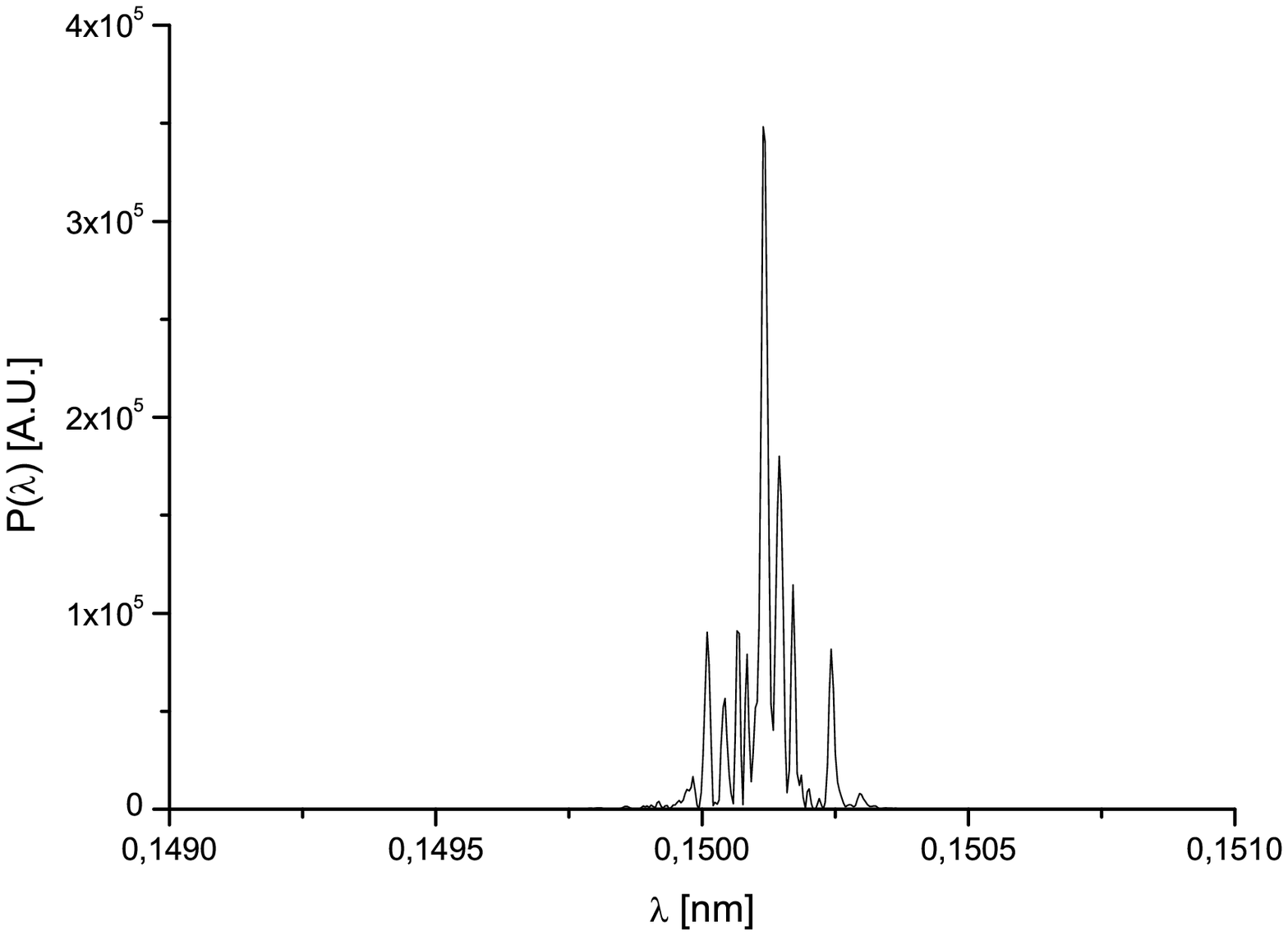}
\caption{Short pulse mode operation, combination of self-seeding
and fresh bunch techniques. Output spectrum at the end of the
second stage, 7 cells long (42 m).} \label{mf3}
\end{figure}
\begin{figure}[tb]
\includegraphics[width=1.0\textwidth]{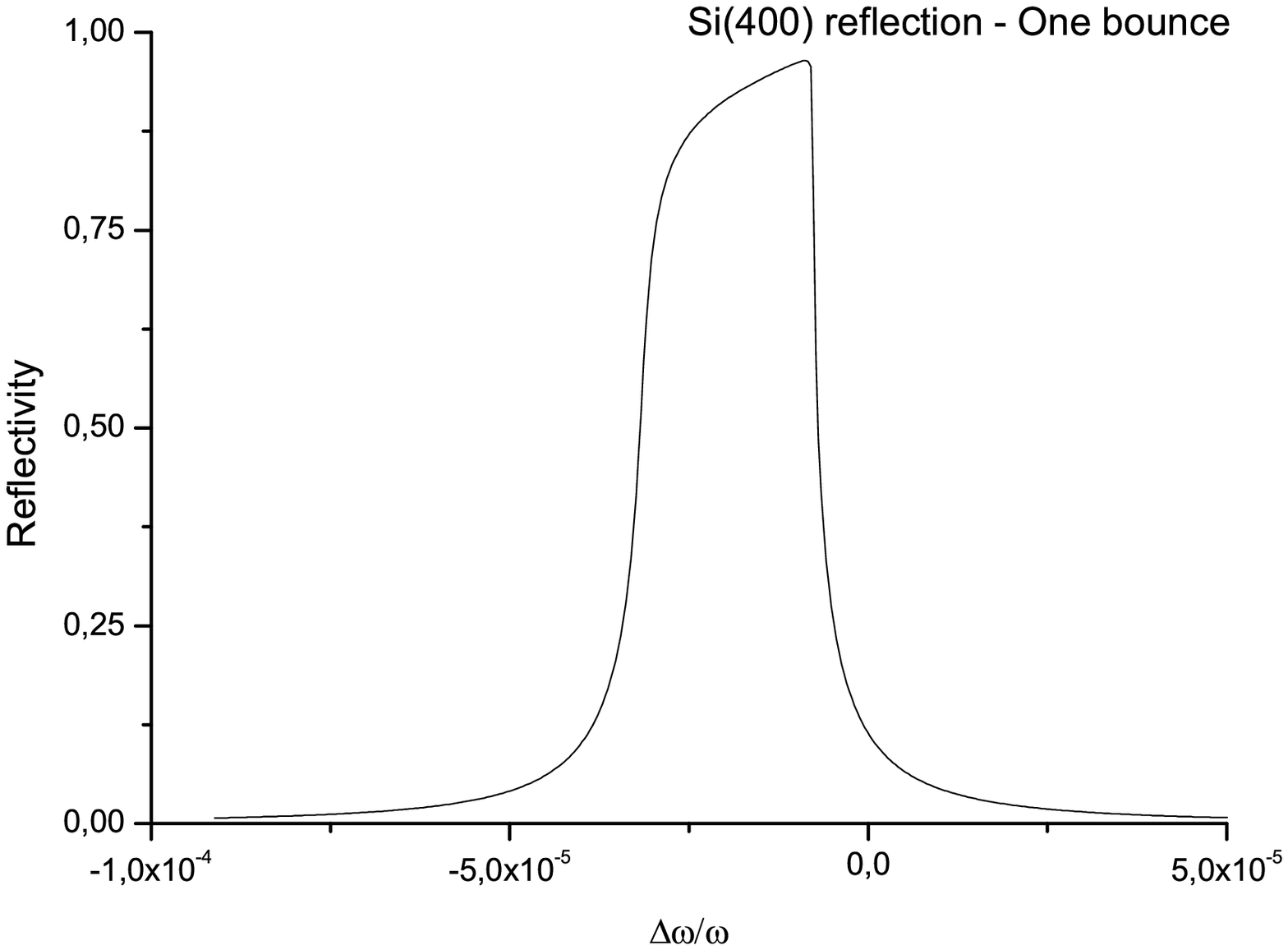}
\caption{Reflectivity curve for a thick absorbing crystal in Bragg
geometry. Si(400) reflection of 0.15 nm X-rays. One bounce.}
\label{mf4}
\end{figure}
\begin{figure}[tb]
\includegraphics[width=1.0\textwidth]{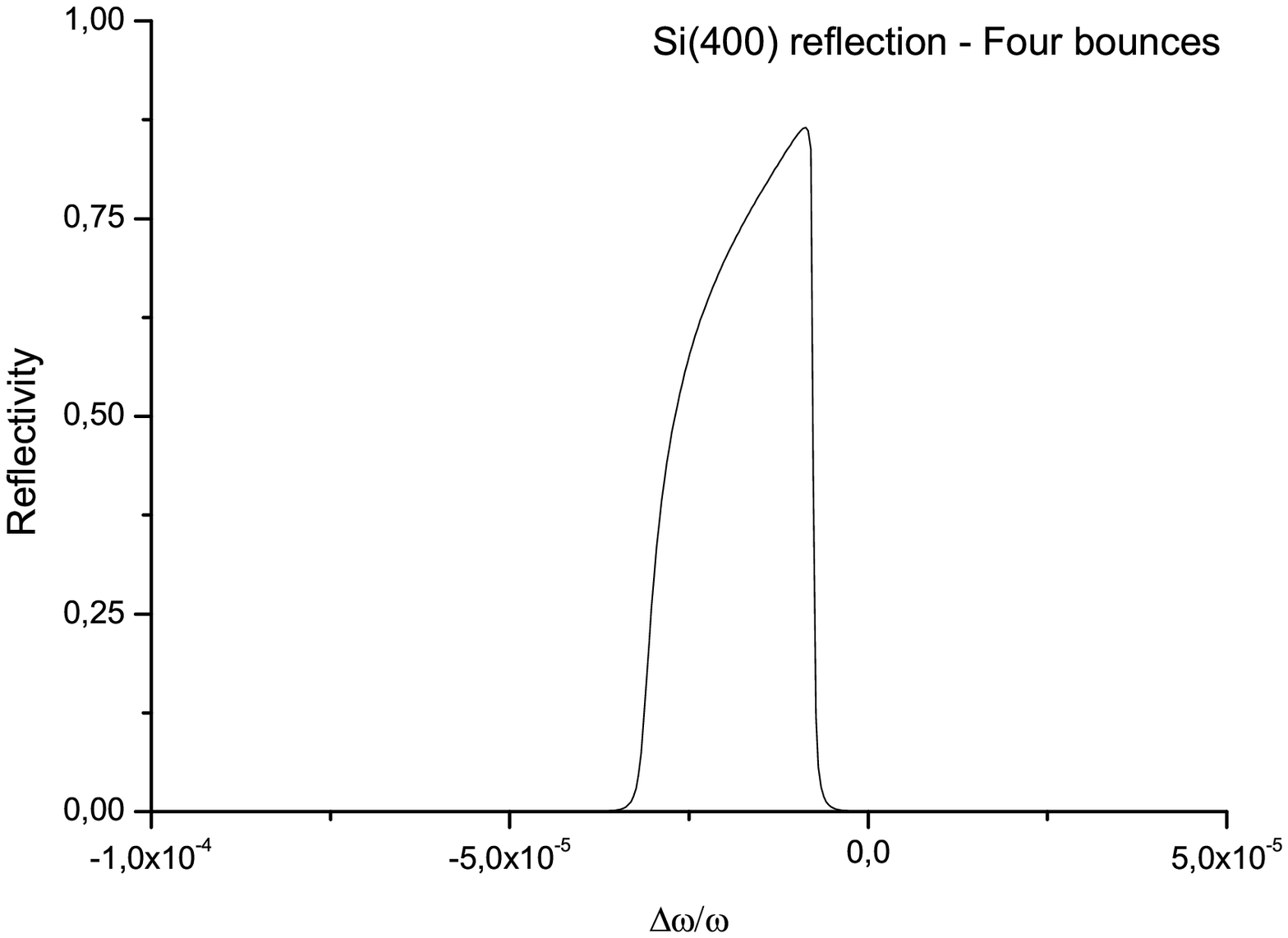}
\caption{Reflectivity curve for a thick absorbing crystal in Bragg
geometry. Si(400) reflection of 0.15 nm X-rays. Four bounces.}
\label{mf5}
\end{figure}
\begin{figure}[tb]
\includegraphics[width=1.0\textwidth]{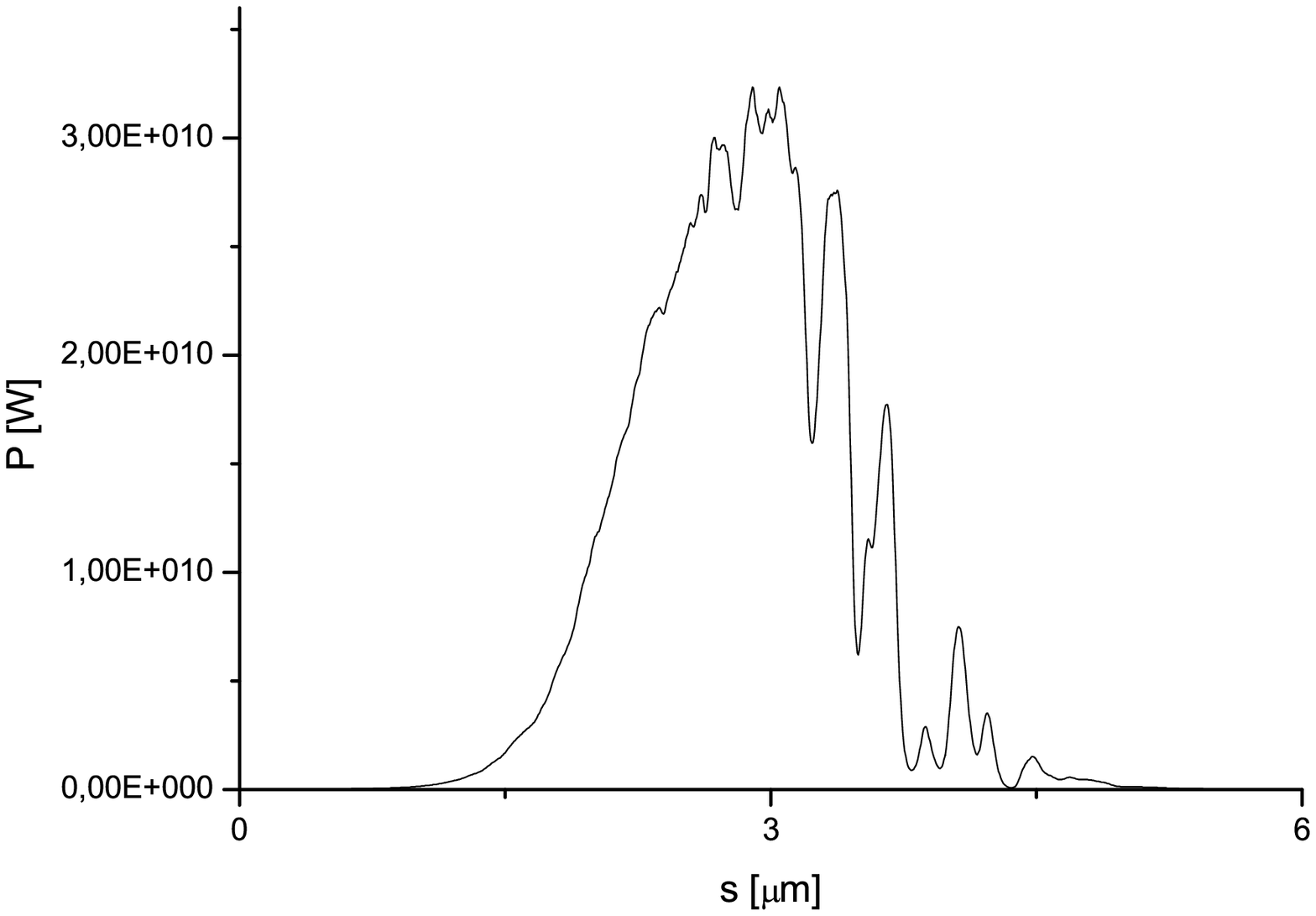}
\caption{Short pulse mode operation, combination of self-seeding
and fresh bunch techniques. Four bounce monochromator, Si(400)
reflection. Output power at the end of the third stage, 8 cells
long (53 m).} \label{mf6}
\end{figure}
\begin{figure}[tb]
\includegraphics[width=1.0\textwidth]{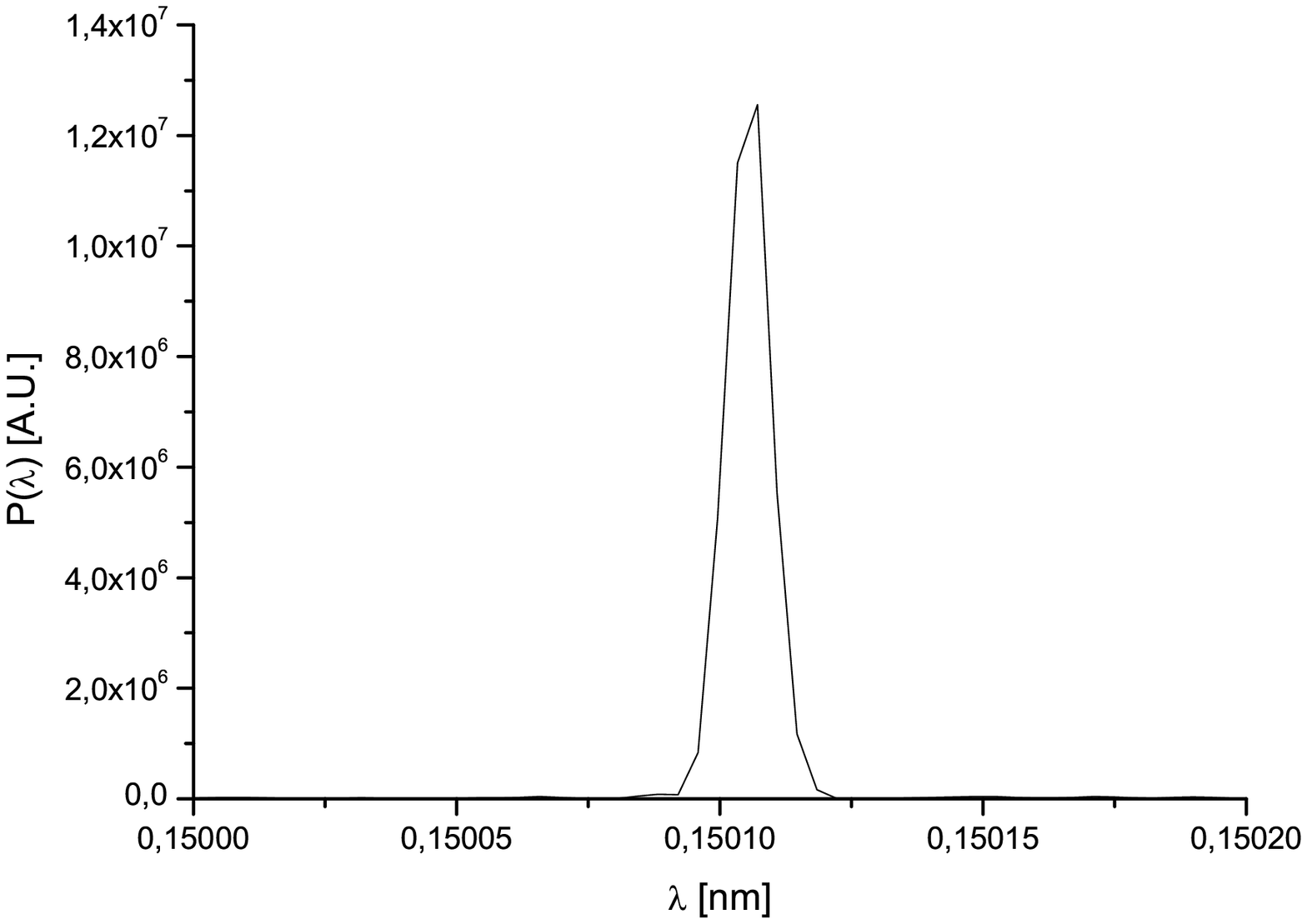}
\caption{Short pulse mode operation, combination of self-seeding
and fresh bunch techniques. Four bounce monochromator, Si(400)
reflection. Output spectrum at the end of the third stage, 8 cells
long (53 m).} \label{mf7}
\end{figure}

The first stage is 7-cells long, and the properties of the
radiation at its exit have already been illustrated in Fig.
\ref{m9} and in Fig. \ref{m14}. In Fig. \ref{mf1} we illustrate
the input power at the second stage. Fig. \ref{mf2} shows the
output power from the second stage, which is $7$-cells long, while
the correspondent spectrum is given in Fig. \ref{mf2}. Following
the second part of the undulator, the electron bunch enters the
self-seeding stage, where radiation is monochromatized. In Fig.
\ref{mf4} and Fig. \ref{mf5} we show the reflectivity curves for a
thick absorbing crystal in the case of the Si(400) reflection, for
a single bounce and for four bounces. With the help of the fresh
bunch technique we now can use, for monochromatization purposes,
reflections like Si(400), Si(440) and Si(444). The field profiles
using these kind of reflections have already been shown in Fig.
\ref{m34}. As noted before, the jitter acceptance is obviously
linked to the FWHM width of these plots, i.e. to the temporal line
width of the monochromator. Since Si(400), Si(440) and Si(444)
reflections are associated to a smaller bandwidth, one can take
advantage of an increased jitter acceptance, which practically
cancels the background from the first electron bunch for the
double bunch scheme. The third undulator part follows, where the
fresh part of the bunch is seeded by the monochromatized
radiation. Output power and spectrum at the end of the $8$-cells
setup are shown in Fig. \ref{mf6} and Fig. \ref{mf7} respectively.
From Fig. \ref{mf7} one can see that radiation is nearly fully
longitudinally coherent.

As regards heat-loading issues, from Fig. \ref{mf2} follows that
one has an average power of about $3$ GW, which is two order of
magnitude larger than the $30$ MW considered in the previous
two-stage scheme. Since the pulse is twice shorter we have a power
density of about $200~ \mathrm{W/mm^2}$, still within an
acceptable value.


%


\section{Feasibility study for the long pulse mode of operation}

We also performed a feasibility study for the long-pulse mode of
operation. We assume that parameters in Table \ref{tt1} are still
valid, except for a ten times larger charge ($0.25$ nC) and a ten
times longer rms bunch length ($10~\mu$m). We considered both the
two-stage scheme and a combination of self-seeding and fresh-bunch
techniques. In the following we will discuss outcomes from a
single-shot simulation.

\subsection{Two stage-scheme}

\begin{figure}[tb]
\includegraphics[width=1.0\textwidth]{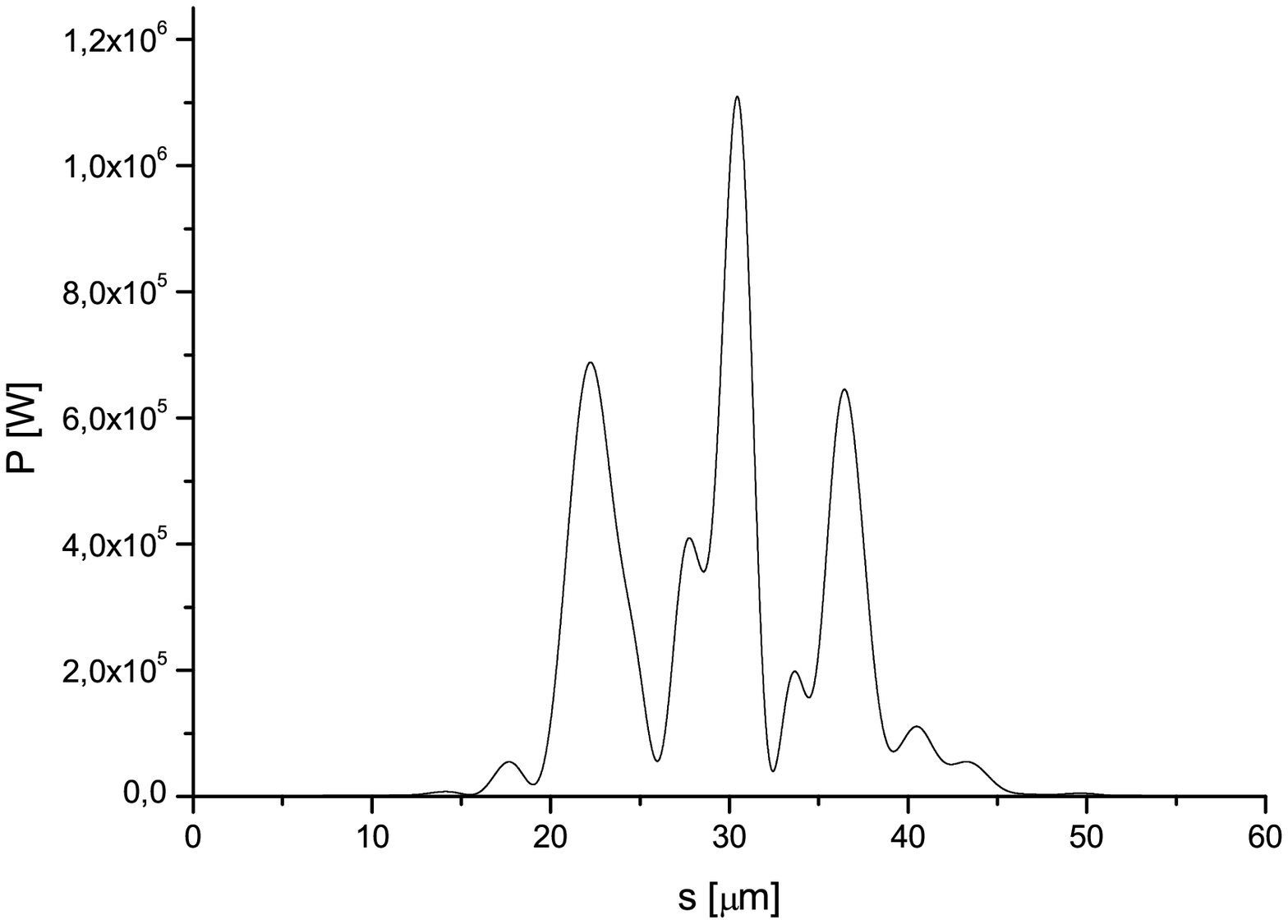}
\caption{Long pulse mode of operation. Power after four-bounce
monochromator, Si(220) reflection.  Radiation is used as input for
the second undulator part.} \label{ml1}
\end{figure}

\begin{figure}[tb]
\includegraphics[width=1.0\textwidth]{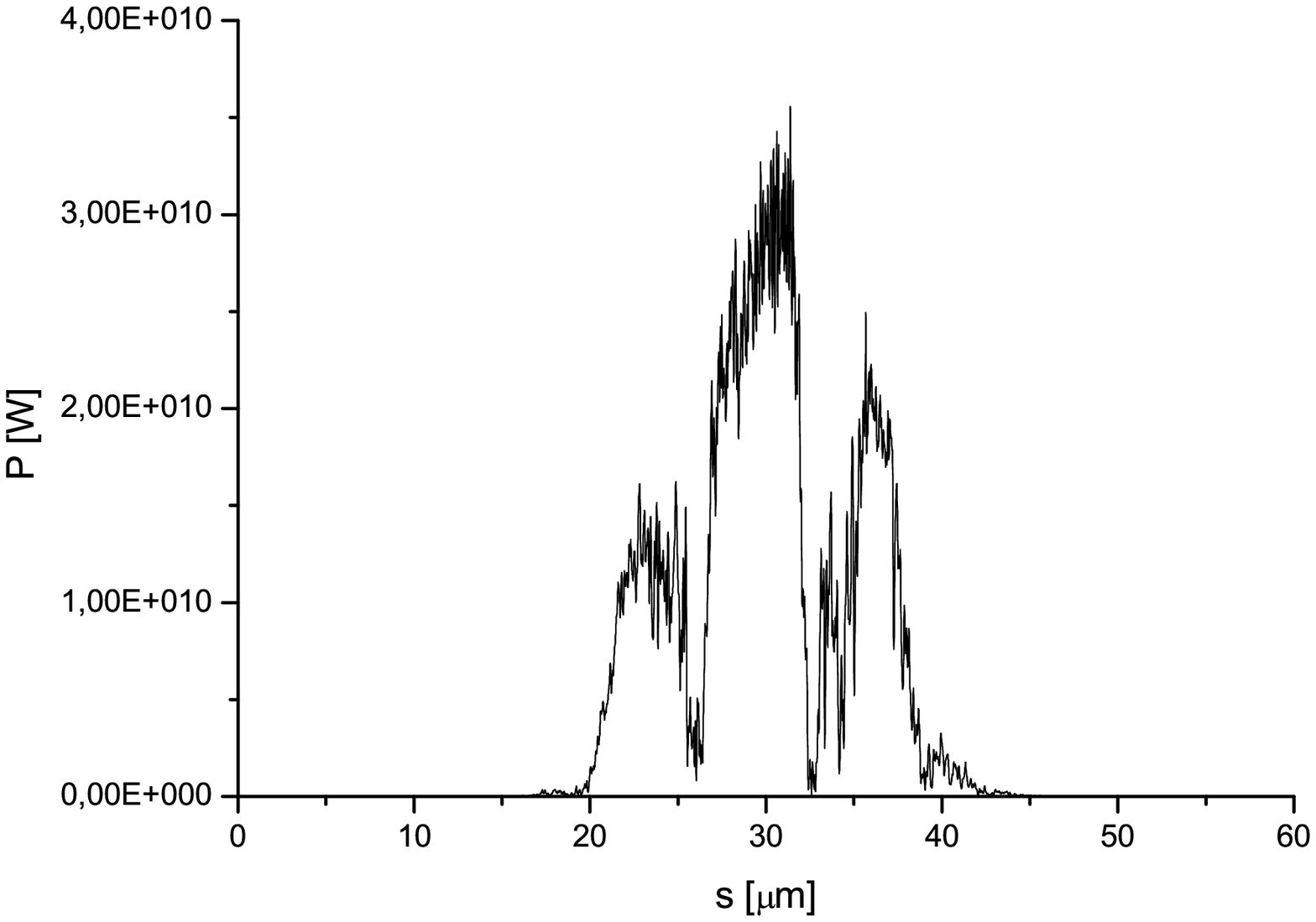}
\caption{Long pulse mode of operation, Si(220) reflection. Power
at exit of the  second undulator, with a length of 72 m (12
cells). } \label{ml3}
\end{figure}

\begin{figure}[tb]
\includegraphics[width=1.0\textwidth]{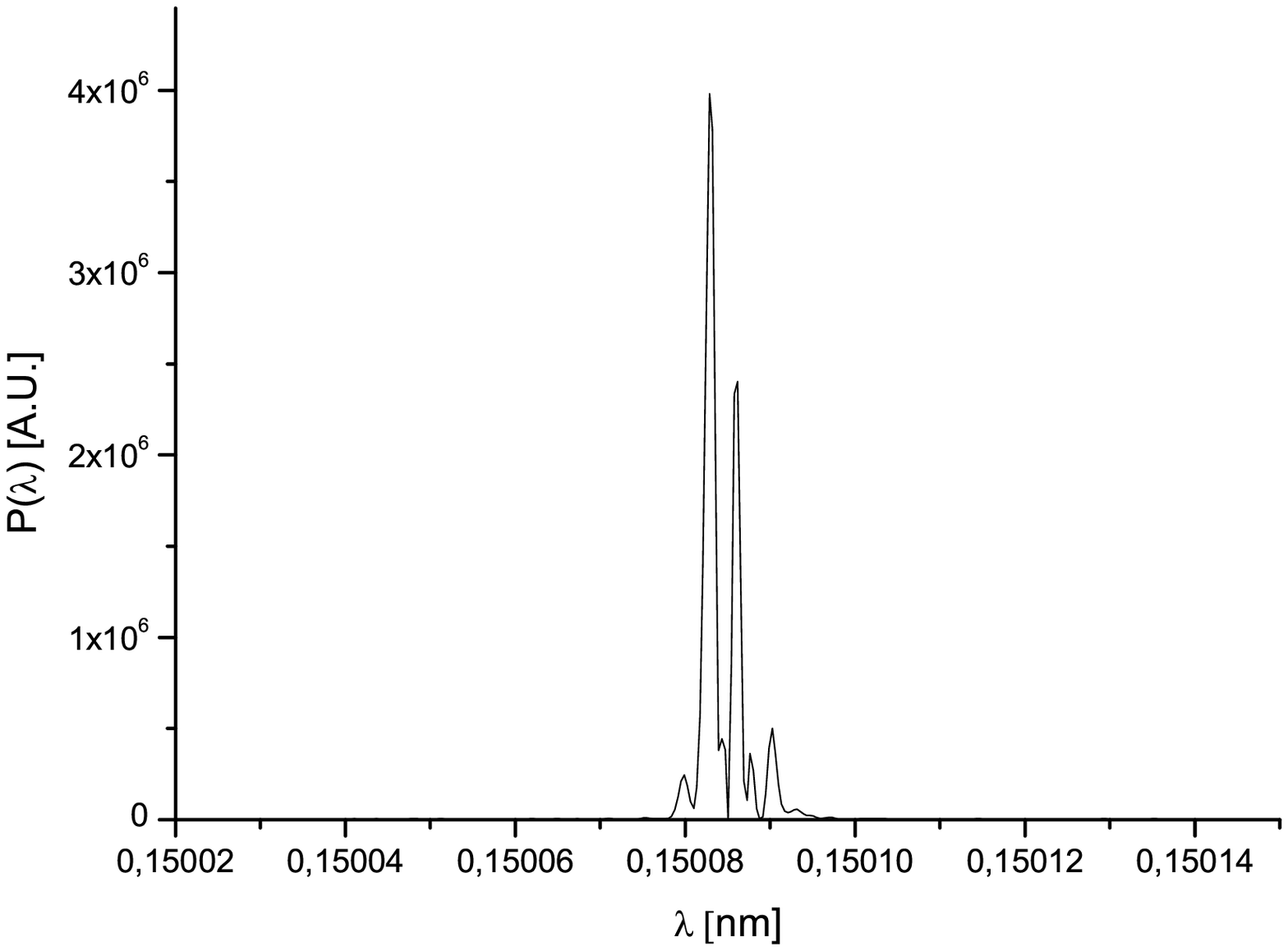}
\caption{Long pulse mode of operation, Si(220) reflection.
Spectrum at exit of the  second undulator, with a length of 72 m
(12 cells). } \label{ml5}
\end{figure}

The first stage is 7-cells long. First we consider the Si(220)
reflection case. The power after the first undulator part and
after monochromatization is shown in Fig. \ref{ml1}. The output of
our scheme is shown in Fig. \ref{ml3}, illustrating the power
after the second undulator part, and in Fig. \ref{ml5},
illustrating the spectrum. The presence of a few longitudinal
modes can be guessed by inspection of Fig. \ref{ml3} and Fig.
\ref{ml5}.


\begin{figure}[tb]
\includegraphics[width=1.0\textwidth]{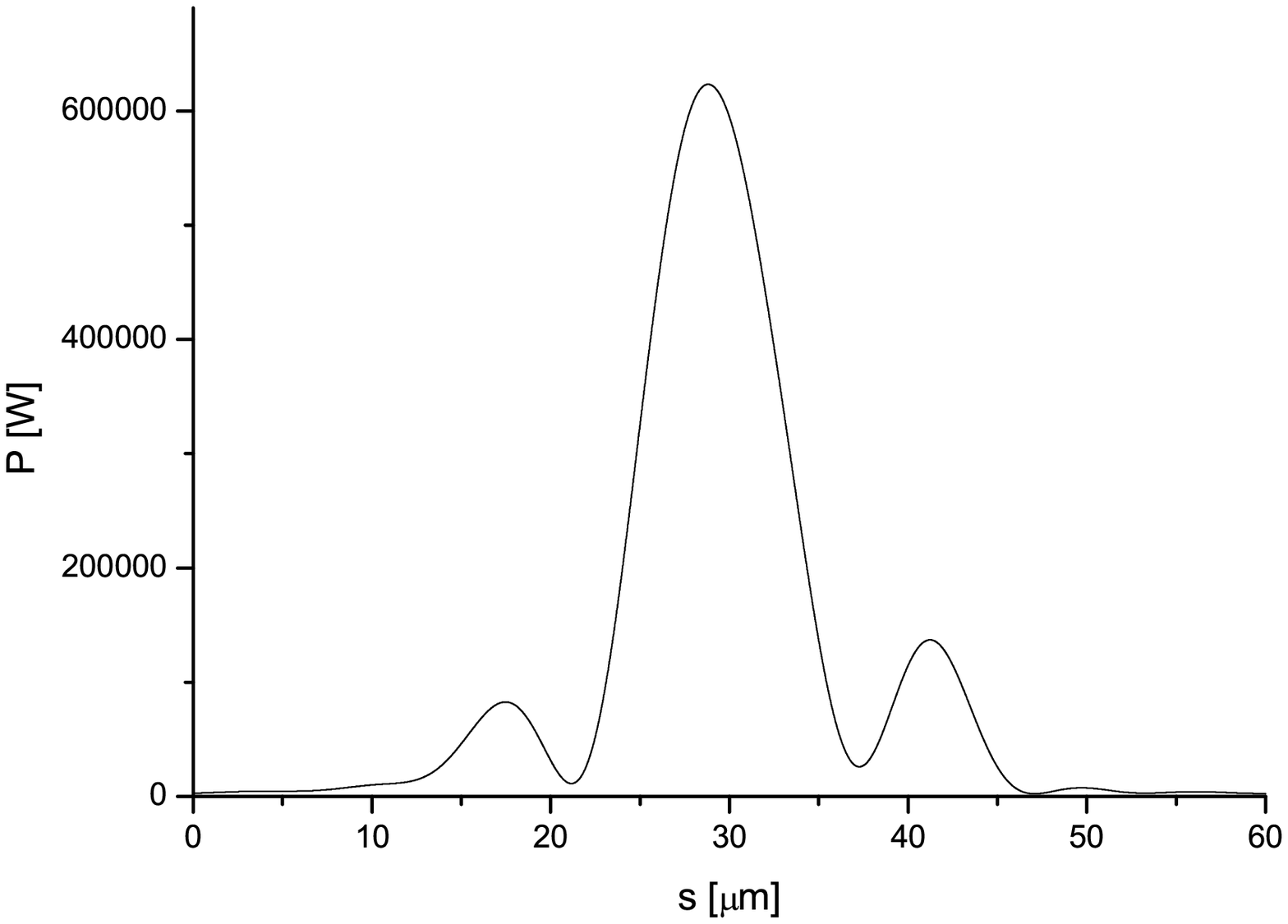}
\caption{Long pulse mode of operation. Power after the four-bounce
monochromator, Si(400) reflection.  Radiation is used as input for
the second undulator part.} \label{ml2}
\end{figure}

\begin{figure}[tb]
\includegraphics[width=1.0\textwidth]{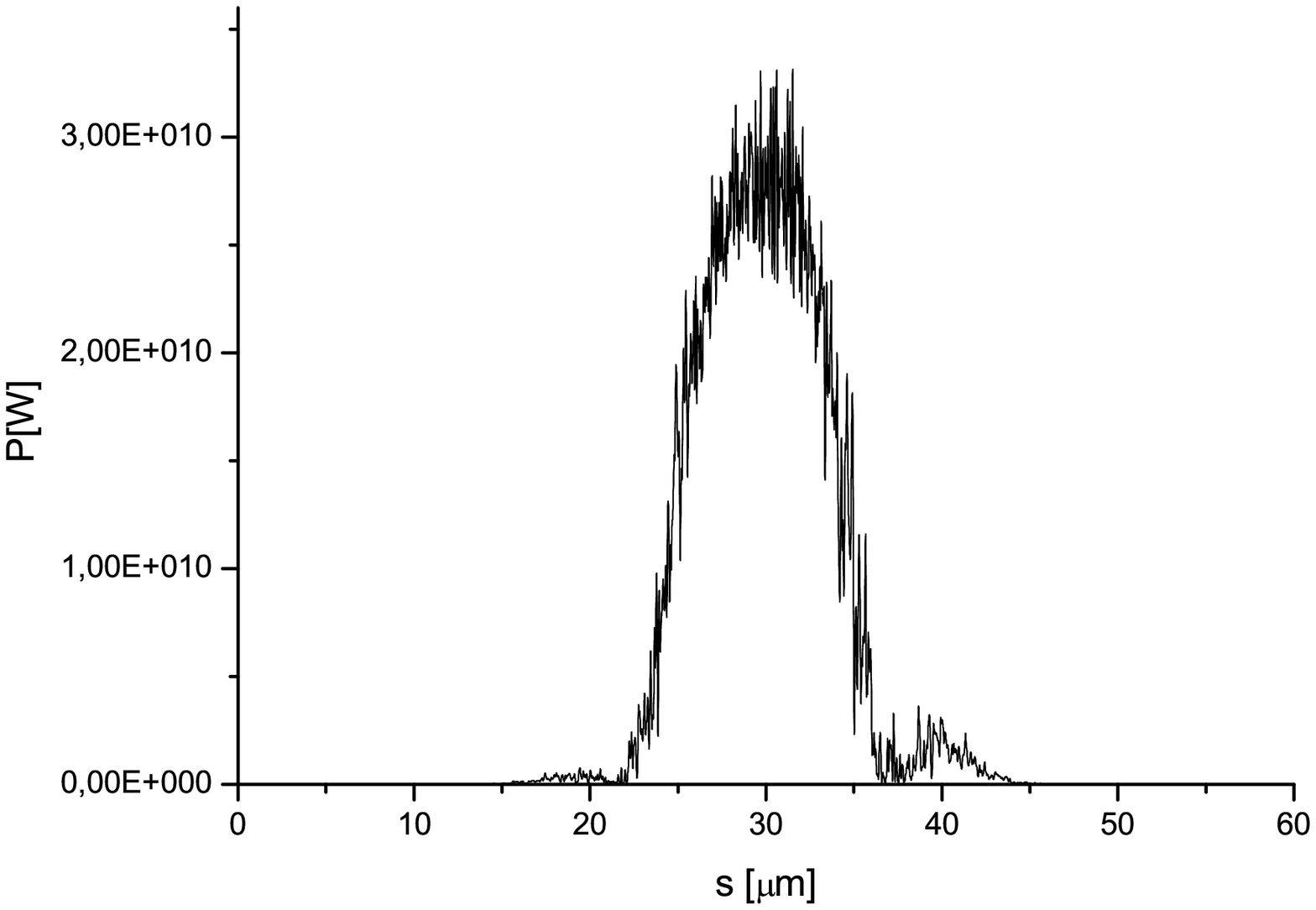}
\caption{Long pulse mode of operation, Si(400) reflection. Power
at the exit of the  second undulator, with a length of 72 m (12
cells). } \label{ml4}
\end{figure}

\begin{figure}[tb]
\includegraphics[width=1.0\textwidth]{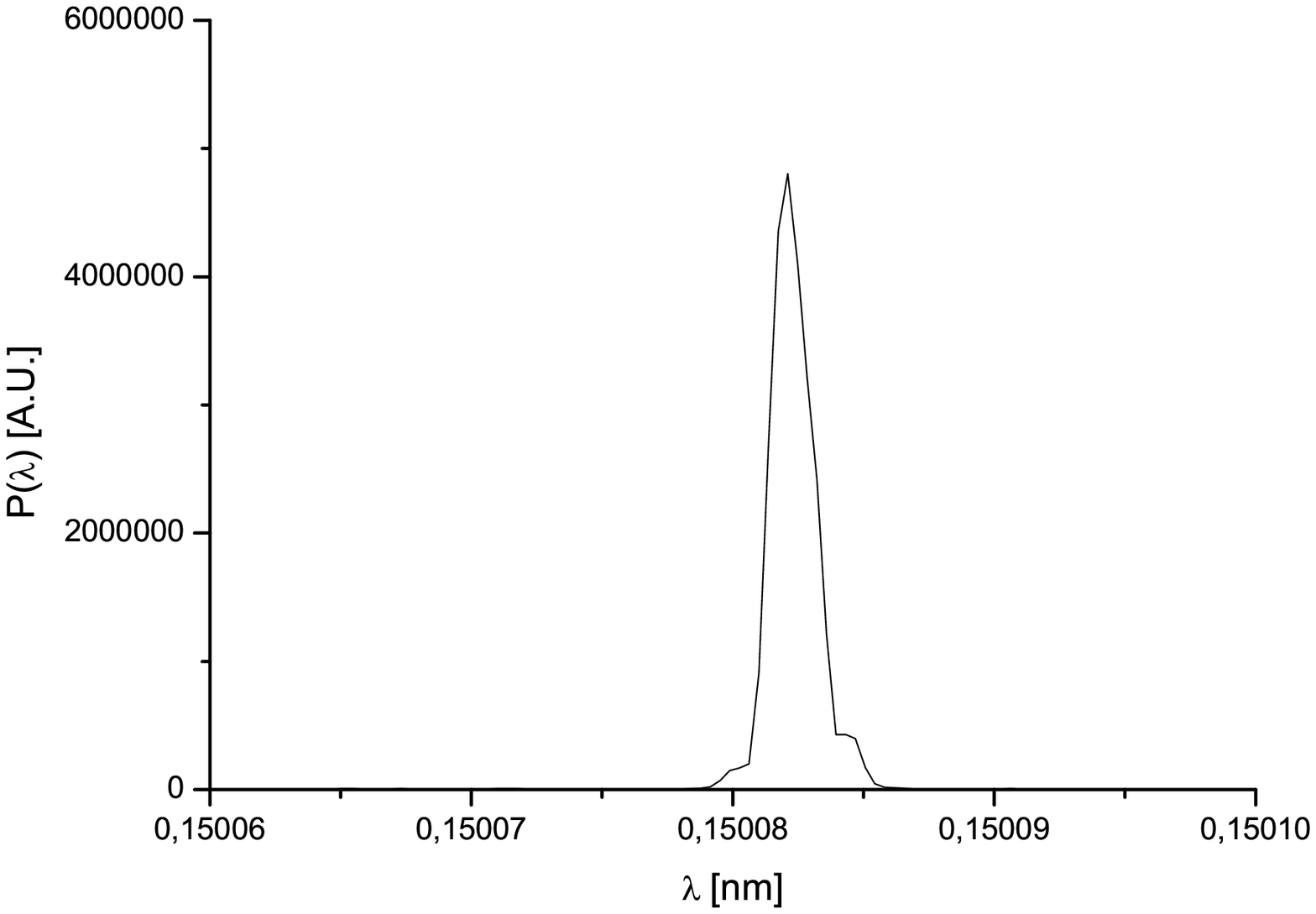}
\caption{Long pulse mode of operation, Si(400) reflection.
Spectrum at the exit of the  second undulator, with a length of 72
m (12 cells). } \label{ml6}
\end{figure}

Subsequently, we analyze the case for the Si(400) reflection. The
power after the first undulator part and after monochromatization
is shown in Fig. \ref{ml2}. The final output is shown in Fig.
\ref{ml4}, illustrating the power after the second undulator part,
and in Fig. \ref{ml6}, illustrating the spectrum. The narrower
monochromator bandwidth results in almost longitudinally coherent
radiation, as it can be guessed by inspection of Fig. \ref{ml4}
and Fig. \ref{ml6}.

Concerning heat-loading problems, it should be noted that the
power density in this case increases of an order of magnitude
compared with the two stage, short pulse mode of operation (as the
rms duration also increases of that magnitude). This results in an
average power density of about $40 \mathrm{W/mm^2}$, which is well
manageable for silicon crystals.

\subsection{Three-stage scheme: combination of self-seeding scheme with fresh bunch technique}

\begin{figure}[tb]
\includegraphics[width=1.0\textwidth]{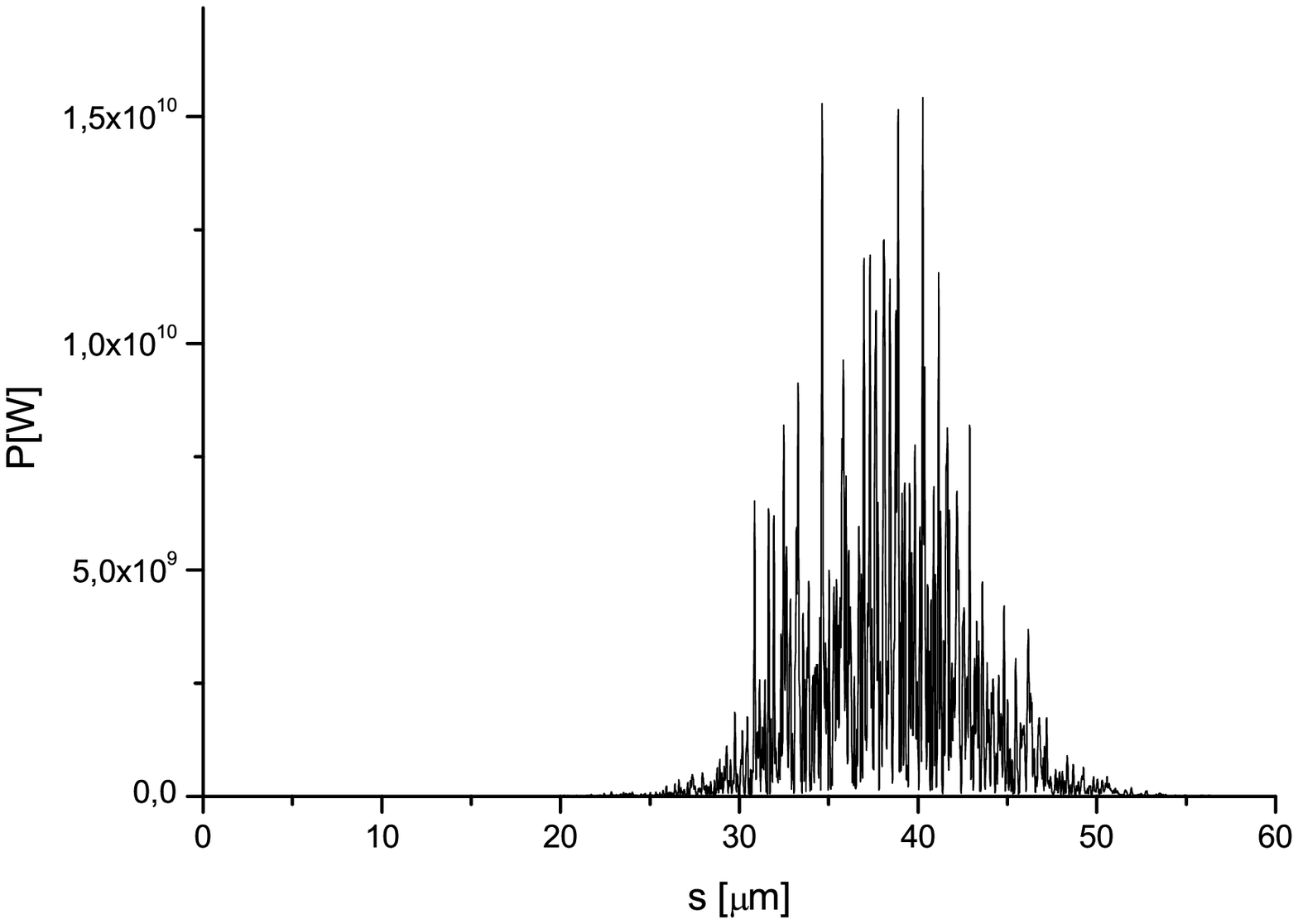}
\caption{Long pulse mode operation. Combination of self-seeding
and fresh bunch techniques. Power at end of the  second stage,
7cells long (42 m).} \label{ml3-1}
\end{figure}

\begin{figure}[tb]
\includegraphics[width=1.0\textwidth]{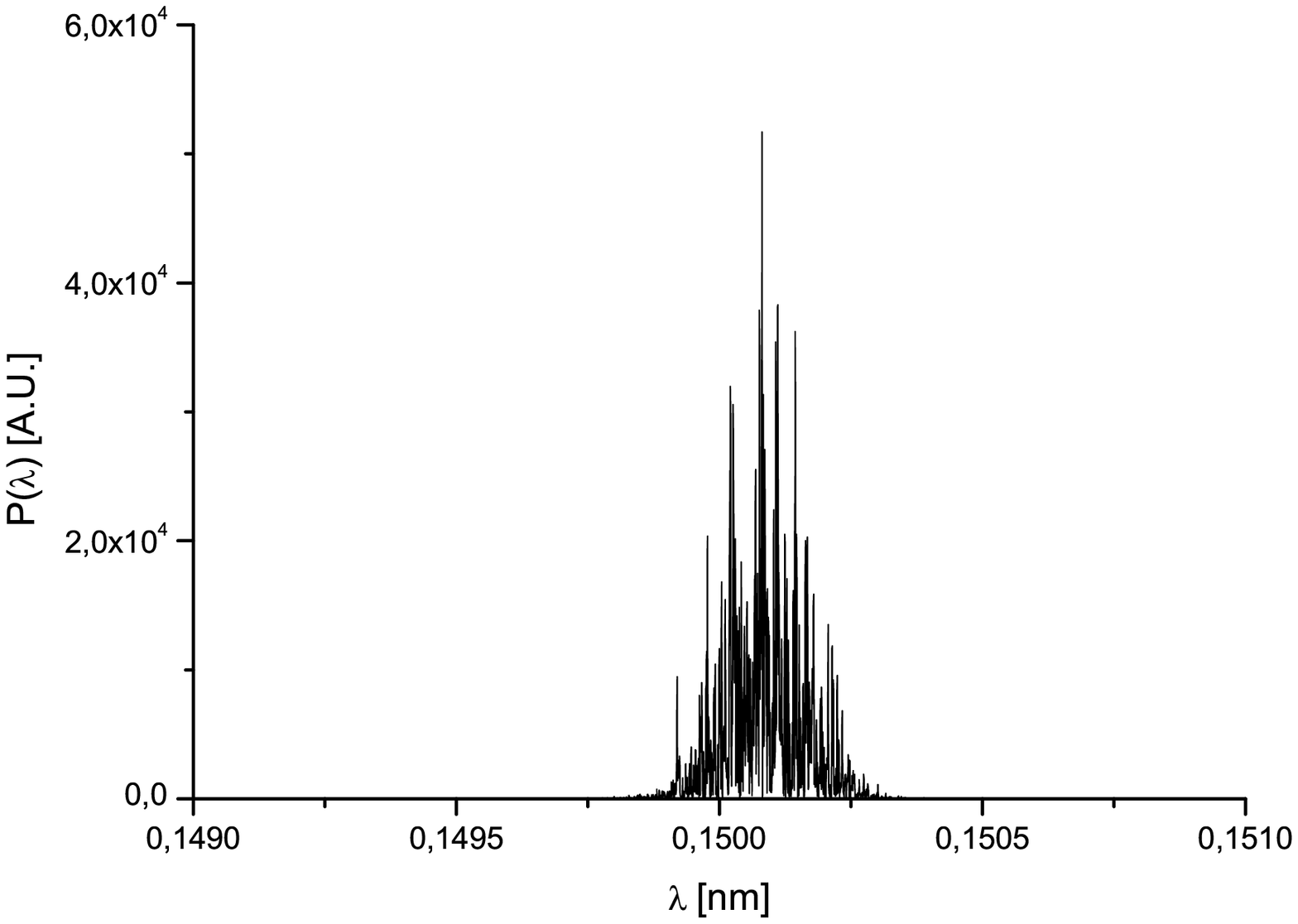}
\caption{Long pulse mode operation. Combination of self-seeding
and fresh bunch techniques. Spectrum at the end of the second
stage, 7 cells long (42 m). } \label{ml3-2}
\end{figure}

\begin{figure}[tb]
\includegraphics[width=1.0\textwidth]{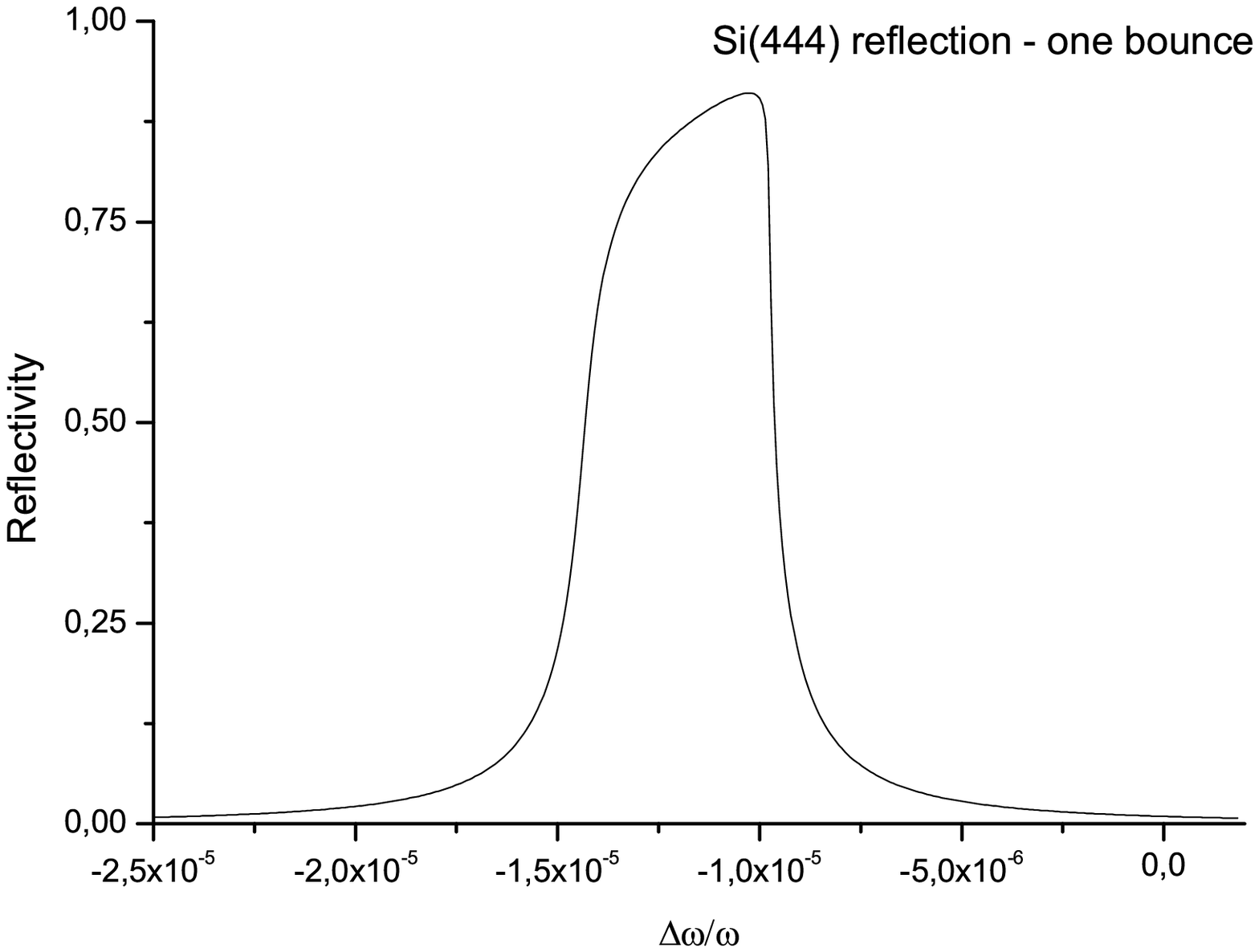}
\caption{Reflectivity curve for a thick absorbing crystal in Bragg
geometry. Si(444) reflection of 0.15 nm X-rays. One bounce.}
\label{m8885}
\end{figure}

\begin{figure}[tb]
\includegraphics[width=1.0\textwidth]{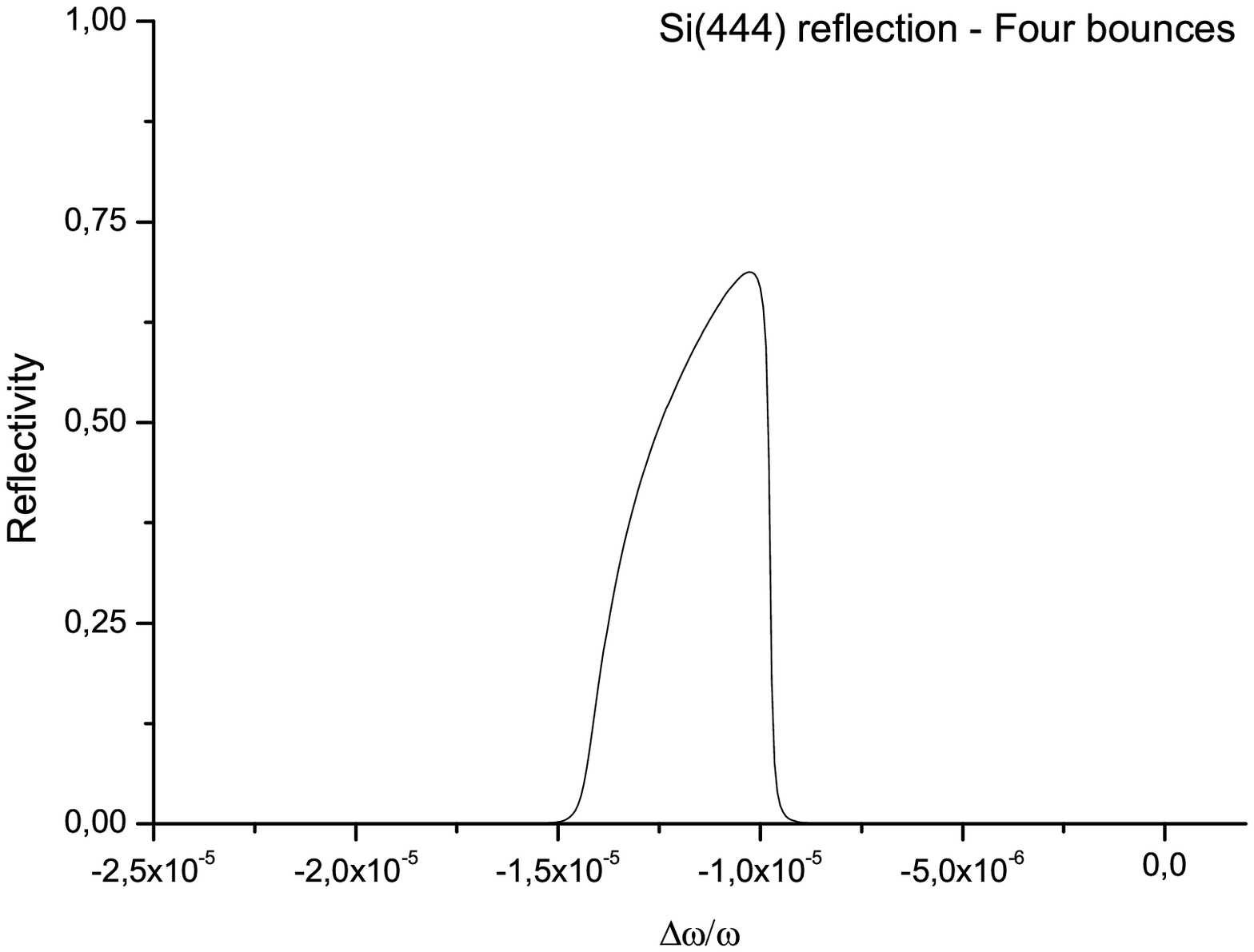}
\caption{Reflectivity curve for a thick absorbing crystal in Bragg
geometry. Si(444) reflection of 0.15 nm X-rays. Four bounces.}
\label{m8886}
\end{figure}

\begin{figure}[tb]
\includegraphics[width=1.0\textwidth]{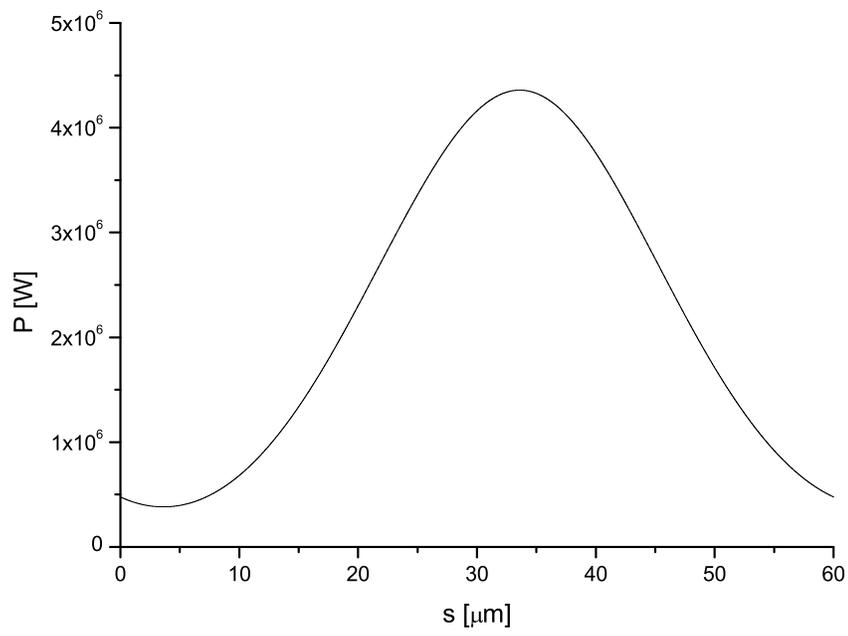}
\caption{Long pulse mode of operation, combination of self-seeding
and fresh bunch techniques. Power after the four-bounce
monochromator, Si(444) reflection. Radiation is used as input for
the third undulator part. } \label{ml3-3}
\end{figure}

\begin{figure}[tb]
\includegraphics[width=1.0\textwidth]{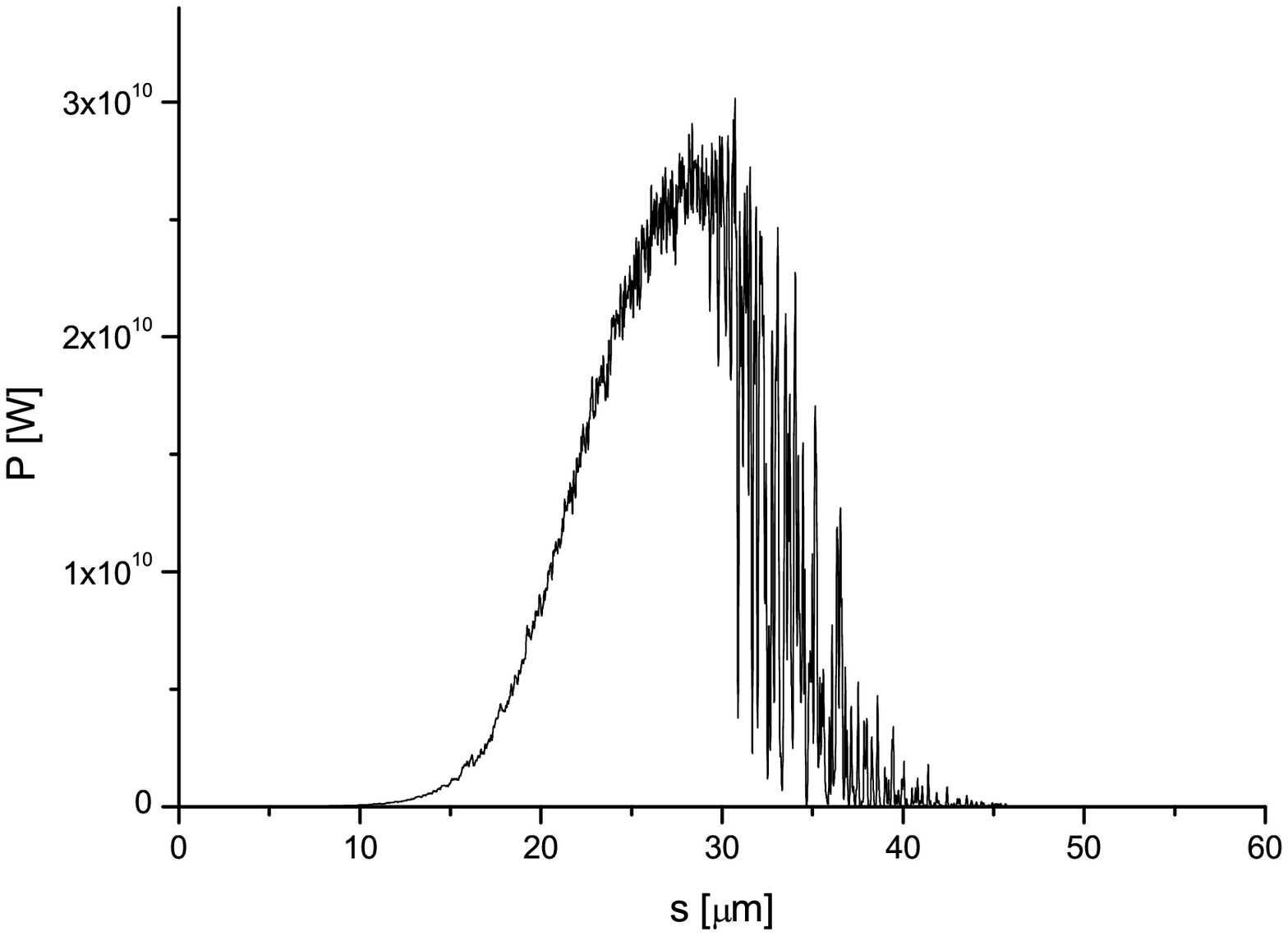}
\caption{Long pulse mode operation, combination of self-seeding
and fresh bunch techniques with a Si(444) reflection, four-bounce
monochromator. Power at the end of the third stage, 7 cells long
(42 m).} \label{ml3-4}
\end{figure}

\begin{figure}[tb]
\includegraphics[width=1.0\textwidth]{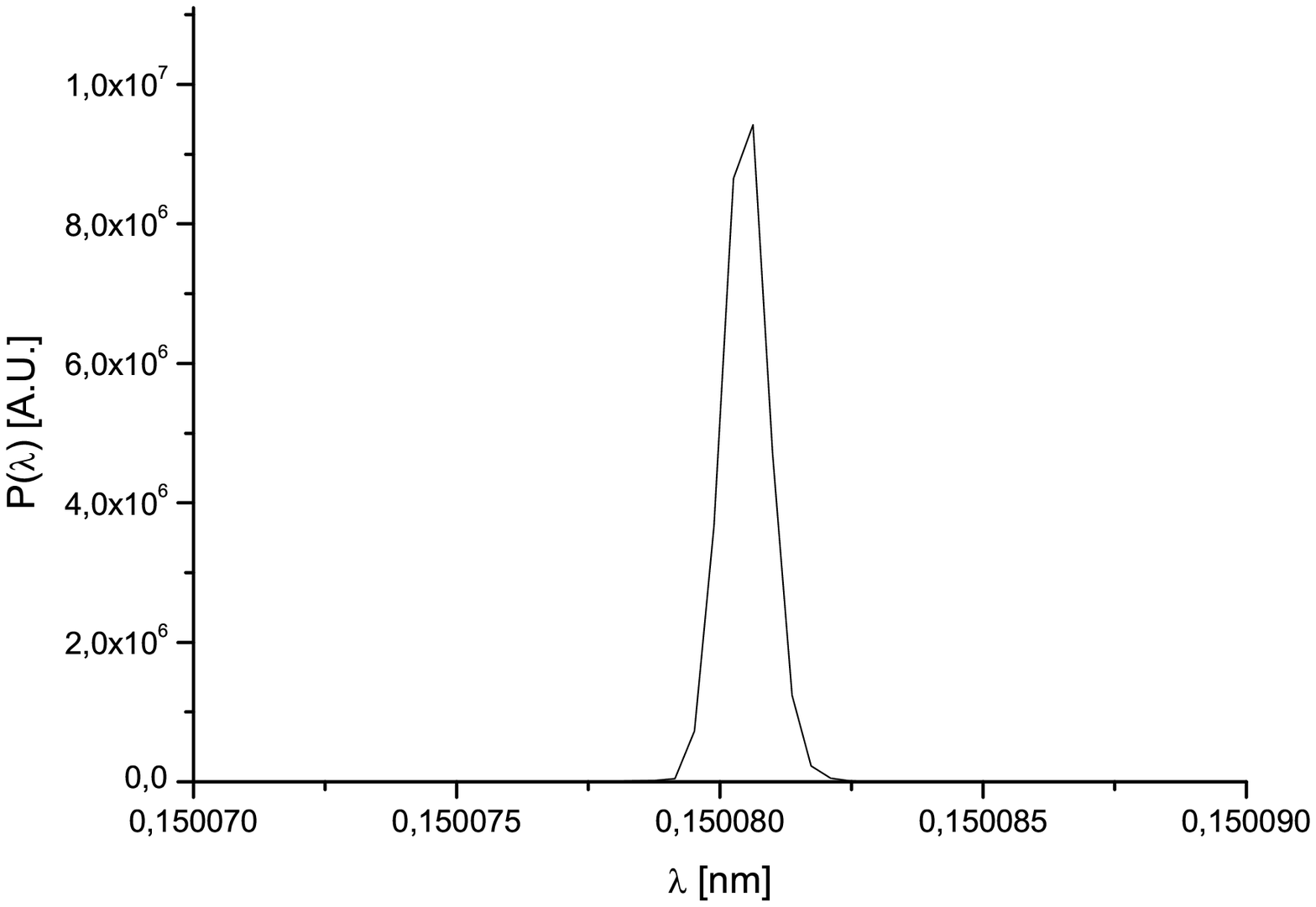}
\caption{Long pulse mode operation, combination of self-seeding
and fresh bunch techniques with a Si(444) reflection, four-bounce
monochromator. Spectrum at the end of the third stage, 7 cells
long (42 m).} \label{ml3-5}
\end{figure}

We finally considered a three-stage scheme, combining the
self-seeding scheme with a fresh-bunch technique. The first  stage
is 7-cells long. Fig. \ref{ml3-1} and and Fig. \ref{ml3-2}
illustrate power and spectrum  at the exit of second undulator
part, before the monochromator,  which is 7-cells long  too. The
monochromatization is performed using the Si(444) reflection. The
reflectivity curves relative to this case are shown in Fig.
\ref{m8885} and Fig. \ref{m8886} for one and four bounces
respectively. Fig. \ref{ml3-3} shows a typical single-shot power
profile after monochromatization. Finally, Fig. \ref{ml3-4} and
Fig. \ref{ml3-5} show the output of our scheme, respectively in
terms of power and spectrum.

\begin{figure}[tb]
\includegraphics[width=1.0\textwidth]{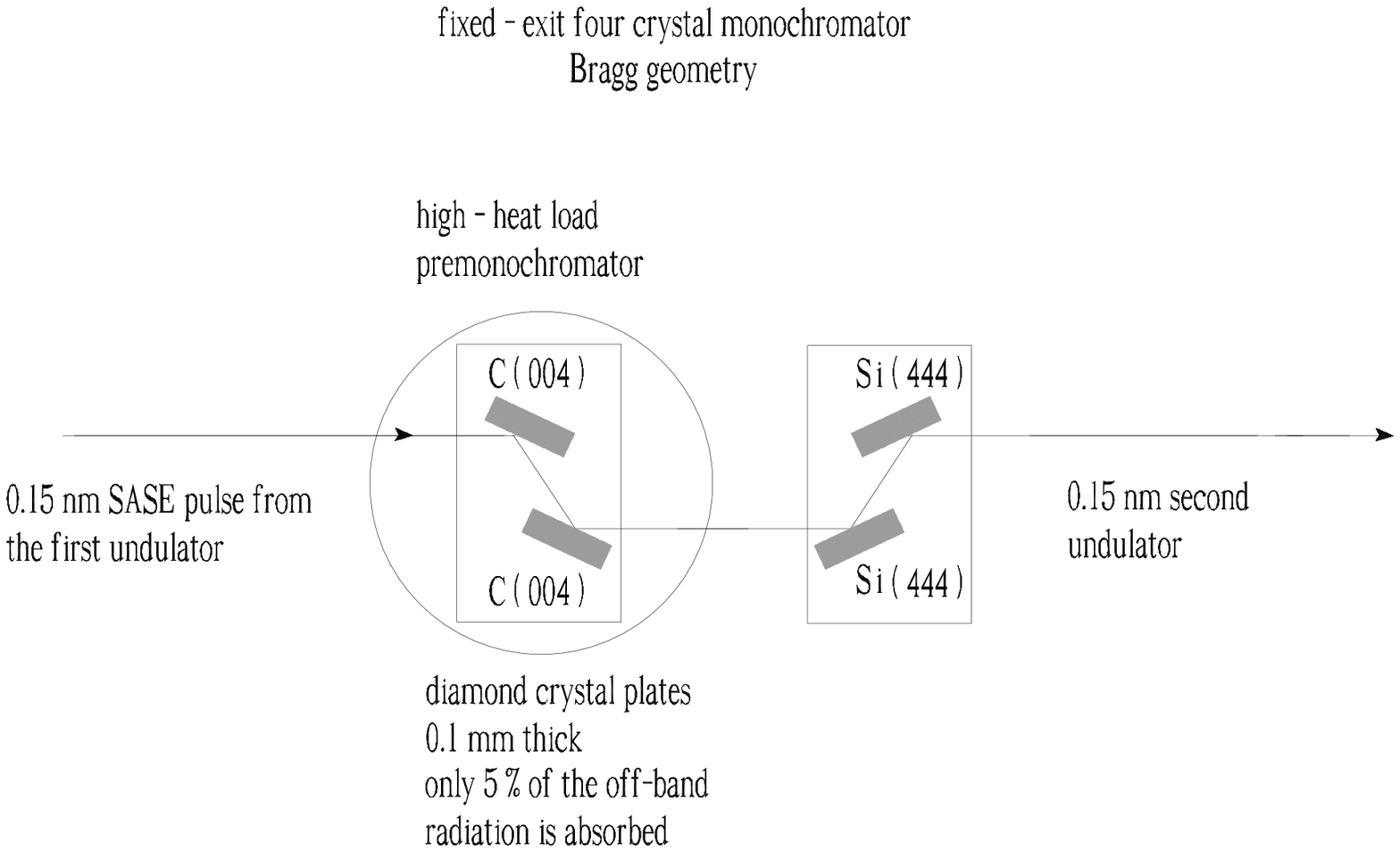}
\caption{Four bounce scheme for the X-ray monochromator. It allows
one to use the first two Bragg reflectors as a high-heat load
premonochromator. In the premonochromator part one can use diamond
crystal plates $100 \mu$m-thick and the Bragg reflection C(004).}
\label{ml117}
\end{figure}

It should be noted, that in this case the power density on the
monochromator is about $2 \mathrm{kW/mm^2}$ which is ten times
more than the typical heat load of monochromators in third
generation sources. To solve such problem, we suggest to use the
first two Bragg reflectors as a high heat-load premonochromator,
which drastically reduces the heat load from the actual high
resolution monochromator, constituted by the third and the fourth
crystal \cite{SXFE}. In the premonochromator one may take
advantage of diamond crystal plates about $100 \mu$m thick and of
the Bragg reflection C(004). Considering a perfect crystal, it
will reflect $99 \%$ of the incident X-rays. Only $5 \%$ of the
off-band radiation will be absorbed, while the rest will pass
through. The absorbed power is thus 20 times less than incident
power, resulting in an absorbed power density of about $100
\mathrm{W/mm^2}$, of the same magnitude of the heat load of
monochromators in third generation sources, Fig. \ref{ml117}.



Similarly to what has been done for the short-pulse case, it is
interesting to estimate the brilliance in the long-pulse case as
well. With the help of Fig. \ref{ml3-4} and Fig. \ref{ml3-5} we
obtain an additional increase of about one order of magnitude with
respect to the short-bunch case (i.e. two orders of magnitude with
respect to the baseline design) due to the narrower spectral width
of the Si(444) reflection, yielding a peak brilliance of the order
of $5 \cdot 10^{35} \mathrm{ph/s/mm^2/mrad^2/0.1\% BW}$.


\section{Conclusions}

In this paper we studied methods to reduce the line width of a
SASE X-ray FEL. One is based on the single bunch self-seeding
scheme. This method consists of two undulators with an X-ray
crystal monochromator located between them. The realization of
this single bunch self-seeding scheme for XFELs requires a 60
m-long electron bypass  to compensate for the X-ray path delay
introduced by the monochromator. This scheme is not compatible
with the baseline mode of operation of XFEL facilities, and as a
result it cannot be implement from the very beginning of the
operation stage. In this paper we propose a method to get around
this obstacle by using a double bunch scheme where radiation from
the first bunch seeds the second. This scheme overcomes the bypass
problem, although research and development is required to develop
a suitable laser-pulse doubler. Combination with a fresh-bunch
technique is discussed too, and will be particularly important
during the commissioning phase. We discussed both short and
long-pulse scenarios, and presented a feasibility study which can
be applied to all the considered techniques.

An estimation of the peak brightness of the pulses produced with
our methods shows an increase up to $5 \cdot 10^{35}
\mathrm{ph/s/mm^2/mrad^2/0.1\% BW}$, i.e. two orders of magnitude
with respect to the baseline parameters. However, it should be
noted that the brilliance may not be an adequate figure of merit
to consider when we deal with fully longitudinally and
transversely coherent x-ray pulses with a duration of $4 - 40$ fs
(respectively short pulse and long pulse mode of operation),
energy per pulse in the order of $0.1-1$ mJ, with shot-to-shot
energy fluctuations within $10\%$ and time jitter of about $50$ fs
(according to the LCLS results). These parameters are near to
those usually reported for high-power fs laser systems, only in
the x-ray regime.

Finally, it should be remarked that the self-seeding based on a
double bunch presented in this paper can be implemented without
disturbing the baseline mode of operation of the XFEL facility.
This study is useful for the European XFEL, as well as for other
present or future XFEL facilities.

\section{Acknowledgements}

We are grateful to Massimo Altarelli, Reinhard Brinkmann, Serguei
Molodtsov and Edgar Weckert for their support and their interest
during the compilation of this work.

\section*{Appendix: Influence of phase
variation of X-ray waves along the reflectivity curve in Bragg
geometry on the temporal profile of the seed X-ray pulse}

In this Appendix we investigate the effects of the monochromator
phase on the temporal shape of the radiation pulse. In fact, up to
now we assumed that the monochromator is composed of real filters,
while in reality these filters include a phase. Here we show that
the effects of such pahse is negligible. In particular, we will
concentrate on the Si(400) reflection. In general, according to
the dynamical theory of X-ray diffraction, the phase $\Phi(\eta)$
and amplitude $Z(\eta)$ introduced by a one-bounce reflection on
any crystal, neglecting absorption, are given by \cite{AUTH}:

\begin{eqnarray}
Z(\eta)  = \left|\eta \pm \sqrt{\eta^2 - 1}\right| \label{ampli}
\end{eqnarray}
\begin{eqnarray}
\Phi(\eta)  = \mathrm{arg}\left[-\eta + \sqrt{\eta^2 -
1}~\right]~.\label{phase}
\end{eqnarray}
Here the positive or negative sign in Eq. (\ref{ampli}) should be
taken such that $Z<1$. In literature $\eta$ is the deviation
parameter. If absorption is neglected as in our case, such
parameter varies from $-1$ to $+1$ when the reflectivity is
$100\%$. Multiplication by the bandwidth $\Delta
\omega_{nkl}/\omega$, which depends on the reflection, yields the
reflectivity curve and the phase as a function of the deviation
from the center of reflectivity curve, $\Delta \omega/\omega =
\eta \Delta \omega_{nkl}/\omega$. Thus, in our case we can say
that $\eta$ represents a reduced frequency deviation.

\begin{figure}[tb]
\includegraphics[width=1.0\textwidth]{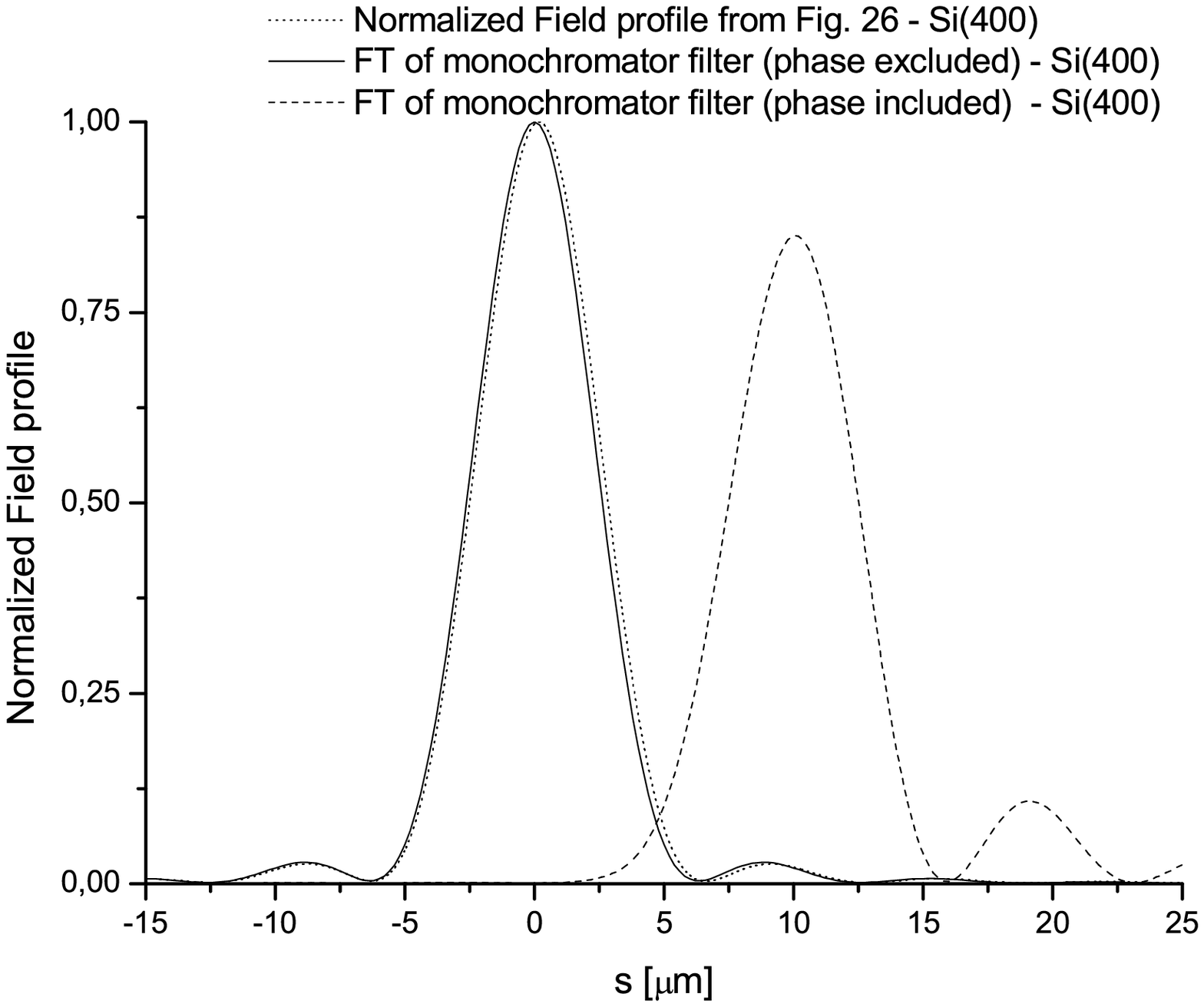}
\caption{Normalized field profile after the four-bounce Si(400)
reflections for the short bunch case, single shot. We compare the
results from Fig. \ref{m34} with the Fourier tansform of the
monochromator filter with and without phase.} \label{a1}
\end{figure}

Starting with the reflectivity curves in Fig. \ref{mf4} we matched
the centers with $\eta$ = 0, and the FWHM with $\eta = \pm 1$. As
a result we fixed the phase change along the curve according to
Eq. (\ref{phase}). Then, for a four-bounce reflection one needs to
multiply the phase by a factor four. Now amplitude and phase are
known, and we can calculate the temporal pulse profile by taking
the Fourier Transform of the complex monochromator line. By this
we neglect effects of absorption on the phase, which are anyway
small. A comparison of the results obtained with and without phase
is shown in Fig. \ref{a1}. The linear component of the phase is
responsible for a temporal shift, and non-linear components modify
the result in a way which may be taken into account in more
detailed studies. We conclude that for the purpose of giving a
feasibility study of our technique it is sufficient to account for
the real part of the filter line only.

\end{document}